\documentclass[fleqn,usenatbib]{mnras}

\usepackage[T1]{fontenc}

\DeclareRobustCommand{\VAN}[3]{#2}
\let\VANthebibliography\thebibliography
\def\thebibliography{\DeclareRobustCommand{\VAN}[3]{##3}\VANthebibliography}

\usepackage{graphicx}	
\usepackage{amsmath}	
\usepackage{amssymb}	
\usepackage{commath}    
\usepackage{physics}    
\usepackage[singlelinecheck=false]{caption}
\usepackage{subcaption} 
\usepackage{mathtools}  
\usepackage{threeparttable}  
\usepackage{booktabs}   
\usepackage{etoolbox}   
\usepackage{tabulary}   
\usepackage{enumitem}   
\setlist[itemize]{wide,leftmargin=\parindent,labelindent=0.3\parindent,itemindent=0\parindent,labelsep=0\parindent,labelwidth=!,listparindent=0\parindent,itemsep=0\parindent,parsep=0.5\topsep,itemsep=\topsep}
\setlist[enumerate]{wide=0pt, widest=99,leftmargin=\parindent, labelsep=*}

\newcommand{\onthermal}[1][]{on-thermal#1 }
\newcommand{\onradiative}[1][]{on-radiative#1 }
\newcommand{\onrelativistic}[1][]{on-relativistic#1 }
\newcommand{\lc}[1][]{light crossing#1 }
\newcommand{\ts}[1][]{time-scale#1 }
\newcommand{\ls}[1][]{length-scale#1 }
\newcommand{\cutoff}[1][]{cut-off#1 }

\newcommand{\nontrivial}[1][]{non-trivial#1 }
\newcommand{\onnegligible}[1][]{on-negligible#1 }
\newcommand{\quoted}[2][]{`#2'#1 }

\newcommand{\mphasiz}[1][]{mphasiz#1 }
\newcommand{\eneraliz}[1][]{eneraliz#1 }
\newcommand{\ormaliz}[1][]{ormalis#1 }
\newcommand{\ocaliz}[1][]{ocalis#1 }
\newcommand{\nalyz}[1][]{nalys#1 }
\newcommand{\crosssection}[1][]{cross-section#1 }
\newcommand{\haracteriz}[1][]{haracteriz#1 }

\newcommand{\ummariz}[1][]{ummariz#1 }

\newcommand{\ptimiz}[1][]{ptimiz#1 }
\newcommand{\ynchroniz}[1][]{ynchronis#1 }
\newcommand{\tabiliz}[1][]{tabilis#1 }
\newcommand{\isk}[1][]{isc#1 }
\newcommand{\pecializ}[1][]{pecialis#1 }
\newcommand{\ealiz}[1][]{ealiz#1 }
\newcommand{\ling}[1][]{lling#1 }
\newcommand{\led}[1][]{lled#1 }
\newcommand{\spellor}[1][]{our#1 }
\newcommand{\speller}[1][]{re#1 }

\newcommand{\zeltron}[1][]{\mbox{\textsc{zeltron}}#1 }
\newcommand{\codename}[2][]{\mbox{\textsc{#2}}#1 }

\newcommand{\me}{\ensuremath{m_e}}
\newcommand{\massp}{\ensuremath{m_p}}
\newcommand{\eph}{\ensuremath{\epsilon_{\rm ph}}}
\newcommand{\uph}{\ensuremath{U_{\rm ph}}}
\newcommand{\nph}{\ensuremath{n_{\rm ph}}}
\newcommand{\gmax}{\ensuremath{\gamma_{\rm max}}}
\newcommand{\gx}{\ensuremath{\gamma_{\rm X}}}
\newcommand{\gcool}{\ensuremath{\gamma_{\rm cool}}}

\newcommand{\tcoolt}{\ensuremath{t_{\rm cool,T}}}
\newcommand{\tcoolk}{\ensuremath{t_{\rm cool,IC}}}
\newcommand{\gradt}{\ensuremath{\gamma_{\rm rad,T}}}
\newcommand{\gradk}{\ensuremath{\gamma_{\rm rad,IC}}}
\newcommand{\gkn}{\ensuremath{\gamma_{\rm KN}}}

\newcommand{\sigc}{\ensuremath{\sigma_{\mathrm{c,}0}}}
\newcommand{\sigh}{\ensuremath{\sigma_{\mathrm{h,}0}}}
\newcommand{\sigcgen}{\ensuremath{\sigma_{\rm c}}}
\newcommand{\sighgen}{\ensuremath{\sigma_{\rm h}}}
\newcommand{\sigcgg}{\ensuremath{\sigma_{\rm c}^{(\gamma\gamma)}}}

\newcommand{\knp}{\ensuremath{q}}
\newcommand{\gknp}{\ensuremath{y}}
\newcommand{\gstitch}{\ensuremath{\gamma_{\rm Th-KN}}}
\newcommand{\rng}{\ensuremath{\mathcal{R}}}
\newcommand{\escat}{\ensuremath{\epsilon_{\rm scat}}}
\newcommand{\ehard}{\ensuremath{\epsilon_{\rm hard}}}
\newcommand{\thinc}{\ensuremath{\psi}}
\newcommand{\einc}{\ensuremath{\epsilon_{0}}}
\newcommand{\pfid}{\ensuremath{p_0}}
\newcommand{\fdrift}{\ensuremath{f_{\rm d}}}
\newcommand{\gamc}{\ensuremath{\gamma_{\rm c}}}
\newcommand{\gamppmax}{\ensuremath{\gamma_{\rm pp,max}}}
\newcommand{\fluxsymb}{\ensuremath{\Phi}}
\newcommand{\eblr}{\ensuremath{\epsilon_{\rm BLR}}}
\newcommand{\ehdr}{\ensuremath{\epsilon_{\rm HDR}}}

\newcommand{\taus}{\ensuremath{\tau_{\gamma\gamma}^{\rm(syn)}}}

\newcommand{\estar}{\ensuremath{\epsilon_{\star}}}
\newcommand{\ustar}{\ensuremath{U_{\star}}}
\newcommand{\gknstar}{\ensuremath{\gamma_{\rm KN,w}}}
\newcommand{\gmaxstar}{\ensuremath{\gamma_{\rm max,w}}}
\newcommand{\gradtstar}{\ensuremath{\gamma_{\rm rad,T}^{\rm (w)}}}
\newcommand{\gcoolstar}{\ensuremath{\gamma_{\rm cool,w}}}
\newcommand{\tauggstar}{\ensuremath{\tau_{\gamma\gamma,\rm w}}}
\newcommand{\ricval}{60}
\newcommand{\sigcstar}{\ensuremath{\sigma_{\rm c,0}^{\rm (w)}}}

\newcommand{\idense}{\ensuremath{n_0}}
\newcommand{\itemp}{\ensuremath{\theta_0}}
\newcommand{\itempt}{\ensuremath{T_0}}

\DeclareMathOperator{\Li}{Li}

\newcommand{\myvec}[1]{\pmb{#1}}
\newcommand{\myunit}[1]{\pmb{\hat{#1}}}

\usepackage{newtxtext,newtxmath}

\title[Klein-Nishina reconnection simulations]{Kinetic simulations and gamma-ray signatures of Klein-Nishina relativistic magnetic reconnection}

\author[J. M. Mehlhaff et al.]{
J. Mehlhaff,$^{1,2}$\thanks{E-mail: john.mehlhaff@univ-grenoble-alpes.fr}
G. Werner,$^{2}$
B. Cerutti,$^{1}$
D. Uzdensky,$^{2}$
M. Begelman$^{3, 4}$
\\
$^{1}$Univ. Grenoble Alpes, CNRS, IPAG, 38000 Grenoble, France\\
$^{2}$Center for Integrated Plasma Studies, Physics Department, 390 UCB, University of Colorado, Boulder, CO 80309, USA\\
$^{3}$JILA, University of Colorado and National Institute of Standards and Technology, 440 UCB, Boulder, CO 80309, USA\\
$^{4}$Department of Astrophysical and Planetary Sciences, 391 UCB, Boulder, CO 80309, USA
}

\date{Accepted XXX. Received YYY; in original form ZZZ}

\pubyear{2023}

\begin{document}
\label{firstpage}
\pagerange{\pageref{firstpage}--\pageref{lastpage}}
\maketitle

\begin{abstract}
    Black hole and neutron star environments often comprise collisionless plasmas immersed in strong magnetic fields and intense baths of low-frequency radiation. In such conditions, relativistic magnetic reconnection can tap the magnetic field energy, accelerating high-energy particles that rapidly cool by inverse Compton (IC) scattering the dense photon background. At the highest particle energies reached in bright gamma-ray sources, IC scattering can stray into the Klein-Nishina regime. Here, the Comptonized photons exceed pair-production threshold with the radiation background and may thus return their energy to the reconnecting plasma as fresh electron-positron pairs. To reliably characterize observable signatures of such Klein-Nishina reconnection, in this work, we present first-principles particle-in-cell simulations of pair-plasma relativistic reconnection coupled to Klein-Nishina and pair-production physics. The simulations show substantial differences between the observable signatures of Klein-Nishina reconnection and reconnection coupled only to low-energy Thomson IC cooling (without pair production). The latter regime exhibits strong harder-when-brighter behaviour; the former involves a stable spectral shape independent of overall brightness. This spectral stability is reminiscent of flat-spectrum radio quasar (FSRQ) GeV high states, furnishing evidence that Klein-Nishina radiative physics operates in FSRQs. The simulated Klein-Nishina reconnection pair yield spans from low to order-unity and follows an exponential scaling law in a single governing parameter. Pushing this parameter beyond its range studied here might give way to a copious pair-creation regime. Besides FSRQs, we discuss potential applications to accreting black hole X-ray binaries, the M87$^*$ magnetosphere, and gamma-ray binaries.
\end{abstract}


\section{Introduction}
The gamma-ray sky is studded with relativistic compact objects -- neutron stars and black holes (of which the most numerous observed varieties are, respectively, pulsars and blazars: \citealt{tevcat, fermi20a}). These systems -- and connected phenomena including winds, jets, and accretion d\isk[s]-- frequently host collisionless highly magnetized plasmas, with magnetic energy density exceeding not just the pressure (i.e.\ small plasma beta) but also the rest-mass energy density of the charge-carrying particles. In such plasmas, relativistic magnetic reconnection \citep{bf94, lu03, l05, wy06} efficiently siphons off the excess magnetic field energy, using it to accelerate relativistic particles and drive relativistic collective motion. The energized particles are then revealed by the light that they shine toward Earth, including in the gamma-ray band. Relativistic magnetic reconnection is, hence, an important candidate mechanism for powering high-energy phenomena linked to the most compact objects in the Universe.

In some astrophysical situations, there is a well-defined \ts[]separation between abrupt reconnection-powered particle acceleration and much slower radiative losses. In this radiatively inefficient regime, observable emission traces particle energization that has occurred in the past. This limit is seldom r\ealiz[ed,]however, in the plasma environments of compact objects, where intense magnetic and radiation fields lead to rapid synchrotron and inverse Compton (IC) cooling. Then, the problem can no longer be cleanly factorized into a sudden acceleration step followed by a more prolonged cooling stage. Instead, radiative cooling couples in real time to reconnection, tracing active (as opposed to past) particle acceleration and feeding back on the reconnection process: a qualitatively distinct radiative reconnection regime \citep{u11, u16, mwu21}.

In the low-energy, optically thin limit where the synchrotron and IC photons freely escape the system -- what we might call \textit{classical radiative reconnection} -- radiative losses, while dynamically important, do not change the fundamental flow of energy from the n\onradiative[]case. Magnetic fields serve as the main energy source for particle acceleration, while the primary energy sink is the emission mechanisms that efficiently and permanently remove liberated magnetic energy from the system. However, for the gamma-ray-bright relativistic compact objects, the photons emitted at the highest energies are above threshold for various quantum electrodynamical (QED) pair production channels. This fundamentally alters the pathways available to the energy in radiative reconnection, allowing radiation to not just carry energy away from the reconnection site, but also to redistribute it in real time in the form of freshly produced electron-positron pairs. Such \textit{QED radiative reconnection} is thus distinguished from merely classical radiative reconnection in its capacity to alter the plasma material composition and in the key role played by photons as a dynamically important particle species \citep{u11, u16, ubb19}.

To interpret observations of systems where QED reconnection may occur, mode\ling[]efforts must employ a self-consistent kinetic plasma description. Such a description is already needed to model collisionless relativistic magnetic reconnection in the n\onradiative[]and classical radiative regimes, because it captures the critical microphysics governing the reconnection rate (i.e.\ the rate of magnetic energy dissipation) as well as the production of n\onthermal[]particle energy distributions and correspondingly n\onthermal[]emission spectra. In the case of QED reconnection, a kinetic paradigm is even more imperative. The QED \crosssection[s]depend sensitively on the energies of both the emitting particles and the pair-producing photons, placing an even greater importance on capturing energization self-consistently.

All of the necessary kinetic physics can be incorporated by augmenting \textit{ab initio} particle-in-cell (PIC) simulations \citep{bl05} with QED physics. The small number of PIC reconnection studies that have done this have focused on a select few QED interactions. For example, \citet{hps19} present a regime, expected in pulsar magnetospheres, where particles suffer strong synchrotron cooling, and the resulting synchrotron photons, with an emission spectrum peaking at~$\gtrsim \, \rm MeV$ energies, collide with one another, leading to copious pair production in the reconnection inflow region \citep[see also][]{hrp23}. \citet{sgu19} and \citet{sgu23} also study reconnection with efficient synchrotron radiation. However, they consider pair production not between colliding synchrotron photons, but from the absorption of single synchrotron photons by an intense electromagnetic field, as may occur in magnetar magnetospheres. The radiative cooling removes particle pressure support in the reconnection layer, leading to strong plasma compression there. This locally amplifies synchrotron emission and pair creation, with the end result that pair production in this regime is concentrated not in the plasma fue\ling[]reconnection, but in the heart of the reconnection layer itself. Lastly, \citet{ccd21} and \citet{ccd22} present global models of reconnection in black hole magnetospheres, where the primary radiation mechanism is IC scattering of low-energy (soft) background photons originating from a larger-scale accretion flow. Pair production then occurs when Comptonized photons collide with the soft parent population from which they were first scattered, supplying the plasma to a luminous equatorial reconnection current sheet in the black hole magnetosphere.

The examples above illustrate two general points. First, while QED reconnection is of general high-energy astrophysical interest, the relevant QED interactions depend on the system under study. Second, s\pecializ[ing]to certain QED interactions over others not only decides the applicable astrophysical sources, but can also lead to divergent qualitative dynamics. These remarks underscore the need to understand QED reconnection -- in all its astrophysical diversity -- as a fundamental physics problem in order to identify its observable signatures in the high-energy Universe. 

In this work, we present PIC simulations run using the \zeltron[]code \citep[][plus needed auxiliary developments detailed here]{cwu13, cw19} of a QED reconnection regime thus far unexplored from first principles as an isolated physics problem. We consider a relativistic magnetic reconnection layer immersed in such an intense bath of soft background radiation that IC scattering strays far into the QED limit. This contrasts the (classical radiative) low-energy, Thomson IC reconnection regime previously studied numerically by \citet{wpu19}, \citet{mwu20}, \citet{sb20}, \citet{ssb21}, and \citet{ssb23} in two respects. First, we account for quantized gamma-ray emission from the highest-energy particles radiating in the Klein-Nishina IC limit \citep{j68, bg70}. Second, because many of the emitted photons lie above pair-production threshold with the soft background \citep{mwu21}, we calculate pair production between the few (low number density) scattered gamma-rays with energies~$\escat \gg \me c^2$, and the abundant (high number density) soft seed photons with energies~$\eph \ll \me c^2$. We refer to this r\ealiz[ation]of QED reconnection as \textit{Klein-Nishina radiative reconnection} (sometimes just \textit{Klein-Nishina reconnection}), omitting explicit reference to pair production since efficient Klein-Nishina IC emission implies pair production in a reconnection context \citep{mwu21}.

The QED interactions studied here are the same as those treated by \citet{ccd21} and \citet{ccd22}. Here, however, we take a complementary approach, stripping away the global morphology and studying reconnection as a local problem. Computationally, this enables us to concentrate resources toward enhancing the separation among the radiative and plasma microscales. Physically, it permits us to remain more agnostic to the host system, focusing instead on the intrinsic reconnection properties that may be generally applicable.

Indeed, Klein-Nishina reconnection may be r\ealiz[ed]in a range of astrophysical systems connected to relativistic compact objects, including: flat-spectrum radio quasars (FSRQs), where reconnection occurring in a relativistic jet launched from an active galactic nucleus (AGN) is likely externally illuminated by large-scale circumnuclear structures \citep[cf.][]{gub09, ngb11, nbc12, g13, sgp16, pgs16, wub18, cps19, cps20, gu19, on20, mwu20, mwu21}; the high/soft states of accreting black hole X-ray binaries (BHXRBs), where reconnection in a highly magnetized collisionless coronal region is illuminated by an underlying geometrically thin, optically thick accretion d\isk[]\citep[cf.][]{grv79, d98, ug08, gu08, hl12, u16, b17, wpu19, sb20, ssb21, ssb23, mwu21, ecc22}; the magnetospheres of supermassive black holes, particularly the one at the center of the M87 galaxy, M87$^*$, wherein reconnection may be bathed in photons from a large-scale radiatively inefficient accretion flow \citep[cf.][]{gub10, bop16, lyw17, rbp20, rlc22, ccd21, ccd22, ecc22, ecc23, sdb22, hrp23, cud23, gpq23}; and gamma-ray binaries, where a plausible scenario involves a pulsar in tight orbit around a bright type O or Be star, which illuminates reconnection occurring near the pulsar in its magnetosphere and striped wind (that is, before the pulsar wind shocks with the stellar wind from the companion; cf.\ \citealt{d06, cdh08, d13, dgp17, cp17, ps18_philippov, cpd20}). The link between Klein-Nishina reconnection and each of these object classes is, in fact, a major result of the present work (section~\ref{sec:discussion}), as further discussed below.

While this study is primarily numerical, analytic and semi-analytic mode\ling[]are also vital for understanding Klein-Nishina reconnection and QED reconnection more broadly. Such theoretical approaches can make targeted, physically motivated arguments for how results from non-QED reconnection may g\eneraliz[e]to the QED case \citep[e.g.][]{b17, mwu21, hrp23, cud23}, even if they cannot treat all of the kinetic physics at play from first principles. This furnishes a useful interpretive framework for \textit{ab initio} simulations. However, the reverse is also true: phenomenological models, which sometimes have the advantage of enhanced physical clarity, can themselves be refined from the findings of simulations. The present study illustrates both directions of this paradigm. Throughout the text, we make frequent reference to our earlier work, \citet{mwu21}, which analytically considers the setup simulated here. As will be seen, that study (besides laying much of the theoretical foundation for the present article) serves both as an interpretive lens for our simulations and as a set of hypotheses that the numerical experiments can check.

We structure this article as follows. In section~\ref{sec:arch}, we detail the QED algorithmic developments that enable our PIC simulations. Then, in section~\ref{sec:setup}, we describe our simulation setup in detail. In section~\ref{sec:results}, we present the results of our simulations, comparing and contrasting Klein-Nishina radiative reconnection to two control cases: one of n\onradiative[]reconnection and one of classical radiative reconnection subject to efficient Thomson IC losses. Section~\ref{sec:pairyield} then provides a second results section. However, there, instead of delving into a detailed analysis of a few simulations with different radiative physics, we conduct parameter scans with all of the QED physics turned on, c\haracteriz[ing]the pair yield of reconnection -- a single number computed per simulation -- as a function of its main contro\ling[]parameters. In section~\ref{sec:discussion}, we survey observations of the four main application systems targeted by this work -- \mbox{FSRQs}, \mbox{BHXRBs}, the~M87$^*$ magnetosphere, and gamma-ray binaries -- discussing connections to our simulation results. We conclude with a complete summary of our findings in section~\ref{sec:conclusions}. In the remaining part of this Introduction, we preview the three principal astrophysical results of this study.

The first concerns the correlated spectral and temporal signatures of Klein-Nishina reconnection (section~\ref{sec:afterglow}). As in the n\onradiative[]and classical (Thomson IC) radiative regimes, Klein-Nishina reconnection powers efficient n\onthermal[]particle acceleration (NTPA) and, hence, n\onthermal[]radiative emission. While the time-averaged observable spectrum is similar to that of Thomson IC reconnection, the relationship between the shape of the output spectral energy density and its luminosity are very different. In Thomson IC reconnection, these two are tightly correlated, with a shallower observed spectrum coinciding with a higher luminosity (i.e.\ \quoted[).]{harder-when-brighter}In Klein-Nishina reconnection, this correlation is broken: the spectrum exhibits a constant shape irrespective of overall brightness. These results, potentially observable during gamma-ray flares, represent an important  distinguishing property of Klein-Nishina reconnection and illustrate the value of temporally resolved observed spectra.

Our second main astrophysical result concerns the electron-positron pair yield of Klein-Nishina reconnection (section~\ref{sec:pairyield}). Using simulations, we derive an empirical formula for the Klein-Nishina reconnection pair yield in terms of a single control parameter. While, in the regime probed by this study, reconnection generally produces, at most, order-unity new pairs per processed pair, our derived scaling law, together with physical arguments for its extrapolation, point to a potential regime where Klein-Nishina reconnection may be a copious pair source. However, even in the case of order-unity pair yield, Klein-Nishina reconnection can still convert an initially electron-ion plasma into a strongly mixed electron-ion-positron plasma. This reconnection regime is thus a potentially important \textit{in-situ} antimatter source in astrophysics.

Our final main astrophysical result is a detailed survey of the four main object classes -- \mbox{FSRQs}, \mbox{BHXRBs}, the~M87$^*$ magnetosphere, and gamma-ray binaries -- where Klein-Nishina reconnection may occur (section~\ref{sec:discussion}). We find potentially strong observational connections to GeV observations of FSRQs, where observed spectral stability during flares is reminiscent of the anticipated spectral-temporal signatures of Klein-Nishina reconnection. We also sketch a roadmap of the theoretical and instrumental developments necessary to link Klein-Nishina reconnection mode\ling[]more rigorously to observations in the other systems. We comment on the potential effect of the Klein-Nishina reconnection pair yield on the global operation and observable aspects of each examined object type.

\section{Simulation architecture}
\label{sec:arch}
We here detail the new QED capabilities that we added to the radiative electromagnetic PIC code \zeltron[]\citep{cwu13,cw19} to enable the simulations presented in this article. Readers wishing to skip these technical details may proceed directly to section~\ref{sec:setup}. Excellent additional references on QED methods in PIC simulations can be found in the literature documenting other PIC codes commonly used in astrophysics, including: \codename{tristan v2}\citep[e.g.][]{hps19, hcs23}, \codename{osiris}\citep[e.g.][]{fst02, dgf20}, and \codename{grzeltron}\citep[e.g.][]{lc18, ccp20}.

\subsection{The QED PIC method}
\label{sec:qedpic}
\begin{figure}
    \centering
    \includegraphics[width=\columnwidth]{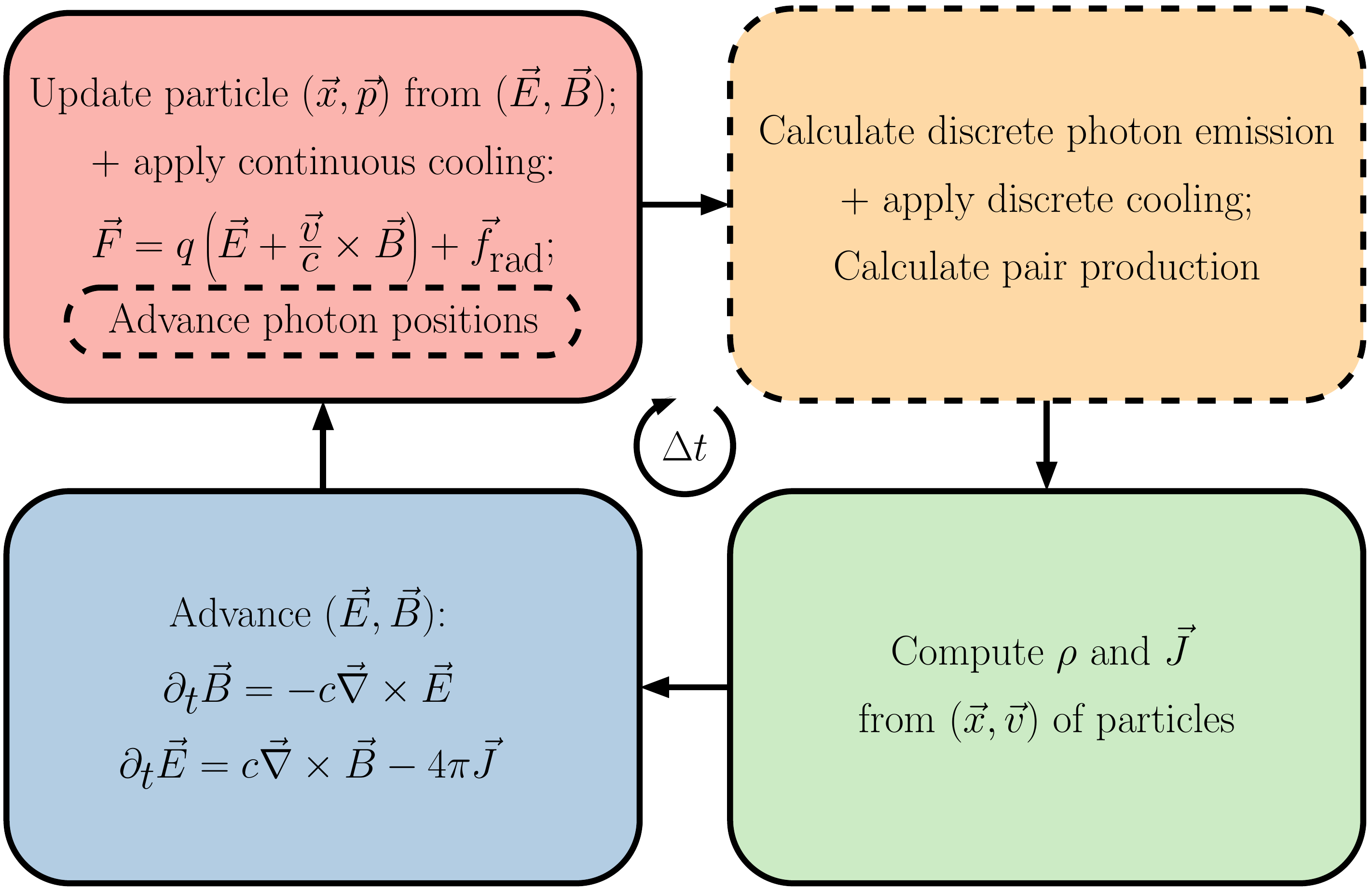}
    \caption{Standard PIC loop augmented with steps (dashed outline) to model QED effects. At each PIC timestep, the electromagnetic field -- plus an optional continuous radiative drag force to model non-QED cooling -- are used to update the particles' positions and momenta according to the Lorentz force law (red panel). Photons are also ballistically propagated at the speed of light in this step. Then (orange panel), QED \crosssection[s]are used to compute Monte Carlo photon emission and pair production. Each time a Monte Carlo photon is emitted, its momentum is self-consistently subtracted from that of its radiating particle. Freshly emitted photons and produced pairs are added to the simulation. In the next step (green panel), the particle positions and velocities are used to calculate the electromagnetic charge and current densities. These are used in the final step (blue panel) to update the electromagnetic fields via the time-dependent Maxwell's equations. Adapted from~\citet{sc17}.}
    \label{fig:qedpicloop}
\end{figure}

To provide some context for the modifications we have made to the \zeltron[]code, we first review salient general features of electromagnetic PIC codes, discussing how they may be extended to include QED effects. Fig.~\ref{fig:qedpicloop} provides a graphic summary of this discussion.

The electromagnetic PIC (hereafter, simply \quoted[)]{PIC}technique is a computational method for simulating first-principles kinetic plasma physics. PIC simulations are kinetic in that they self-consistently describe the full phase space (position+momentum) plasma distribution function (in contrast to, for example, fluid plasma frameworks, which track bulk quantities -- such as spatial density and local mean velocity -- in real space only). The PIC technique is, furthermore, a first-principles method because it evolves physical equations (the Maxwell-Vlasov system) requiring minimal approximations. Owing to these properties, PIC simulations can probe detailed microscopic plasma physical effects, while furnishing vital, self-consistent astrophysical observables such as lightcurves and spectra.

The PIC method represents the simulated plasma as a large number of discrete charged particles coupled to electric and magnetic fields,~$\myvec{E}$ and~$\myvec{B}$, respectively. The~$\myvec{E}$ and~$\myvec{B}$ fields are tracked on a spatial grid, while the particles' positions,~$\myvec{x}$, and momenta,~$\myvec{p}$, can vary continuously. At each timestep, the gridded field values are interpolated to the positions of the particles, allowing their momenta to be evolved via the Lorentz force law,
\begin{align} 
    \frac{\dif \myvec{p}}{\dif t} = q \left( \myvec{E} + \frac{\myvec{v}}{c} \times \myvec{B} \right) \,
    \label{eq:lorentzforce}
\end{align}
(Fig.~\ref{fig:qedpicloop}, red panel;~$\myvec{v}$ is the 3-velocity for a particle of momentum~$\myvec{p}$).
Then, to capture the response of the fields to the particles, the particles' positions and momenta are used to calculate the bulk charge,~$\rho$, and current,~$\myvec{J}$, densities on the computational field grid (Fig.~\ref{fig:qedpicloop}, green panel). This allows~$\myvec{E}$ and~$\myvec{B}$ to be advanced via the time-dependent Maxwell's equations,
\begin{align}
    \partial_t \myvec{B} &= -c \myvec{\nabla} \times \myvec{E} \quad \mathrm{and} \notag \\
    \partial_t \myvec{E} &= c \myvec{\nabla} \times \myvec{B} - 4 \pi \myvec{J} \,
    \label{eq:timedepmax}
\end{align}
(Fig.~\ref{fig:qedpicloop}, blue panel).
It is worth noting that equations~(\ref{eq:lorentzforce}) and~(\ref{eq:timedepmax}) imply a collisionless PIC method, since the particles only interact collectively through the long-range fields~$\myvec{E}$ and~$\myvec{B}$ rather than individually through short-range two- or few-body encounters.

By itself, the procedure described so far does not model the high-energy radiation (with photon frequencies unresolved by the simulation timestep) that is often important in the environments of relativistic compact objects, as in the present work. The simplest way that such radiative effects can be incorporated is by adding a cooling radiative drag term,~$\myvec{f}_{\rm rad}$, to equation~(\ref{eq:lorentzforce}) such that it becomes
\begin{align}
    \frac{\dif \myvec{p}}{\dif t} = q \left( \myvec{E} + \frac{\myvec{v}}{c} \times \myvec{B} \right) + \myvec{f}_{\rm rad} \,
    \label{eq:lorentzforcerad}
\end{align}
(Fig.~\ref{fig:qedpicloop}, red panel). The radiative PIC code \zeltron[]implements this term after the method of \citet{tpd10}.

Treating radiative cooling as a continuous drag force is suitable when the radiating particles do not lose a significant fraction of their energy to any single photon emission event. However, when particles begin to emit photons at energies that rival their own, radiation becomes inherently discrete, and a more general approach is necessary. As an additional concern, when the emitted photons are above the threshold energy for one or more pair-production processes, their propagation and absorption must be handled self-consistently.

The needed additional QED operations can be mostly consolidated into one extra step beyond the standard PIC loop (Fig.~\ref{fig:qedpicloop}, orange panel). Here, the particle positions and velocities are used -- perhaps in conjunction with the electromagnetic fields -- to evaluate QED \crosssection[s]for the photon emission and pair-production processes of interest, yielding probabilities for these events to occur. A subset of the possible events are triggered by comparing their probabilities with randomly drawn numbers -- a Monte Carlo procedure -- and the resulting photons and particles are added to the simulation. Besides this additional QED Monte Carlo step (Fig.~\ref{fig:qedpicloop}, orange panel), the positions of photons also need to be evolved. This is typically done alongside the particles (Fig.~\ref{fig:qedpicloop}, red panel) and is much easier by comparison because photons follow straight lines (in flat spacetime, as in the present work). We term the PIC method, expanded to include these extra operations as diagrammed in Fig.~\ref{fig:qedpicloop}, the \textit{QED PIC} method.

\subsection{QED effects implemented in \zeltron}
To enable the present work, we have g\eneraliz[ed]the~3D Cartesian version of the PIC code \zeltron to include QED physics as sketched in section \ref{sec:qedpic}. The emission and pair-production processes are: IC scattering (including the high-energy Klein-Nishina limit) of a soft seed photon background by ultrarelativistic particles; and pair production when the high-energy Comptonized photons are absorbed by the soft background. In the following discussion, we provide a sketch of the algorithm used to model these processes. 

\subsubsection{Monte Carlo inverse Compton emission procedure}
\label{sec:icsimple}
We begin by describing algorithmic details of the IC emission mechanism. In this section, we present a conceptually simple but computationally expensive implementation. In section~\ref{sec:codeopt}, we discuss the modifications we made to the simpler procedure of this section for the sake of o\ptimiz[ation.]

Our simulations are immersed in a homogeneous, static, isotropic, and monochromatic photon radiation bath of energy density (per unit energy interval)
\begin{align}
    u(\epsilon) = \uph \delta (\epsilon - \eph) \, .
    \label{eq:umono}
\end{align}
These background photons are not tracked by the simulation; their energy density is prescribed by hand. Electrons and positrons, through the IC process, can upscatter these background photons. If an upscattered photon attains final energy above the (gamma-ray) threshold for pair production with the background radiation bath, then it is promoted to a tracked photon whose position is evolved explicitly by the simulation.

For a particle with ultrarelativistic Lorentz factor~$\gamma \gg 1$ traversing the radiation field~(\ref{eq:umono}), the number of photons scattered per unit time is given as
\begin{align}
    \frac{\dif N}{\dif t} = c \sigma_{\rm T} \nph \, g_{\rm KN}\left( \gamma / \gkn \right) \, ,
    \label{eq:icscatrate}
\end{align}
where~$\sigma_{\rm T}$ is the Thomson cross section,~$\nph \equiv \uph / \eph$ is the background photon number density,
\begin{align}
    \gkn \equiv \me c^2 / 4 \eph
    \label{eq:gamkn}
\end{align}
is the critical Lorentz factor above which IC scattering transitions to the Klein-Nishina regime, and~$g_{\rm KN}(\knp) \leq 1$ is the dimensionless function \citep{mwu21}
\begin{align}
    g_{\rm KN}(\knp) &= \frac{3}{2 \knp^2} \left[ \left( \knp + 9 + \frac{8}{\knp} \right) \ln( 1 + \knp ) \right. \notag \\
    &\left. - \frac{1}{1 + \knp} \left( \frac{\knp^2}{2} + 9 \knp + 8 \right) + 4 \Li_2 (-\knp) \right]  \, .
    \label{eq:gkn}
\end{align}
Here,~$\Li_2(\knp)$ is the dilogarithm. As needed to recover the Thomson regime,~$g_{\rm KN}(\gamma / \gkn)$ tends to unity in the low-energy limit,~$\gamma \ll \gkn$. Over a simulation timestep~$\Delta t$, equation~(\ref{eq:icscatrate}) yields a probability
\begin{align}
    p_{\rm emit} = \Delta t \dif N / \dif t
    \label{eq:pemit}
\end{align}
for the particle to scatter a photon. We operate in the regime~$\pfid \equiv c \sigma_{\rm T} \nph \Delta t \ll 1$, implying~$p_{\rm emit} = \pfid g_{\rm KN}(q) \leq \pfid \ll 1$. 

Computationally, the Monte Carlo photon emission process can be accomplished, for each particle at each timestep, by evaluating the probability~(\ref{eq:pemit}) and drawing a random number to determine whether the emission event occurs. Then, for the subset of particles that actually emit photons, the scattered photon energy~$\escat$ must be assigned. This demands drawing a second random number~$\rng$ and inverting the probability distribution over~$\escat$: that is, finding~$\escat$ such that
\begin{align}
    \rng = \frac{\int_0^{r(\escat)} K(r',\knp) \dif r'}{\int_0^1 K(r', \knp) \dif r'} \, .
    \label{eq:escatcdf}
\end{align}
Here,~$\knp \equiv \gamma / \gkn$ as above,~$r(\escat)$ is a proxy for the photon energy defined by
\begin{align}
    r(\escat) \equiv \frac{\escat / \gamma \me c^2}{\knp \left( 1 - \escat / \gamma \me c^2 \right)} \, ,
    \label{eq:rdef}
\end{align}
and~$K(r,\knp)$ is the single-particle Klein-Nishina scattering kernel as reported by \citet{j68} and \citet{bg70}:
\begin{align}
    K(r,\knp) = \frac{3}{(1 + \knp r)^2} \left[ 2 r \ln r + (1 + 2 r)(1 - r) + \frac{1}{2} \frac{(\knp r)^2}{1 + \knp r}(1 - r) \right] \, .
    \label{eq:scatkern}
\end{align}
That is, the number of photons scattered per unit time by a particle with~$\gamma = \knp \gkn$ to final proxy photon energy between~$r$ and~$r + \dif r$ is
\begin{align}
    \frac{\dif N}{\dif t \dif r} &\dif r = c \sigma_{\rm T} \nph K(r, \knp) \dif r \, .
    \label{eq:spectralscatrate}
\end{align}

With the photon energy~$\escat$ known, the momentum of the scattered photon has magnitude~$\escat/c$ and points, for the ultrarelativistic~$\gamma \gg 1$ approximation relevant to our simulations, along the direction of the radiating particle's motion. This momentum is subtracted from that of the radiating particle. Furthermore, if~$\escat$ is above pair-production threshold with the background photons -- i.e., if~$\escat \eph / (\me c^2)^2 \geq 1$ -- the scattered photon is promoted to a tracked simulation particle, allowing its later potential absorption to be self-consistently calculated.

Though the Monte Carlo emission scheme described so far relies on the ultrarelativistic approximation~$\gamma \gg 1$, some of our simulations contain trans- or even sub-relativistic particles. Thus, at lower energies, we need to stitch to a cooling procedure that is n\onrelativistic[ally]valid. To that end, we restrict ourselves to the main regime of astrophysical interest where~$\gkn = \me c^2 / 4 \eph \gg 1$ (i.e.~$\eph \ll 130 \, \rm keV$). This means that there is a broad range of particle energies~$\gamma$ that are ultrarelativistic~($\gamma \gg 1$) but still well below the threshold where quantum Klein-Nishina effects kick in~($\gamma \ll \gkn$): a shared applicability range where either the Monte Carlo emission scheme, which demands~$\gamma \gg 1$, or a continuous Thomson radiative drag force, which needs~$\gamma \ll \gkn$, could be used. Within this range, we select a threshold \quoted{stitching}particle energy~$\gstitch$. Above~$\gstitch$, we employ the IC Monte Carlo emission procedure. Below~$\gstitch$, we switch to the continuous Thomson radiative drag force \citep[cf.][]{bg70, rl79, pss83, u16, wpu19, sb20, mwu20, mwu21, ssb21, ssb23},
\begin{align}
    \myvec{f}_{\rm rad} = -(4/3) \sigma_{\rm T} \gamma^2 \uph \myvec{\beta},
    \label{eq:fradt}
\end{align}
that enters into the particle push through equation~(\ref{eq:lorentzforcerad}) and is, importantly, n\onrelativistic[ally]correct. Regarding the choice of~$\gstitch$, we find that artifacts of the stitching generally disappear when~$\gstitch \leq 0.1 \gkn$ -- well inside of the Thomson regime. At the same time, we find that~$\gstitch$ should be at least of order a few, limiting us to~$\gkn$ values that are above~$20$ or so.

\subsubsection{Monte Carlo pair production procedure}
\label{sec:pp}
We next describe how pair production is processed in the code. For a collision between a gamma-ray and a background photon with angle~$\theta$ between their velocity vectors and respective energies~$\ehard$ and~$\eph$, the center of mass energy is~$s=s_0 (1 - \cos{\theta})/2 \leq s_0$, where~$s_0 \equiv \ehard \eph / \me^2 c^4$ must exceed unity for pair production to be possible. Integrating over all possible collision angles~$\theta$ such that~$s>1$ for a given~$s_0$ yields an overall \crosssection[]presented to a propagating gamma-ray by the background radiation \citep{gs67} of\footnote{The parenthesized term in~(\ref{eq:siggg}),~$4 \left( \pi^2/12 + \Li_2(-W_0) \right)$, is equal to the last term,~$-4 L(W_0)$, in equation~10 of \citep{gs67} but is corrected for the missing factor of~$4$ in that work.} 
\begin{align}
    \sigma_{\gamma\gamma}(s_0) &= \frac{3}{8} \frac{\sigma_{\rm T}}{s_0^2} \left[ \frac{1 + \beta_0^2}{1 - \beta_0^2} \ln W_0 - \beta_0^2 \ln W_0 - \ln^2 W_0 - \frac{4 \beta_0}{1 - \beta_0^2} \right. \notag \\
    &+ \left. 2 \beta_0 + 4 \ln W_0 \ln \left(1 + W_0 \right) + 4 \left( \frac{\pi^2}{12} + \Li_2 (-W_0) \right) \right] \, ,
    \label{eq:siggg}
\end{align}
where~$\beta_0$ and~$W_0$ are both~$s_0$-dependent, reading, respectively,
\begin{align}
    \beta_0^2(s_0) \equiv 1 - \frac{1}{s_0}
    \label{eq:b0def}
\end{align}
and
\begin{align}
    W_0(s_0) \equiv \frac{1 + \beta_0(s_0)}{1 - \beta_0(s_0)} \, .
    \label{eq:w0def}
\end{align}
Thus, the pair-production probability accumulated by a gamma-ray of energy~$\ehard$ in one simulation timestep~$\Delta t$ is
\begin{align}
    p_{\rm abs} = c \nph \sigma_{\gamma\gamma}(s_0) \Delta t \, .
    \label{eq:pabs}
\end{align}
Because,~$\max(\sigma_{\gamma\gamma}) \simeq \sigma_{\rm T} / 5$, our choice to operate in the regime~$\pfid \equiv c \sigma_{\rm T} \nph \Delta t \ll 1$ implies~$p_{\rm abs} \leq \pfid / 5 \ll 1$.

Just as equation~(\ref{eq:pemit}) determines which potential IC scattering events occur, equation~(\ref{eq:pabs}) determines which photons produce new pairs. Ordinarily, like the additional steps necessary to determine the final scattered photon energy in the case of IC emission -- e.g. equation~(\ref{eq:escatcdf}) -- one would also need to proceed beyond equation~(\ref{eq:pabs}) to fix the energy of the newborn electron and positron. First, the angle~$\theta$, which is integrated out when interested only in the total \crosssection[~(\ref{eq:siggg})]and corresponding absorption probability~(\ref{eq:pabs}), needs to actually be sampled to determine the center-of-mass energy~$s$. Then, one must also sample the angle of one of the newborn particles' momenta with respect to that of the collision axis in the center-of-mass frame. We have indeed implemented both steps in \zeltron[,]but they are unnecessary when~$\ehard \gg \me c^2 \gg \eph$ (i.e., when~$4 \gkn \gg 1$), which is all that concerns us in this study. Then, the collision energy budget is entirely dominated by the incoming gamma-ray, and the electron and positron each simply inherit half of the absorbed gamma-ray's momentum. 

\subsubsection{O\ptimiz[ation]methods for inverse Compton emission}
\label{sec:codeopt}
The Monte Carlo implementation of IC emission sketched in section~\ref{sec:icsimple} suffers from two performance bottlenecks. First, the loop computing the IC \crosssection[,]equations~(\ref{eq:gkn}) and~(\ref{eq:pemit}), for every simulation particle is costly. Second, within this loop, the inversion of the cumulative distribution function in equation~(\ref{eq:escatcdf}) is also expensive (though only necessary for the subset of particles that actually scatters photons). We mitigate these issues using two techniques, which we discuss in turn below. We note that, while similar techniques could be applied to the pair production procedure, our simulations are optically thick to pair production, and hence the steady-state number of tracked photons is much smaller than the (always growing) number of particles. Thus, the QED physics involving photons, for us, can tolerate a less rigorously o\ptimiz[ed]implementation.

First, we speed up the assignment of the scattered photon energy -- done in the context of section~\ref{sec:icsimple} through equation~(\ref{eq:escatcdf}). To do this, we consider the IC emission in the rest frame of the scattering particle. As we show below, this requires a larger number of random number draws per particle, but enables the use of simpler expressions. These, in turn, furnish an approximation to the scattering \crosssection[]that effectively replaces equation~(\ref{eq:escatcdf}), relieving its performance bottleneck. Our handling of the problem this way, presented in detail below, follows closely \citet{lc18} and \citet{ccp20}.

Before transforming to the rest frame of the potentially scattering particle, we isolate to an interaction with a single photon rather than with the entire isotropic radiation bath. This is accomplished by first drawing a random photon angle~$\psi$ with respect to the direction of the particle's velocity~$c \beta$ according to the probability density function
\begin{align}
    p(\thinc) \propto 1 - \beta \cos \thinc \, ,
    \label{eq:thincpdf}
\end{align}
which accounts for the relative lab-frame rate of encounters of the particle with photons incident from different directions. The selected photon is then boosted to the particle's rest frame, where it has energy
\begin{align}
    \einc' = \gamma \eph (1 - \beta \cos \thinc) \, .
    \label{eq:eincp}
\end{align}
Particle rest-frame quantities are primed in our convention. 

We now evaluate the spectral (per unit final photon energy) and total (integrated over final photon energies) scattering rate for this interaction in the particle's rest frame. We discuss after this derivation how these two quantities are used by the code. In the primed frame, the scattering \crosssection[]is that of ordinary Compton scattering \citep[e.g.][]{bg70}: 
\begin{align}
    \frac{\dif \sigma}{\dif \escat' \dif \Omega'} = \frac{3\sigma_{\rm T}}{16 \pi} &\left( \frac{\escat'}{\einc'} \right)^2 \left( \frac{\einc'}{\escat'} + \frac{\escat'}{\einc'} - \sin^2 \Psi' \right) \notag \\
    \times \, &\delta \left[ \escat' - \frac{\einc'}{1 + (\einc' / \me c^2) (1 - \cos \Psi')} \right] \, ,
    \label{eq:comptoncross}
\end{align}
where~$\Psi'$ is the angle through which the photon is scattered in the particle frame. The total scattering rate into final photon energies between~$\escat'$ and~$\escat' + \dif \escat'$ is then
\begin{align}
    \frac{\dif N}{\dif t \dif \escat'} = \frac{1}{\gamma} \frac{\dif N}{\dif t' \dif \escat'} = c \nph \int \dif \Omega' \frac{\dif \sigma}{\dif \escat' \dif \Omega'} \, ,
    \label{eq:specscatrate}
\end{align}
where we used~$\nph' = \gamma \nph$. Defining~$x\equiv\escat'/\einc'$ and~$\gknp\equiv \me c^2/\einc'$, the delta function only activates for~$1/(1+2/\gknp)\leq x \leq 1$, in which case the integral evaluates to
\begin{align}
    \frac{\dif N}{\dif t \dif x} &= \einc' \frac{\dif N}{\dif t \dif \escat'} = \frac{3 \sigma_{\rm T}}{8} \notag \\
    \times \, & \gknp \left[ x + \frac{1}{x} + 2 \gknp \left( 1 - \frac{1}{x} \right) + \gknp^2 \left( \frac{1}{x^2} - \frac{2}{x} + 1 \right) \right] \, .
    \label{eq:specscatrateeval}
\end{align}
The total scattering rate is then
\begin{align}
    \frac{\dif N}{\dif t} \equiv \int \dif x \frac{\dif N}{\dif t \dif x} = \frac{3 \sigma_{\rm T} c \nph}{8} \left[ F\left(1,\gknp\right) - F\left(\frac{1}{1+2/\gknp},\gknp \right) \right] \, ,
    \label{eq:icscatratep}
\end{align}
where
\begin{align}
    F(x,\gknp) \equiv y \left[ -\frac{y^2}{x} + x y \left( 2 + y \right) + \frac{x^2}{2} + \left(1 - 2 y - 2 y^2 \right) \ln x \right] \, .
    \label{eq:fcdf}
\end{align}

These results are used by the code as follows. First, equation~(\ref{eq:icscatratep}) is evaluated and multiplied by~$\Delta t$ to determine the probability~$p_{\rm emit} = \Delta t \dif N / \dif t$ that the scattering occurs during the simulation timestep. This step replaces the evaluation of equation~(\ref{eq:icscatrate}). On average, these two procedures are completely equivalent, but in this second method, we have traded the analytic integral over incident photon directions [which yields equation~(\ref{eq:icscatrate})] for a random Monte Carlo sampling over these directions. 

Then, for particles that scatter photons, the final photon energy is obtained by drawing a random number~$\rng$ and inverting the cumulative distribution function: i.e.\ finding~$x$ such that
\begin{align}
    \rng &= \frac{3 \sigma_{\rm T} c \nph }{8}\frac{\left[ F\left(x,y\right) - F\left(1/(1+2/y),y\right) \right]}{\dif N / \dif t} \notag \\
    &= \frac{F\left(x,y\right)-F\left(1/(1+2/y),y\right)}{F\left(1,y\right) - F\left(1/(1+2/y),y\right)} \, .
    \label{eq:xcdf}
\end{align}
This replaces the inversion of the equivalent lab-frame cumulative distribution function~(\ref{eq:escatcdf}) and, as we now show, constitutes the main advantage of this method. The issue with equation~(\ref{eq:escatcdf}) is that its solution cannot be expressed analytically. Instead, the cumulative distribution function (or its inverse) must be stored as a table and consulted for each emitting particle -- a costly procedure. However, the function~$F(x,y)$ can be approximated, when~$x\ll1$, by only its~$\ln x$ term. This approximation breaks down as~$x$ approaches unity but, as it turns out, does not very strongly disturb the cumulative distribution function on the right-hand-side of equation~(\ref{eq:xcdf}) even when used across all~$x$. If one then approximates
\begin{align}
    F(x,y) \simeq y \left(1 - 2 y - 2 y^2 \right) \ln x \, ,
    \label{eq:fapprox}
\end{align}
the solution to equation~(\ref{eq:xcdf}) can be found as
\begin{align}
    x = \exp \left[ -\left(1-\rng\right) \ln \left(1 + 2/y\right) \right] \, .
    \label{eq:xexplicit}
\end{align}
We have checked that the use of this approximation does not change any discernible aspects of our simulations. We have also conducted experiments to show that it yields nearly the same average emitted photon energy as the exact cumulative distribution function. It therefore provides a powerful speed-up to the code by facilitating the otherwise impossible analytic evaluation of~$x$ without compromising the important physics.

Once the rest-frame scattered photon energy~$\escat' = x \einc'$ is known, it can be boosted back to the lab frame as follows. First, one notes that in the ultrarelativistic approximation~$\gamma \gg 1$ relevant to this work, the incident photon approaches the particle nearly head-on in the primed frame. Thus,~$\cos \Psi' = - \cos \psi'_{\rm scat}$, where~$\psi'_{\rm scat}$ is the angle between the emerging photon's momentum and the particle's lab velocity. Then, the delta function in equation~(\ref{eq:comptoncross}) can be used to write
\begin{align}
    \cos \psi'_{\rm scat} = - \cos \Psi' = \frac{\me c^2}{\escat'} - \frac{\me c^2}{\einc'} - 1 \, ,
    \label{eq:cosscatp}
\end{align}
which yields the lab-frame emerging photon energy via
\begin{align}
    \escat = \gamma \escat' (1 + \beta \cos \psi'_{\rm scat}) \, .
    \label{eq:escatlab}
\end{align}

In addition to considering the IC scattering in the rest frame of each particle, which facilitates the powerful approximations~(\ref{eq:fapprox}) and~(\ref{eq:xexplicit}), we also leverage the following second strategy to reduce the cost of the QED module. This o\ptimiz[ation]is known in the plasma physics literature as the \textit{null collision method} \citep{r68, lb77, bm82, b91}. It speeds up the code drastically by avoiding the need to loop over all of the particles during the IC Monte Carlo emission step.

The technique exploits the fact that the probability~$p_{\rm emit}$ that an individual particle emits a photon in a given timestep is capped to a global maximum, given by the Thomson limit of equation~(\ref{eq:icscatrate}), of~$\pfid = c \sigma_{\rm T} \nph \Delta t$. This is a small number in our simulations: of order~$10^{-3}$. Thus, instead of looping over all of the particles to determine whether fewer than~$1$ in every~$1/\pfid$ of them emits a photon, we randomly select a small fraction~$\pfid$ of all the particles, loop over this reduced subset, and exactly compensate the limited particle sample by enhancing the per-particle emission probability by the inverse factor~$1/\pfid$.

These two techniques -- the approximation of the cumulative distribution function on IC emission energies in equation~(\ref{eq:fapprox}) and the use of the null collision method -- allow us to run QED-PIC simulations with negligible added cost per timestep (of order~$10$ per cent) taken by the QED module (orange panel in Fig.~\ref{fig:qedpicloop}). The main costs are instead the accumulation of particles and photons in the simulation and the typically larger amount of data that one wishes to dump in QED runs. We note, however, that for the regime of reconnection studied in this work, the growth in the total number of simulated particles is never more than a factor of a few, and thus we do not need, at this stage, additional algorithmic strategies to regulate such growth \citep[cf.][]{vgm15, hps19}.

\section{Simulation setup}
\label{sec:setup}
We perform pair-plasma simulations of relativistic collisionless magnetic reconnection using the radiative QED-outfitted (section~\ref{sec:arch}) electromagnetic PIC code \zeltron[]\citep{cwu13, cw19}. The simulations are in a 2D periodic box of physical dimensions~$L_x \times L_y = L \times 2L$ and grid size~$N_x \times N_y = N \times 2N = 7680 \times 15360$. Spatial dependence is only tracked in the~$x$- and~$y$-directions, but all vectors may have an out-of-plane~$z$-component.

Our four main runs share the same setup and parameter values, but differ in their mode\led[]radiative effects. In particular, we present one case without any radiative cooling; one with purely continuous Thomson IC cooling \citep[similar to, e.g.,][]{wpu19, sb20, mwu20, ssb21, ssb23}; one with fully general IC cooling (including the Klein-Nishina regime) but with pair production artificially turned off; and one with general IC cooling and self-consistent pair production. Synchrotron cooling is ignored in all runs. We describe the non-radiative aspects of our setup in section~\ref{sec:nonradsetup} and move on to the radiative details in section~\ref{sec:radsetup}.

\subsection{Non-radiative aspects of the setup}
\label{sec:nonradsetup}
Our simulations begin with zero electric field. The initial magnetic field is force-free and undergoes reversals via current sheets located at~$y_1=L/2$ and~$y_2=3L/2$. It has the form \citep[cf.][]{gld14, gld15, gll16, gld19, lgl18, lgl19, gld21, zlg18, zlg22, gl22, fgz23}
\begin{align}
    \myvec{B} &= B_x(y)\myunit{x} + B_z(y)\myunit{z} \notag \\
&= \pm B_0 \tanh \left( \frac{y - y_{1,2}}{\delta} \right) \myunit{x} + B_0 \sqrt{\sech^2 \left( \frac{y - y_{1,2}}{\delta} \right) + \left( \frac{B_g}{B_0} \right)^2 } \, \, \myunit{z} \, ,
    \label{eq:binit}
\end{align}
where~$\myunit{x}$ and~$\myunit{z}$ are unit vectors pointing in the respective~$x$ and~$z$ directions, the~$+$~($-$) sign is taken at~$y_1$~($y_2$), and~$\delta$ is the half-thickness of the current sheets. In the reconnection upstream region far away from the current sheets~($|y-y_{1,2}|\gg \delta$), equation~(\ref{eq:binit}) reduces to a uniform field with in-plane component~$\pm B_0 \myunit{x}$ and out-of-plane guide-field component~$B_g \myunit{z} = 0.15 B_0 \myunit{z}$.
In addition to this modest guide field, which accompanies the upstream plasma into the reconnection layer, there is also a strong l\ocaliz[ed]($|y-y_{1,2}|\ll \delta$) initial out-of-plane field of peak strength~$\sqrt{B_0^2 + B_g^2}$ that supplies the force-free magnetic field reversal. Both this strong l\ocaliz[ed]out-of-plane field and the upstream guide field provide some pressure support to the plasma energized by reconnection as it radiatively cools down, which helps ensure the Debye length in the simulations remains well-resolved. 

Our initial fields satisfy the force-free condition,~$\myvec{J} \times \myvec{B} / c = -\myvec{\nabla} \myvec{B}^2 / 8 \pi + \myvec{B} \cdot \myvec{\nabla} \myvec{B} / 4 \pi = 0$. Thus, and unlike the case of a Harris equilibrium \citep{ks03}, no added plasma pressure is needed inside the initial current layers. This allows us to start the simulations with a plasma of completely homogeneous initial (electron+positron) number density,~$\idense$, and temperature,~$\itempt = \itemp \me c^2 = 24 \me c^2$. Specifically, we use a relativistic Maxwell-Jüttner initial plasma distribution function.

The number density,~$\idense$, and initial reconnecting field strength,~$B_0$, together define the \textit{cold magnetization},
\begin{align}
    \sigc \equiv \frac{B_0^2}{4 \pi \idense \me c^2} \, ,
    \label{eq:sigc}
\end{align}
equal to about twice the reconnecting magnetic field energy per particle. While~$\sigc$ defines how much energy a typical reconnection-accelerated particle may acquire, another similar quantity, the \textit{hot magnetization},
\begin{align}
    \sigh \equiv \frac{B_0^2}{4 \pi w_0} \, ,
    \label{eq:sigh}
\end{align}
defines the magnetic dominance of the upstream region. Here, the initial plasma enthalpy density,~$w_0$, can be written as~$w_0 = u_0 + P_0$, where~$u_0$ and~$P_0$ are, respectively, the initial plasma internal energy density and pressure. For a non-relativistically cold initial temperature,~$\itemp \ll 1$, the enthalpy density is dominated by rest-mass energy,~$w_0 \simeq u_0 = n_0 \me c^2$, and, hence,~$\sigh \simeq \sigc$. For a relativistically hot plasma,~$\itemp \gg 1$, the thermal kinetic energy and pressure dominate the enthalpy,~$w_0 \simeq 4 P_0 = 4 \itemp n_0 \me c^2$, and, in this case,~$\sigh \simeq \sigc / 4 \itemp = 1/(2 \beta_{\rm pl})$, where~$\beta_{\rm pl}$ is plasma beta. For all regimes,~$\sigh \leq \sigc$, meaning that~$\sigh \gg 1$ is a stronger condition than~$\sigc \gg 1$. In fact, because~$\sigh$ sets the plasma Alfvén speed~$v_A = c \sqrt{\sigh/(1+\sigh)}$, a high~$\sigh$ means that the energy budget permits not just relativistic individual particles (which merely requires high~$\sigc$), but also for the collective bulk motion of the plasma itself to become highly relativistic. In our simulations, we set~$\sigc=1.2\times10^5$ and~$\sigh=\sigc/4 \itemp=1250$.

Though the force-free initial condition does not require any extra plasma density or temperature in the current sheets to balance the upstream magnetic pressure, the field-reversing currents must still be supplied. Correspondingly, we set a local fraction
\begin{align}
    \fdrift(y) = \sech ( y/\delta ) \sqrt{\frac{1+(B_g/B_0)^2}{1 + [B_g \cosh(y/\delta)/B_0]^2}} 
    \label{eq:driftfrac}
\end{align}
of the positrons in motion at a drift velocity
\begin{align}
    c \myvec{\beta}_{\rm d,i}(y) &= \frac{c \beta_{\rm d}}{\sqrt{1+(B_g/B_0)^2}} \notag \\
    &\times \left[ \mp \sqrt{\sech^2(y/\delta)+(B_g/B_0)^2} \myunit{z} - \tanh(y/\delta) \myunit{x} \right]
    \label{eq:driftvel}
\end{align}
so that they carry half the field-reversing current
\begin{align}
    \myvec{J} = c \myvec{\nabla} \times \myvec{B} / 4 \pi = e c \idense \fdrift(y) \myvec{\beta}_{\rm d,i}(y) \, ,
    \label{eq:driftj}
\end{align}
where the~$\mp$ corresponds to equation~(\ref{eq:binit}) and~$e$ is the positron charge.
The other half of the current is carried by the initial electrons, which are assigned~$\myvec{\beta}_{\rm d,e} = -\myvec{\beta}_{\rm d,i}$. The drifting particles follow a drifting Maxwell-Jüttner distribution function with initial temperature,~$\itempt$, defined in their local [boosted by~$\myvec{\beta}_{\rm d,i/e}(y)$] rest frame.

The force-free setup ties the current sheet half-thickness~$\delta$ to the other \ls[s]in the problem as follows. The drifting plasma supplies a current density~$e c \idense \beta_{\rm d} \leq e c \idense$, whereas the current needed at the heart of the layer is, according to Ampère's law~(\ref{eq:driftj}),~$c B_0 / 4 \pi \delta$. This means that
\begin{align}
    \delta = \frac{B_0}{4 \pi e \idense \beta_{\rm d}} = \frac{\sigc \, \rho_0}{\beta_{\rm d}} \geq \sigc \, \rho_0 \, ,
    \label{eq:deltamax}
\end{align}
where we have introduced the \textit{nominal gyroradius},
\begin{align}
    \rho_0 \equiv \frac{\me c^2}{e B_0} \, .
    \label{eq:rhonaught}
\end{align}
Equation~(\ref{eq:deltamax}) demands that the current sheet half-thickness be limited to the typical gyroradius,~$\sigc \, \rho_0$, of reconnection-energized particles. Therefore, in order to achieve a high aspect ratio,~$L/\delta$, while also complying with the other demanding radiative constraints described later (section~\ref{sec:radsetup}), we assign~$\beta_{\rm d} = 0.9$. This corresponds to~$\delta = 1.1 \sigc \rho_0$ and to an initial drifting-particle Lorentz factor of~$2.3$ (still much less than~$\itemp$).

The current sheet aspect ratio,~$L/\delta = L \beta_{\rm d} / \sigc \, \rho_0 \simeq L / \sigc \, \rho_0$, represents not just a ratio of \ls[s]but also one of particle energy scales. This is because, during reconnection, some particles are swept into the vicinity of an X-point (X-line in 3D), which is a region wherein the in-plane magnetic field reconnects. There, they become unmagnetized and linearly accelerated by the out-of-plane reconnection electric field,~$E_{\rm rec} = \beta_{\rm rec} B_0 \simeq 0.1 B_0$, where~$\beta_{\rm rec} \equiv 0.1 v_A / c \simeq 0.1$ is the fiducial collisionless reconnection rate.
As described by \citet{wuc16}, in systems with modest aspect ratios, for which the reconnection layer is dominated by a single X-point, particles can be accelerated by~$E_{\rm rec}$ until, after traveling an out-of-plane distance~$\sim L$, they escape the acceleration region with final \textit{system-size-limited Lorentz factor}
\begin{align}
    \gmax \equiv \frac{0.1 e B_0 L}{\me c^2} = \frac{0.1L}{\rho_0} \, .
    \label{eq:gmax}
\end{align}
However, for larger systems, the reconnection layer tears into a hierarchical chain of plasma-filled magnetic islands, or \textit{plasmoids}, studded with many X-points. Then, particles cannot travel the whole system size before escaping a given acceleration zone. Instead, at least in~2D, they may travel a distance comparable to that between the smallest-scale plasmoids, which subsequently capture the particles, limiting them to energies of order the intrinsic reconnection \textit{X-point acceleration Lorentz factor} \citep[cf.][]{wuc16, sgp16, u22},
\begin{align}
    \gx \equiv 4 \sigc \, .
    \label{eq:gx}
\end{align}
The aspect ratio becomes large enough that the X-point acceleration channel is intrinsically limited by the self-consistent evolution of the plasmoid chain rather than by the modest size of the system when these two energy scales cross each other:~$\gmax / \gx \geq 1 \Rightarrow L / \sigc \, \rho_0 \geq 40$ \citep{wuc16}. Thus, the requirement to have a large aspect ratio~$L/\delta \gg 1$ is synonymous with having a healthy separation between~$\gmax$ and the particle energies,~$\sim \gx$, at which reconnection X-point acceleration saturates. We adopt~$\gmax = 9.1 \sigc$, which corresponds to~$L=91 \sigc \, \rho_0$. This meets the fiducial~$\gmax > \gx$ criterion but, for reasons described below, yields~$\gmax / \sigc$ smaller than typical for simulations of the numerical size,~$N_x \times N_y = 7680 \times 15360$, that we present.

Our grid resolution~$\Delta x = \Delta y$ is set by the need to resolve the initial Debye length,
\begin{align}
    \lambda_{\rm D,0} = \sqrt{\frac{\itemp \me c^2}{4 \pi e^2 n_0}} = \sqrt{\itemp \sigc} \, \rho_0 \, ,
    \label{eq:debyeinit}
\end{align}
and we set~$\Delta x = \lambda_{\rm D,0} / 1.2$ in all runs. Because we operate in the highly relativistic regime,~$\sigc \geq \sigh \gg 1$, the Debye length is much smaller than the typical energized particle's gyroradius,~$\sigc \rho_0 \sim \delta$. This is largely why we cannot afford a larger~$\gmax / \sigc$ -- our choice of~$\sigh = 1250$ induces a large separation between the plasma microscales,~$\sigc \, \rho_0 / \lambda_{\rm D,0} = \sqrt{\sigc/\itemp} \simeq 2 \sqrt{\sigh} \sim 70$, which occupies much of our grid resolution to resolve.\footnote{While~$\sigc \, \rho_0 / \lambda_{\rm D,0} = \sqrt{\sigc/\itemp}$ is generally true, the approximation,~$\sqrt{\sigc/\itemp} \simeq 2 \sqrt{\sigh}$, assumes~$\itemp \gg 1$.} We do, however, underresolve the gyroradii of particles with energies less than~$\gamma_{\Delta x} \equiv (\itemp \sigc)^{1/2}/1.2$, including the upstream particles (energies~$\sim \itemp \ll \gamma_{\Delta x}$). We do not observe any strong artifacts of this in our results, and the energy in all of our simulations is conserved to~$1$ per cent or better.

\begin{table*}
\centering
\begin{threeparttable}
\caption{Simulation parameters used in this study. Non-radiative parameters (described in section~\ref{sec:nonradsetup}) are presented in the upper part of the table and radiative parameters (described in section~\ref{sec:radsetup}) below. The radiative parameters~$\gcool$ and~$\gradt$ apply only to the three simulations with radiative cooling,~$\gkn$ applies only to the two simulations with fully general IC cooling (including Klein-Nishina effects), and~$\tau_{\gamma\gamma}$ applies only to the one simulation with pair production. The expression for~$\sigh$ assumes~$\itemp \gg 1$.
}
\label{table:params}
\begin{tabular}{l r l r l l}
  \hline
  Parameter & Symbol & (=definition) & Value & (=equivalent) \\
  \hline \hline
  Upstream reconnecting field strength & $B_0$ & & \\
  Nominal gyroradius & $\rho_0$ & $= \me c^2 / e B_0$ & \\ 
  Initial upstream density & $\idense$ & & \\
  Initial cold magnetization & $\sigc$ & $= B_0^2 / 4 \pi \idense \me c^2$ & $120000$ \\
  Initial hot magnetization & $\sigh$ & $= B_0^2 / 16 \pi \idense \itemp \me c^2$ & $1250$ \\
  Initial upstream temperature & $\itemp$ & $= \itempt / \me c^2$ & $2 \times 10^{-4} \sigc$ & $=24$ \\
  System size & $L_x$ & $=L$ & $91 \sigc \rho_0$ \\
  System-size-limited Lorentz factor & $\gmax$ & $= 0.1 L / \rho_0$ & $9.1 \sigc$ & $= 1.1 \times 10^6$ \\
  Guide field & $B_g$ & & $0.15 B_0$ \\
  Layer drift velocity & $\beta_{\rm d} c$ & & $0.9 c$ \\
  Layer half-thickness & $\delta$ & $= \sigc \rho_0 / \beta_{\rm d}$ & $1.1 \sigc \rho_0$ \\
  Cell size & $\Delta x , \, \Delta y$ & & $\sigc \rho_0 / 85$ \\
  Time step & $\Delta t$ & & $3 \Delta x / 4 \sqrt{2} c$ \\
  Grid size & $N_x$ & $=N$ & $7680$ \\
  Computational particles per cell & & & $20$ \\ 
  \hline \hline
  Soft photon energy density & $\uph$ & & \\
  Soft photon energy & $\eph$ & & \\
  Soft photon number density & $\nph$ & $= \uph / \eph$ & \\
  Nominal cooling Lorentz factor & $\gcool$ & $= 3 \me c^2 / 4 \uph \sigma_{\rm T} L$ & $3.6 \times 10^{-3} \sigc$ & $= 430$ \\
  Thomson IC-limited Lorentz factor & $\gradt$ & $= (0.3 e B_0 / 4 \sigma_{\rm T} \uph)^{1/2}= (\gmax \gcool)^{1/2} $ & $0.18 \sigc$ & $= 2.2 \times 10^4$ \\
  Klein-Nishina Lorentz factor & $\gkn$ & $= \me c^2 / 4 \eph$ & $0.046 \sigc$ & $= 5500$ \\
  Box pair-production optical depth & $\tau_{\gamma\gamma}$ & $= \nph \sigma_{\rm T} L/5 = 3\gkn/5\gcool$ & $7.7$ \\
  \hline
\end{tabular}
\end{threeparttable}
\end{table*}
Given the cell size,~$\Delta x$, we employ a timestep~$\Delta t = 3 \Delta x / 4 \sqrt{2} c \simeq 0.5 \Delta x / c$. This is slightly smaller than the maximum allowed Courant-Friedrichs-Lewy step~$\Delta x / \sqrt{2} c$, a choice that we have found to slightly improve the simulations' energy conservation. We initially place~$20$ computational particles (electrons+positrons) per cell, except within a few~$\delta$ of~$y_{1,2}$ where we place~$40$ per cell. The principal n\onradiative[]simulation parameters described in this section and their values are s\ummariz[ed]in the top part of Table~\ref{table:params}.

\subsection{Radiative aspects of the setup}
\label{sec:radsetup}
The two principle radiative parameters in our simulations are the total energy density,~$\uph$, and monochromatic single-photon energy,~$\eph$, of the IC seed photons [equation~(\ref{eq:umono})]. These photons are not tracked, but provide a static, homogeneous, isotropic target population for the charged particles to scatter. Though only the two numbers~$\uph$ and~$\eph$ need to be prescribed to fully specify the radiative physics, they introduce a variety of derived energy- and length-scales into the problem, and, hence, divide the parameter space into many different regimes. We first provide a brief overview of these regimes in section~\ref{sec:radscalesintro}, s\ummariz[ing]the more detailed discussion from our earlier work, \citet{mwu21}. Afterwards, in section~\ref{sec:radscaleschoose}, we discuss how these parameters are chosen for our simulations (as in Table~\ref{table:params}).

\subsubsection{Reconnection scales introduced by radiative physics}
\label{sec:radscalesintro}
For particles with Lorentz factors~$\gamma \ll \gkn \equiv \me c^2 / 4 \eph$, IC emission proceeds in the Thomson regime, where typical scattered photons are enhanced to energies~$\sim \gamma^2 \eph$, up to a maximum of~$4 \gamma^2 \eph$. Because no individual photon robs the particle of a significant fraction of its energy, cooling proceeds continuously and is modeled (section \ref{sec:qedpic}) as a radiative drag force,~$\myvec{f}_{\rm rad}$. The total power radiated by a given particle in the Thomson regime is [cf. equation~(\ref{eq:fradt})]
\begin{align}
    P_{\rm T}(\gamma) = \abs{c \myvec{f}_{\rm rad} \cdot \myvec{\beta}} = \frac{4}{3} \sigma_{\rm T} c \gamma^2 \beta^2 \uph \, ,
    \label{eq:pthom}
\end{align}
where~$c \myvec{\beta}$ is the 3-velocity of the particle.

However, at sufficiently high~$\gamma$, the maximum Thomson emission energy,~$4 \gamma^2 \eph$, rivals the scattering particle's energy~$\gamma \me c^2$ (the two are formally equal at~$\gkn = \me c^2 / 4 \eph$). Then, particles pass into the Klein-Nishina regime where they lose energy in discrete quanta,~$\sim \gamma \me c^2$. Here, the \crosssection[][e.g. equations~(\ref{eq:gkn}) and~(\ref{eq:comptoncross})] needs to be described by QED and gives rise to an average power radiated per particle of
\begin{align}
    P_{\rm IC}(\gamma) = P_{\rm T}(\gamma) f_{\rm KN}(\gamma / \gamma_{\rm KN}) \, ,
    \label{eq:pic}
\end{align}
where~$f_{\rm KN}(\knp)$, with~$\knp = \gamma / \gkn$, is the dimensionless function \citep[cf.][]{j68, nyc18, mwu21}
\begin{align}
    f_{\rm KN}(\knp) &= \frac{9}{\knp^3} \left[ \left( \frac{\knp}{2} + 6 + \frac{6}{\knp} \right) \ln \left( 1 + \knp \right) \right. \notag \\
    &- \left. \frac{1}{\left(1 + \knp\right)^2} \left( \frac{11}{12}\knp^3 + 6 \knp^2 + 9 \knp + 4 \right) - 2 + 2 \Li_{2}(-\knp) \right] \, 
    \label{eq:fkn}
\end{align}
and~$\Li_2(\knp)$ is the dilogarithm. As necessary to recover the Thomson limit,~$f_{\rm KN}(\knp \ll 1) \to 1$. In the opposite, deep Klein-Nishina regime,~$f_{\rm KN}(\knp \gg 1) \simeq (9/2 \knp^2) [ \ln(\knp) - 11/6 ]$.

Equations~(\ref{eq:pthom}) and~(\ref{eq:pic}) define the respective Thomson-limit and general-case IC cooling times,
\begin{align}
    \tcoolt(\gamma) &= \frac{\gamma \me c^2}{P_{\rm T}(\gamma)} \simeq \frac{3 \me c}{4 \sigma_{\rm T} \uph \gamma} \notag \\
    &= \frac{L}{c} \frac{\gcool}{\gamma}
    \label{eq:tcoolt}
\end{align}
and
\begin{align}
    \tcoolk(\gamma) &= \frac{\gamma \me c^2}{P_{\rm IC}(\gamma)} \simeq \frac{3 \me c}{4 \sigma_{\rm T} \uph \gamma f_{\rm KN}(\gamma / \gkn)} \notag \\
    &= \frac{L}{c} \frac{\gcool}{\gamma f_{\rm KN}(\gamma / \gkn)} \, ,
    \label{eq:tcoolic}
\end{align}
where we make the relativistic~$\beta \simeq 1$ approximation in both cases. The above expressions also invoke the \textit{nominal efficient-cooling Lorentz factor}
\begin{align}
    \gcool \equiv \frac{3 \me c^2}{4 \sigma_{\rm T} \uph L} \, ,
    \label{eq:gcool}
\end{align}
corresponding, in the Thomson regime, to the minimum Lorentz factor for a particle to cool on \ts[s]shorter than the system light/Alfvén-crossing time,~$L/c$. One may express~$\gcool$ in terms of the radiative compactness,~$\ell \equiv \uph \sigma_{\rm T} L / \me c^2$, as~$\gcool = 3 / 4 \ell$. Unlike a real particle Lorentz factor, the formal parameter~$\gcool$ can be less than~$1$. This corresponds to the highly compact case,~$\ell > 1$, and signals that all particles cool to n\onrelativistic[]energies in less than~$L/c$.

In addition to~$\gcool$ -- and ignoring Klein-Nishina effects for the moment -- another radiative Lorentz factor scale may be defined by equating the radiative cooling time,~$\tcoolt(\gamma)$, to the linear acceleration time for particles experiencing the reconnection electric field~$E_{\rm rec} = 0.1 B_0$ near reconnection X-points,
\begin{align}
    t_{\rm X}(\gamma) = \frac{\gamma \me c^2}{0.1 e c B_0} = \frac{10 \gamma \rho_0}{c} \, .
    \label{eq:tx}
\end{align}
Putting~$t_{\rm X}(\gradt) \equiv \tcoolt(\gradt)$ yields the \textit{nominal radiatively-limited Lorentz factor} \citep[cf.][]{n16, u16, wpu19, sb20, mwu20, mwu21, ssb21, ssb23},
\begin{align}
    \gradt \equiv \sqrt{\frac{0.3 e B_0}{4 \sigma_{\rm T} \uph}} \, .
    \label{eq:gradt}
\end{align}
The energy~$\gradt$ is radiatively limited because a particle with Lorentz factor~$\gamma > \gradt$ experiences a Thomson radiative drag force stronger than the acceleration force from the reconnection electric field. Exceeding this energy is impossible in the Thomson regime in the absence of other, faster (i.e. operating on \ts[s~$< t_{\rm X}$)]acceleration mechanisms.

\begin{figure}
    \centering
    \includegraphics[width=\columnwidth]{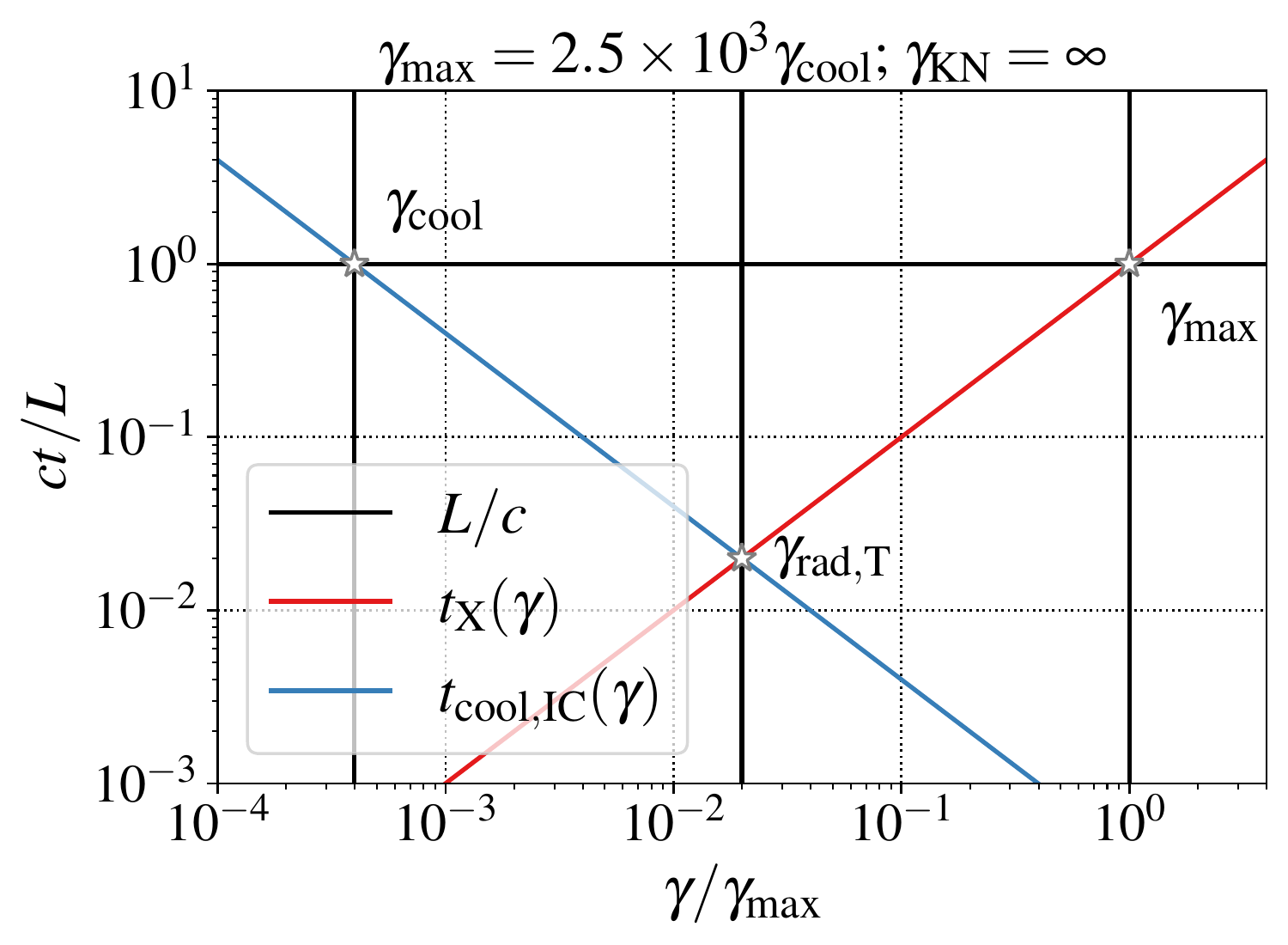}
    \caption{The Lorentz factors~$\gcool$,~$\gradt$, and~$\gmax$ define the topology of the \ts[s~$L/c$,~$t_{\rm X}(\gamma)$,]and~$\tcoolt(\gamma)$ viewed as functions of~$\gamma$. These \ts[s]depend, respectively, on~$L$,~$B_0$, and~$\uph$. Thus,~$\gcool$, at the intersection of~$\tcoolt(\gamma)$ and~$L/c$, depends on~$\uph$ and~$L$ but not on~$B_0$;~$\gradt$, at the intersection of~$\tcoolt(\gamma)$ and~$t_{\rm X}(\gamma)$, depends on~$\uph$ and~$B_0$ but not on~$L$; and~$\gmax$, at the intersection of~$t_{\rm X}(\gamma)$ and~$L/c$, depends on~$B_0$ and~$L$ but not on~$\uph$ (it is the only n\onradiative[]parameter of the three). The condition,~$\gradt^2 = \gcool \gmax$ [equation~(\ref{eq:gradgeomean})], follows from the isosceles geometry. This diagram corresponds to the values of~$\gcool$,~$\gradt$, and~$\gmax$ adopted for all three of our simulations with radiative cooling.}
    \label{fig:tstop_thom}
\end{figure}

The Lorentz factors~$\gcool$ and~$\gradt$ are not independent; they are both set by the same underlying radiative parameter~$\uph$. Hence,~$\gradt$ is fixed by~$\gcool$ and~$\gmax$ \citep{mwu21}:
\begin{align}
    \gradt^2 = \gcool \gmax \, .
    \label{eq:gradgeomean}
\end{align}
Equation~(\ref{eq:gradgeomean}) reflects the fact that, since the X-point acceleration time -- which defines~$\gradt$ -- is faster than the system \lc[]time -- which defines~$\gcool$ -- one generally has~$\gcool < \gradt < \gmax$. The exception is the n\onradiative[]regime, where~$\gcool > \gradt > \gmax$ and all particles cool on \ts[s]exceeding~$L/c$. The topology of the \ts[s~$L/c$,]~$t_{\rm X}(\gamma)$, and~$\tcoolt(\gamma)$ induced by the presence of Thomson IC cooling and c\haracteriz[ed]by the energy scales~$\gcool$,~$\gradt$, and~$\gmax$ is shown in Fig.~\ref{fig:tstop_thom}.

Let us now summarize the situation for Thomson IC cooling. In this regime, according to equation~(\ref{eq:pthom}), cooling depends only on~$\uph$. The seed photon energy,~$\eph$, completely drops out of the dynamics (though it still influences the observed photon energies). The parameter~$\uph$ can then be recast in terms of the particle energy scales~$\gcool$ and~$\gradt$, which c\haracteriz[e]intersection points of important \ts[s]in the reconnection problem (Fig.~\ref{fig:tstop_thom}).

\begin{figure}
    \centering
    \begin{subfigure}{\columnwidth}
        \centering
        \includegraphics[width=\linewidth]{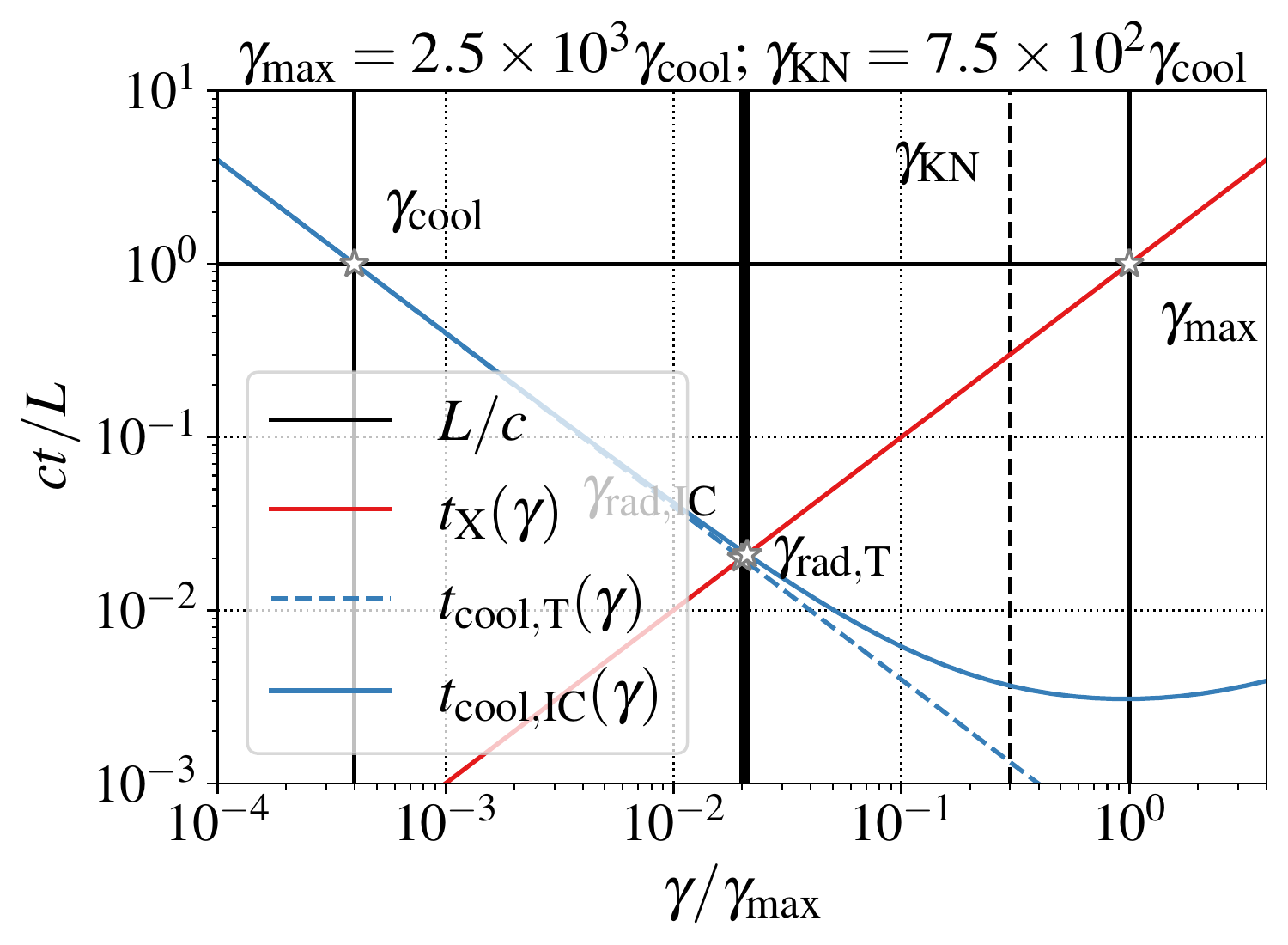}
    \end{subfigure}
    \begin{subfigure}{\columnwidth}
        \centering
        \includegraphics[width=\linewidth]{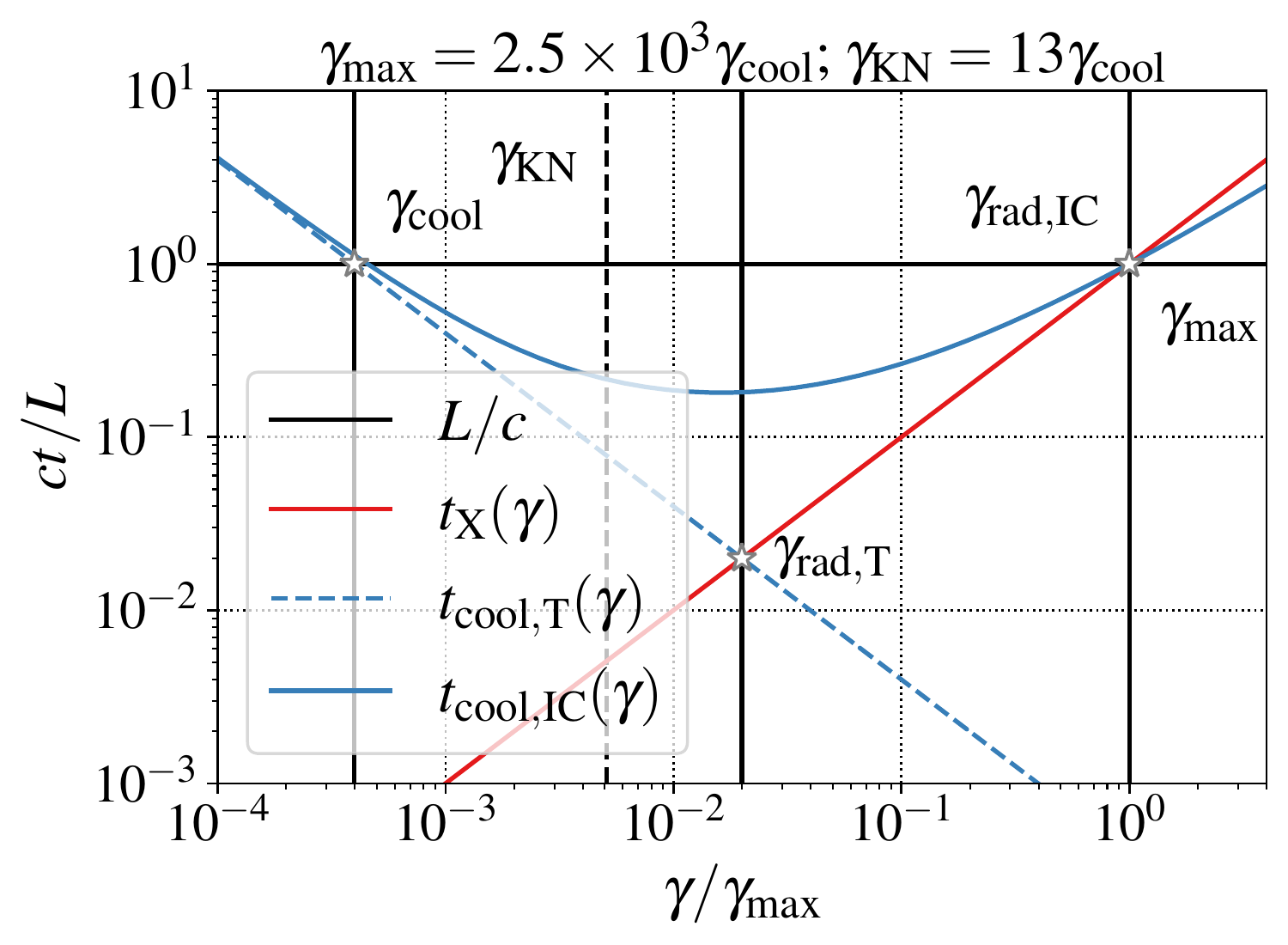}
    \end{subfigure}
    \begin{subfigure}{\columnwidth}
        \centering
        \includegraphics[width=\linewidth]{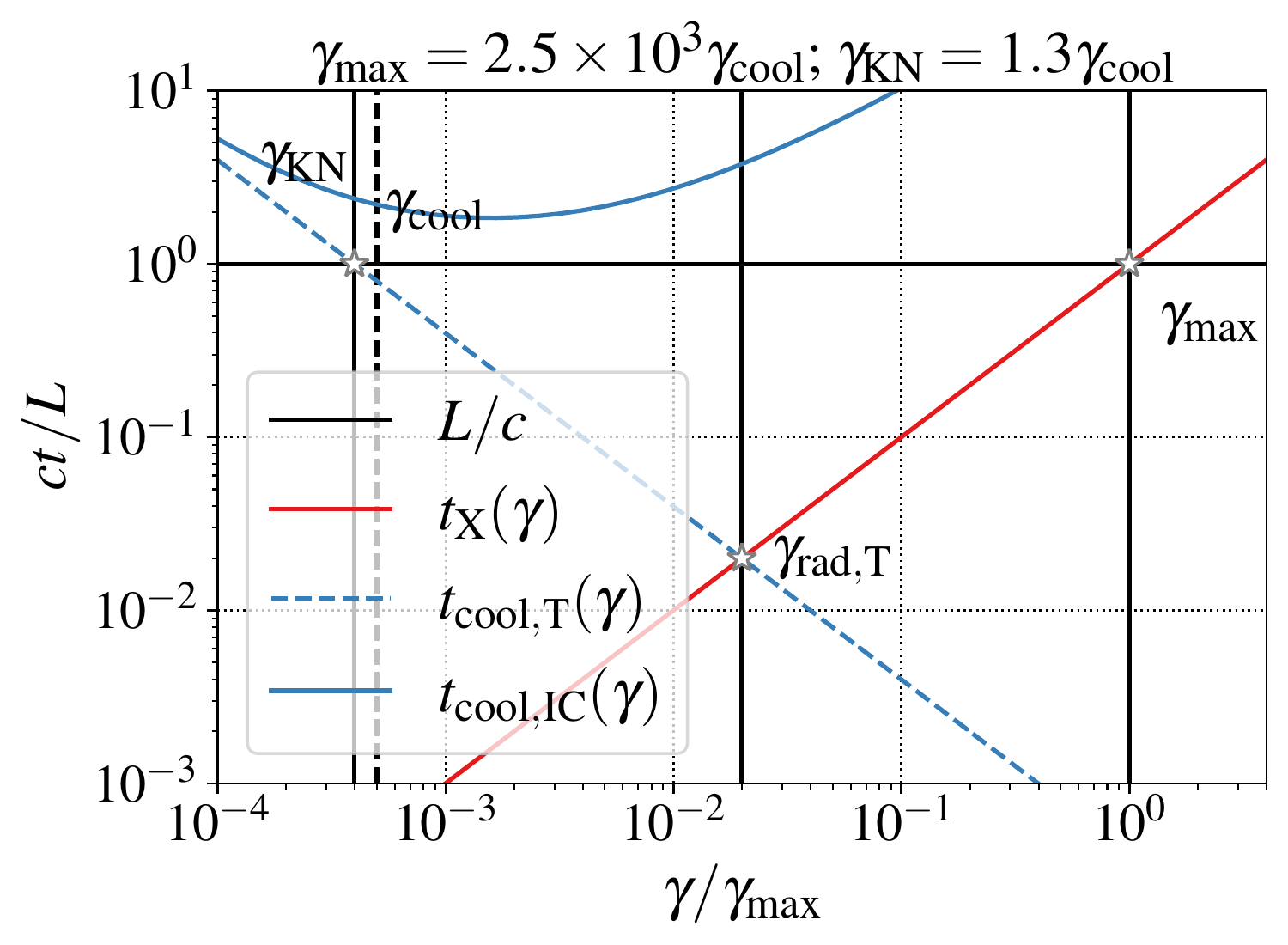}
    \end{subfigure}
    \caption{The addition of~$\gkn$ to the scales~$\gcool$,~$\gradt$,~$\gmax$ qualitatively changes the topology of the \ts[s,~$L/c$,~$t_{\rm X}(\gamma)$,]and~$\tcoolk(\gamma)$. When~$\gkn > \gradt > \gcool$ (top panel), the situation reduces essentially to that of Fig.~\ref{fig:tstop_thom}, where particles are cooled exclusively in the Thomson regime and none of the Comptonized photons can produce electron-positron pairs through absorption by the soft background. When~$\gcool < \gkn < \gradt$ (middle panel), the main case of interest to this study, particles may efficiently radiate in the deep Klein-Nishina regime and emit pair-producing gamma-rays with mean free paths~$< L$. Finally, when~$\gkn < \gcool < \gradt$ (bottom panel), particles cool on \ts[s]longer than~$L/c$. Though their radiation may be above pair-threshold, the pair production optical depth~$\tau_{\gamma\gamma} \sim \gkn / \gcool$ is small. The Lorentz factors illustrated in the middle panel correspond to those of our~$2$ simulations with Klein-Nishina radiative cooling.}
    \label{fig:tstop_kn}
\end{figure}

To illuminate the influence of the seed photon energy,~$\eph$, in the general IC case (i.e.\ including the Klein-Nishina regime), a similar procedure can be employed as for the Thomson limit. Here, the relevant energy scale in terms of which~$\eph$ is recast,~$\gkn$, has already been introduced; it is the Lorentz factor above which particles lose their energies in discrete photon quanta. We then need only to repeat the comparison of \ts[s]as done above, but now replacing the Thomson-regime~$\tcoolt(\gamma)$ with the more general~$\tcoolk(\gamma)$. This is done graphically in Fig.~\ref{fig:tstop_kn}. In the figure, we define the new auxiliary energy scale~$\gradk$ as the g\eneraliz[ation]of~$\gradt$ to include Klein-Nishina effects:~$\tcoolk(\gradk) \equiv t_{\rm X}(\gradk)$.

The diagrams of Fig.~\ref{fig:tstop_kn} illustrate three basic \ts[]topologies as~$\gkn$ is brought in from infinity (equivalently, as the seed photon energy,~$\eph$, is increased). First, when~$\gkn > \gradt$ (Fig.~\ref{fig:tstop_kn}, upper panel), the g\eneraliz[ed]radiative Lorentz factor,~$\gradk$, remains approximately equal to~$\gradt$. Thus, radiative losses inhibit particles from accessing energies~$\geq \gkn$ where they would experience Klein-Nishina effects. This regime therefore essentially reduces to the Thomson limit of radiative reconnection.

The next regime occurs once~$\gkn$ falls below~$\gradt$ (Fig.~\ref{fig:tstop_kn}, middle panel). Then, Klein-Nishina effects [entering through the function~$f_{\rm KN}(\gamma/\gkn)$ falling appreciably below unity] create a pronounced departure of~$\tcoolk(\gamma)$ from the Thomson limit. As a result,~$\gradk$ rises rapidly \citep[it depends super-exponentially on~$\gradt/\gkn > 1$;][]{mwu21} until it eventually crosses~$\gmax$. The middle panel of Fig.~\ref{fig:tstop_kn} illustrates the particular case where~$\gradk = \gmax$, which already occurs when~$\gkn$ is smaller than~$\gradt$ by just a factor of a few. As~$\gkn$ continues to diminish from this point,~$\gradk$ comes to exceed~$\gmax$, meaning that Klein-Nishina effects effectively remove the radiative limit on direct X-point acceleration (though of course the intrinsic and system-size limits,~$\gx$ and~$\gmax$, are still present). Nevertheless, as long as~$\gkn \gtrsim \gcool$, there remains a range of particle energies c\haracteriz[ed]by relatively fast cooling, with~$\tcoolk(\gamma) < L/c$.

Finally, as~$\gkn$ approaches~$\gcool$ (Fig.~\ref{fig:tstop_kn}, lower panel), the~$\tcoolk(\gamma)$ curve is lifted above the line~$L/c$ for all~$\gamma$. Here, Klein-Nishina effects suppress cooling to the point that all particles cool on times longer than~$L/c$, an effectively n\onradiative[]regime.

Of these three cases, the one where Klein-Nishina effects are expected to influence the reconnection dynamics is the second; only there can particles access energies~$> \gkn$ while maintaining a relatively rapid cooling time. We hence call this the \textit{Klein-Nishina radiative regime}, c\haracteriz[ed]by the scale hierarchy
\begin{align}
    \gcool < \gkn < \gradt < \gmax \, .
    \label{eq:knregime}
\end{align}
As shown next, this hierarchy is not just important from the standpoint of radiative cooling; it is also where pair production from the emitted photons may play an important role.

Particles with Lorentz factor~$\gkn$ scatter photons to typical energies~$\escat \sim \gkn \me c^2 = (\me c^2)^2 / 4 \eph$: i.e. close to pair-production threshold with the seed-photon background. This means that particles must necessarily be accelerated above~$\gkn$ in order for a significant fraction of their radiated energy to be recaptured as fresh pairs. In addition, the fiducial pair-production optical depth is [using the peak \crosssection[~$\sigma_{\gamma\gamma} \simeq \sigma_{\rm T} / 5$;]equation~(\ref{eq:siggg}); see also \citet{mwu21}]
\begin{align}
    \tau_{\gamma\gamma} = \frac{\uph \sigma_{\rm T} L}{5 \eph} = \frac{3 \gkn}{5 \gcool} \, ,
    \label{eq:taugg}
\end{align}
meaning that placing~$\gkn$ between~$\gcool$ and~$\gradt$ enables both the emission of above-threshold gamma-rays and their absorption inside the system on \ts[s~$L/c\tau_{\gamma\gamma}<L/c$.]

In summary, the Klein-Nishina radiative scale hierarchy,~$\gcool < \gkn < \gradt < \gmax$, triggers three simultaneous and important QED effects:
\begin{enumerate}
    \item it permits particles to reach energies~$> \gkn$, where their IC cooling transitions to the quantized Klein-Nishina regime;
    \item it permits particles to reach energies~$> \gkn$, where their IC-emitted gamma-rays exceed pair-production threshold with the background radiation bath; and
    \item it guarantees almost all of these gamma-rays to be absorbed inside the system on sub-dynamical~($< L/c$) \ts[s,]allowing the resulting pairs to feed back on reconnection.
\end{enumerate}
This is the target regime of this study, and how we r\ealiz[e]it in numerical simulations is the topic of the next section.

\subsubsection{Selection of the radiative reconnection parameters}
\label{sec:radscaleschoose}
We now discuss our choices of the radiative reconnection parameters,~$\uph$ and~$\eph$, recast, as described in the preceding section~\ref{sec:radscalesintro}, in terms of the energy scales~$\gcool$,~$\gradt$, and~$\gkn$, plus the pair-production optical depth~$\tau_{\gamma\gamma}$. We set as a first goal the Klein-Nishina scale hierarchy,~$\gcool < \gkn < \gradt < \gmax$, discussing afterward the necessary placement of the remaining n\onradiative[]energy scale,~$\sigc$ (or, equivalently,~$\gx \equiv 4 \sigc$) within this base ordering. A key issue is that as much space as possible needs to be opened up between each successive energy scale in the Klein-Nishina hierarchy. The reason for this is that a high optical depth,~$\tau_{\gamma\gamma} = 3 \gkn / 5 \gcool$, demands that~$\gkn$ substantially exceed~$\gcool$. At the same time,~$\gkn$ cannot be as large as~$\gradt$ without, as previously described, leading to a prohibitively small radiatively limited Lorentz factor,~$\gradk$. Thus, the gaps from~$\gcool$ to~$\gkn$ and, then, from~$\gkn$ to~$\gradt$ both need to be as wide as resolution requirements permit. In effect, the separation from~$\gcool$ to~$\gradt$ needs to be maximized, which is equivalent, through equation~(\ref{eq:gradgeomean}), to maximizing the ratios~$\gmax / \gradt$ and~$\gmax / \gcool$. Hence, resolving the necessary scale hierarchy demands pushing the outer scales,~$\gcool$ and~$\gmax$, as far away from each other as possible.

Let us now examine the implications this has on numerical cost. We begin by noting that~$\gcool$ cannot be made arbitrarily small. Otherwise, the ambient upstream plasma becomes efficiently radiative, appreciably cooling down over the course of the simulation. Physically, this renders the initial background plasma temperature,~$\itempt$, meaningless and, hence, makes interpreting the simulation results more difficult; it also means, numerically, that the upstream Debye length,~$\lambda_{\rm D,0}$, quickly becomes unresolved (we only resolve it marginally to begin with:~$\lambda_{\rm D,0} = 1.2 \Delta x$), leading to spurious numerical heating. If, to avoid this, one requires the upstream plasma cooling time to be at least some factor~$M > 1$ longer than~$L/c$, then equation~(\ref{eq:tcoolt}) implies~$\gcool > M \langle \gamma \rangle = 3 M \itemp$. Here,~$\langle \gamma \rangle = 3 \itemp = 72$ is the initial mean upstream Lorentz factor, and~$\tcoolt$ can be used instead of~$\tcoolk$ because upstream particles cool in the Thomson regime. We find empirically that our simulations take~$\simeq 3L/c$ for the reconnected magnetic flux to saturate, and thus we conservatively set~$M = 6$, giving~$\gcool = 430 = 3.6 \times 10^{-3} \sigc = \gmax / 2500$.

What the preceding paragraph shows, importantly, is that, regardless of the particular choice for~$M$, the minimum value of~$\gcool$ is inevitably tied, through~$\itemp$, to the Debye-length resolution requirement. The ratio~$\gmax / \gcool$ thus becomes a proxy for~$L / \lambda_{\rm D,0} \simeq L / \Delta x \equiv N$, the number of cells (in the $x$-direction) across the simulation box. Specifically
\begin{align}
    N &= L / \Delta x = 1.2 L / \lambda_{\rm D,0} = 1.2 L / \sqrt{\itemp \sigc} \, \rho_0 \notag \\
    &\simeq 12 \gmax / 2 \itemp \sqrt{\sigh} = 18 M \gmax / \gcool \sqrt{\sigh} = 7680 \, ,
    \label{eq:simcost}
\end{align}
where we used the~$\itemp \gg 1$ approximation,~$\sigh \simeq \sigc / 4 \itemp$. Our available resources limit us to~$N=7680$, and, hence, for~$\sigh = 1250$, to~$\gmax / \gcool \simeq 2500$. This ratio is, nevertheless, sufficient to realize a healthy Klein-Nishina energy-scale hierarchy.

With~$\gmax/\gcool = \sqrt{\gmax / \gradt} = \sqrt{\gradt / \gcool}$ set, we must now decide where to place~$\gkn$. We empirically find that a good fiducial choice is to set~$\gkn / \gradt$ such that~$\gradk = \gmax$. Raising~$\gkn$ from here lowers~$\gradk$, somewhat inhibiting reconnection-powered NTPA and, hence, the production of high-energy gamma-rays and pairs. On the other hand, lowering~$\gkn$ limits the pair-production optical depth,~$\tau_{\gamma\gamma} \propto \gkn / \gcool$, and, simultaneously, the range of energies where particles are efficiently cooled,~$\tcoolk(\gamma) < L/c$ (an extreme case of which is the lower panel of Fig.~\ref{fig:tstop_kn}). Overall, this slows the pair-production response of the system: high-energy particles take longer to radiate pair-producing photons, and those photons travel farther before being absorbed to create pairs. Choosing, then,~$\tcoolk(\gradk) = t_{\rm X}(\gradk) = L/c$ to set~$\gradk = \gmax$, gives~$\gkn = 5500 = 0.046 \sigc = 0.26 \gradt = 13 \gcool$ and~$\tau_{\gamma\gamma} = 3 \gkn / 5 \gcool = 7.7$.

Although our fiducial~$\gkn$ is finely tuned, this parameter is much more flexible in real astrophysical systems where numerical requirements do not limit the scale separation. With much larger values of~$\gmax / \gcool$, astrophysical reconnection may have~$\gkn$ different from our numerical sweet spot while still preserving the important features: copious particle acceleration above~$\gkn$, efficient cooling of high-energy particles, and short mean-free-paths of emitted gamma-rays. A much more detai\led[]discussion of this point for two concrete astrophysical systems -- FSRQ jets and black hole accretion disc coronae -- is presented by \citet{mwu21}.

Having set~$\gkn$,~$\gcool$,~$\gmax$, and~$\tau_{\gamma\gamma}$, we are at last equipped to justify our choice of~$\sigc$. Even though~$\sigc$ is technically a n\onradiative[]reconnection parameter (that we have already discussed in section~\ref{sec:nonradsetup}), it has unique consequences in the presence of Klein-Nishina radiative physics. Namely,~$\sigc$ needs to be at least a factor of several higher than~$\gkn$ in order for the intrinsic X-point particle acceleration limit,~$\gx = 4 \sigc$, to lie deeply in the regime where reconnection-energized particles emit pair-producing photons. As discussed by \citet{mwu21}, since the pair-production \crosssection[,~$\sigma_{\gamma\gamma}$][equation~(\ref{eq:siggg})], peaks at gamma-ray energies~$\simeq 3.6 (\me c^2)^2 /\eph \simeq 14 \gkn \me c^2$, and because a particle in the Klein-Nishina limit tends to donate about half its energy to a Comptonized photon, particles emitting photons at peak pair-production \crosssection[]have typical Lorentz factors~$\gamma \simeq 30 \gkn$. To permit reconnection-energized particles to easily pass this limit, we have set~$\sigc = 22 \gkn$~($\gx = 88 \gkn$). This is why our simulations feature such a high~$\sigc$ and, consequently, a relatively low~$\gmax / \sigc = 0.1 L / \sigc \rho_0 = 9.1$ given their numerical size. The issue is that~$\sigc$ needs to be on the upper end of our energy-scale hierarchy, which is already strapped to maximize~$\gmax / \gcool$, and so~$\sigc$ ends up being somewhat close to~$\gmax$.

\begin{figure*}
    \centering
    \includegraphics[width=\linewidth]{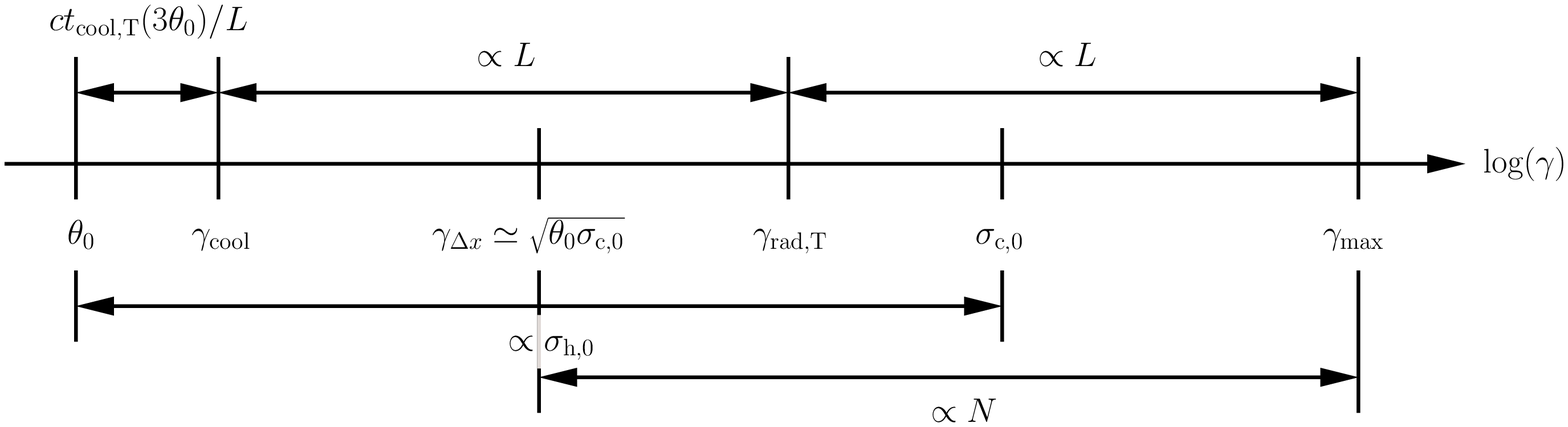}
    \caption{Important radiative~($\gcool$ and $\gradt$) and non-radiative~($\itemp$,~$\sigc$, and~$\gmax$) energy scales in reconnection connected with lines indicating how their relative magnitudes depend on: the system size,~$L$; the number of simulation cells in the~$x$-direction,~$N$; the characteristic cooling time,~$c\tcoolt(3 \itemp)/L$, of the initial upstream particles~($M$ in the text); and the hot magnetization,~$\sigh$. The labels~$\sigh$ and~$c\tcoolt(3 \itemp)/L$ follow only in the relativistically hot case,~$\itemp \gg 1$. The Klein-Nishina Lorentz factor~$\gkn$ is not drawn because it can be freely selected independently of all other diagrammed quantities (since it depends only on~$\eph$, but not any of the other parameters, like~$\uph$ and~$L$, that the other scales depend on).}
    \label{fig:scalenumline}
\end{figure*}
A summary of all our radiative and n\onradiative[]simulation parameters is given in Table~\ref{table:params}. In addition, we present a graphical description of how the various reconnection energy scales relate to key physical and numerical quantities in Fig.~\ref{fig:scalenumline}. The figure also illustrates how changing one quantity in this high-dimensional parameter space affects the others. For example, a simple system-size scan requires added care in the presence of all of these radiative effects, because changing~$L$ (for example, by changing~$N$), displaces not only~$\gmax$ higher, but also~$\gcool$ lower, reducing the cooling time of the upstream particles with respect to~$L/c$. Thus, to conduct such a scan, one would need to take care to set~$\itemp$ low enough such that, even at the end of the scan (highest~$L$), the upstream particles are still sufficiently cold that they do not radiate appreciably.

\section{Klein-Nishina impact on reconnection}
\label{sec:results}
Here we present and compare the four main simulations whose setup is discussed in section~\ref{sec:setup}. We stress that, all parameters being the same (Table~\ref{table:params}), the simulations differ only in their mode\led[]physics. In our n\onradiative[]run, \quoted[,]{no rad.}radiative cooling and pair production are completely turned off. In our Thomson-radiative run, \quoted[,]{IC(Th)}we apply continuous Thomson radiative cooling, but pair production remains absent. In the runs \quoted[]{IC(KN)}and \quoted[,]{IC(KN)+PP}we employ our general IC cooling scheme (including the high-energy Klein-Nishina regime; sections~\ref{sec:icsimple} and \ref{sec:codeopt}), but only self-consistently calculate pair production (section \ref{sec:pp}) in the IC(KN)+PP run. In this sense, the IC(KN) run is artificial. In it, we simply pretend that all radiation emitted by particles, even the part above pair-threshold, is permanently lost from the simulation.

In each of the following subsections, we a\nalyz[e]one aspect of the simulations, starting with those that are more similar among the four and moving to those that are more different. We finish by addressing unique properties of the run IC(KN)+PP that only exist in the context of pair production.

\subsection{Global spatial evolution}
\label{sec:globaldynamics}
\begin{figure*}
    \centering
    \includegraphics[width=\linewidth]{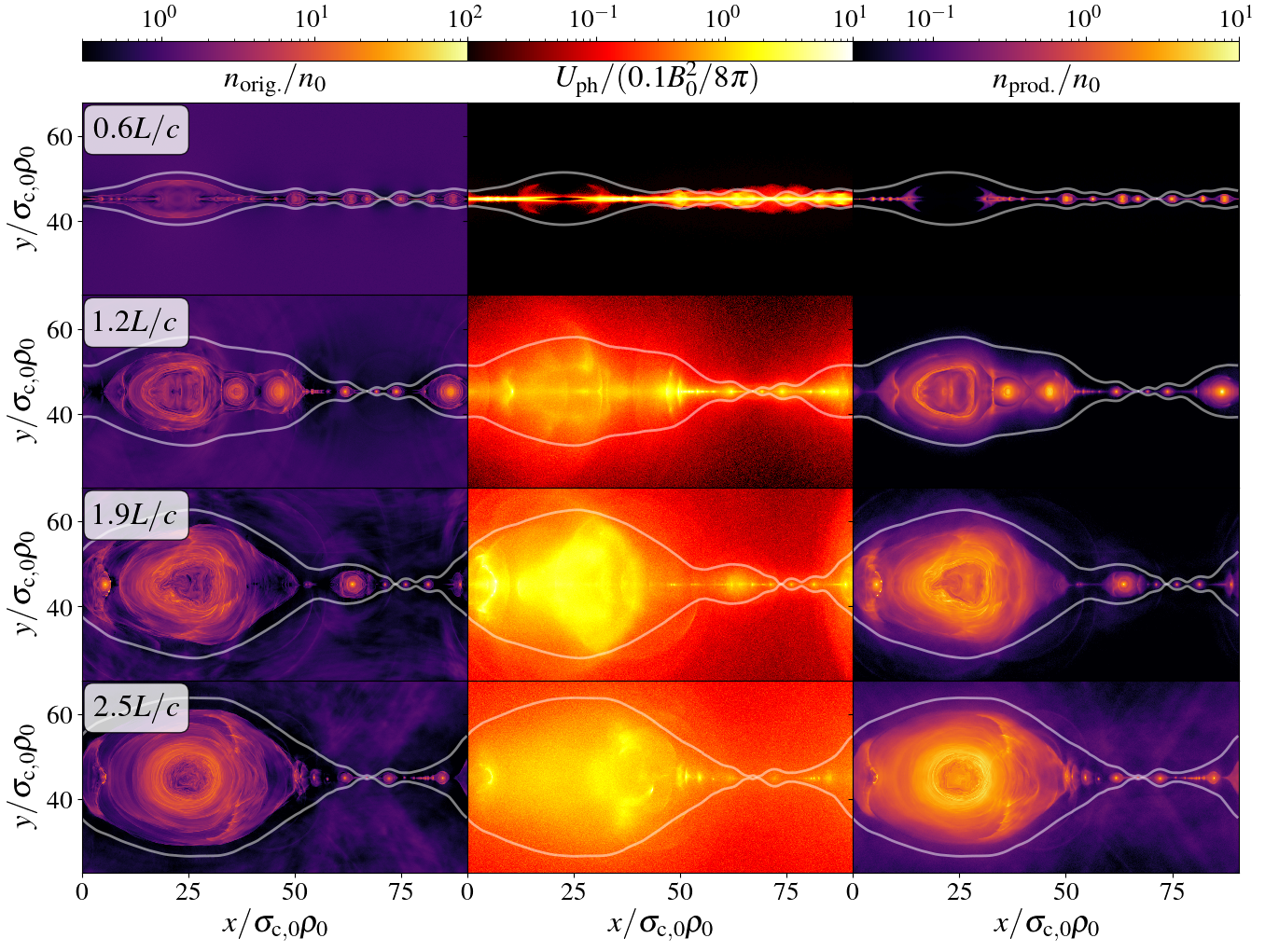}
    \caption{Time snapshots of the original pair number density (left column), above-threshold photon energy density (middle column), and produced pair number density (right column) for the main Klein-Nishina reconnection simulation with pair production, IC(KN)+PP. In each panel, the reconnection separatrix is drawn in white and intersects at the main reconnection X-point. All runs exhibit similar large-scale temporal evolution to that depicted: the initial current sheet quickly tears into a hierarchical chain of merging plasmoids (e.g.\ left column). In addition to this familiar picture, in the IC(KN)+PP run, regions of active particle acceleration are signa\led[]by flashes of gamma-rays (middle column). These photons are eventually absorbed to build up the population of particles born \textit{in situ} (right column).}
    \label{fig:moviestills}
\end{figure*}

We present the large-scale temporal evolution of the simulations in Fig.~\ref{fig:moviestills}. For generality, the figure depicts the run, IC(KN)+PP, with general IC cooling and pair production, but the temporal evolution in terms of the spatial plasma number density (left column) is similar in all runs. 

At the simulation onset, we apply a~$1$ per cent perturbation to the in-plane magnetic field. This seeds the tearing instability in the initial current sheet, rapidly disrupting it into a chain of plasmoids separated by smaller current sheets, themselves tearing-unstable. The successive tearing of these current sheets down to smaller and smaller scales yields a self-similar hierarchy of plasmoids and inter-plasmoid current layers (\citealt{uls10}; e.g.\ Fig.~\ref{fig:moviestills} snapshots at~$t=0.6L/c$ and~$t=1.2L/c$). As the plasmoids merge with one another, they also create separate miniature reconnection sites with current sheets oriented perpendicular to those of the main plasmoid chain (parallel to the~$yz$-plane instead of to the~$xz$-plane). These reconnection sites -- for example at~$x \simeq 35 \sigc \rho_0$ in the~$t=1.2L/c$ snapshot and at~$x \simeq 5 \sigc \rho_0$ in the~$t=1.9L/c$ snapshot -- beget their own plasmoid hierarchies.\footnote{In a real instance of astrophysical reconnection, the outer scale,~$L$, would dwarf the plasma microscales (e.g.~$\sigc \, \rho_0$) by many orders of magnitude, leading to a deep self-similar hierarchy both in the main plasmoid chain and in the recursive ones birthed between merging plasmoids. However, in our simulations with limited computational resources, we only witness the primary chain and the first secondary plasmoid-merger chains.}

In both cases -- whether in the midplane plasmoid chain or at vertical reconnection regions between colliding plasmoids -- reconnecting current sheets are prominent sites of particle acceleration. This fact is underscored in the depicted IC(KN)+PP run by the copious emission of gamma-ray (above pair-production threshold) radiation from these regions in the middle column of Fig.~\ref{fig:moviestills}.

Given the periodic boundary conditions of our setup, the large plasmoid with center at~$x \simeq 25 \sigc \rho_0$ in our simulations serves as an exhaust for the plasma processed by reconnection. Reconnection slows down and eventually stalls once about half of the initial magnetic flux in the box is reconnected. This corresponds to a state where the \textit{separatrix} -- the topological boundary between the domains of reconnected and unreconnected flux -- crosses the midplane at an angle of about~$45^\circ$ (crosses itself at~$90^\circ$ angles) at the dominant X-point and opens up around the large exhaust plasmoid, which is then the only plasmoid remaining in the layer.

Finally, even though the runs are quite similar in their global spatial evolution, one unique aspect of the IC(KN)+PP run is the difference in spatial coherence between the original particles (Fig.~\ref{fig:moviestills}, left column) and those produced \textit{in situ} (Fig.~\ref{fig:moviestills}, right column). Original particles are bound to magnetic field lines (except at reconnection X-points), and thus the striations in the original-particle number density follow closely the wrapping of magnetic field lines around large plasmoids. This forms a tree-ring pattern, with density striations and corresponding field lines tracing the history of magnetic flux accumulation onto each plasmoid. In contrast, for the produced particles, such an effect, while still discernible, is much less pronounced. This owes to the added channel through which produced particles can take up residence in plasmoids: they can be born there directly. They are not constrained, like the original particles, to essentially follow the reconnection of a given field line onto a plasmoid. This pollutes, for the produced particles, what would otherwise be pristine tree-ring plasmoid density striations.

\subsection{Reconnection rate}
\label{sec:recrate}
\begin{figure}
    \centering
    \includegraphics[width=\linewidth]{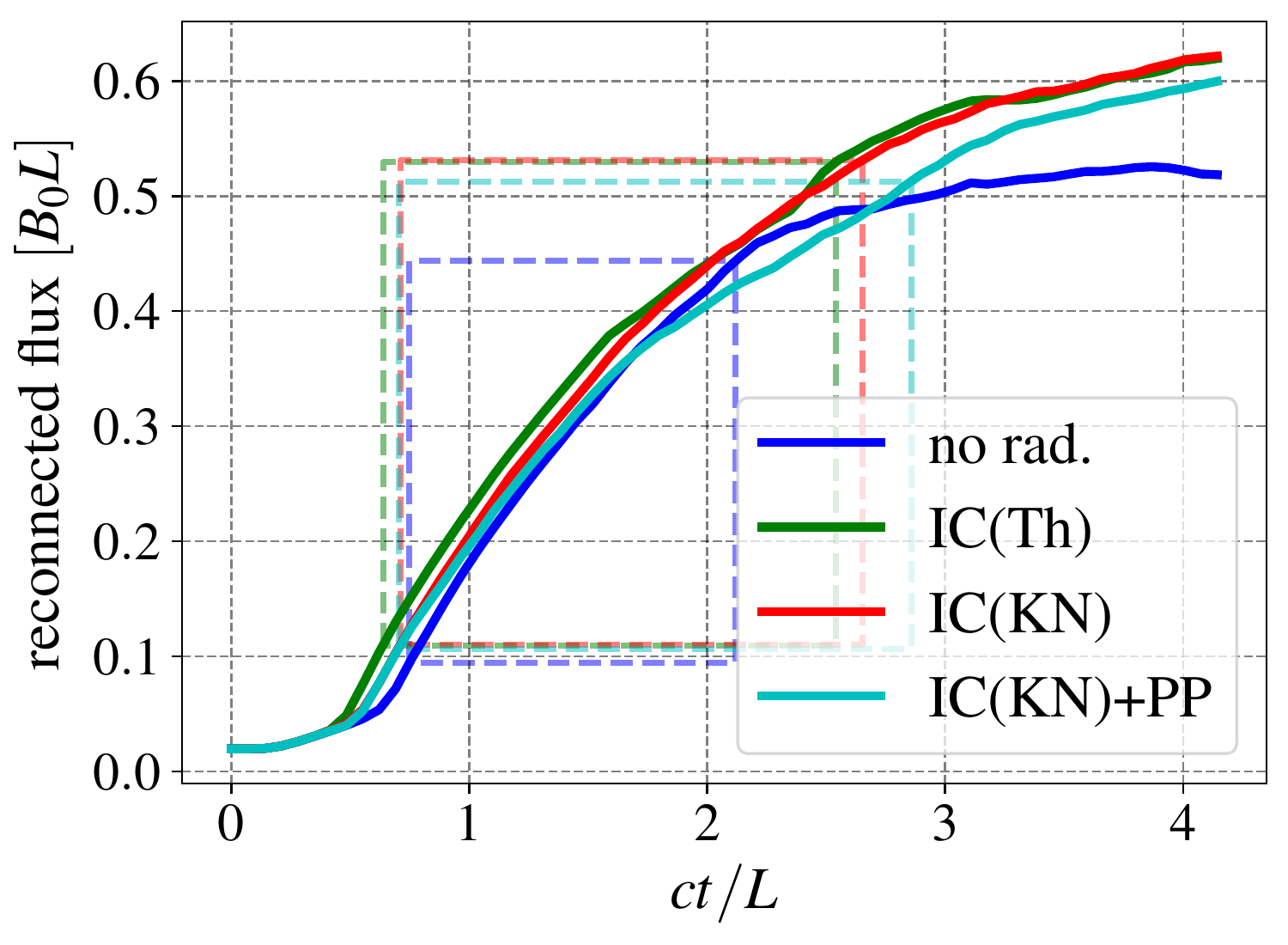}
    \caption{Reconnected magnetic flux as a function of time in our four main runs: no rad., IC(Th), IC(KN), and IC(KN)+PP. In this plot, we include information from both the top and bottom reconnection layers in our simulations. Thus, the saturation of each simulation's reconnected flux at~$\sim 0.5 B_0 L$ indicates that the initially available flux in the box,~$B_0 L_y / 2 = B_0 L$, has been approximately half consumed. The active phase of each run is displayed as a dashed box during which the middle~$70$ per cent of the simulation's total change in reconnected flux occurs. Each run's average reconnection rate (see Fig.~\ref{fig:recrate}) is the slope of the secant line through this box.}
    \label{fig:recflux}
\end{figure}

\begin{figure}
    \centering
    \includegraphics[width=\linewidth]{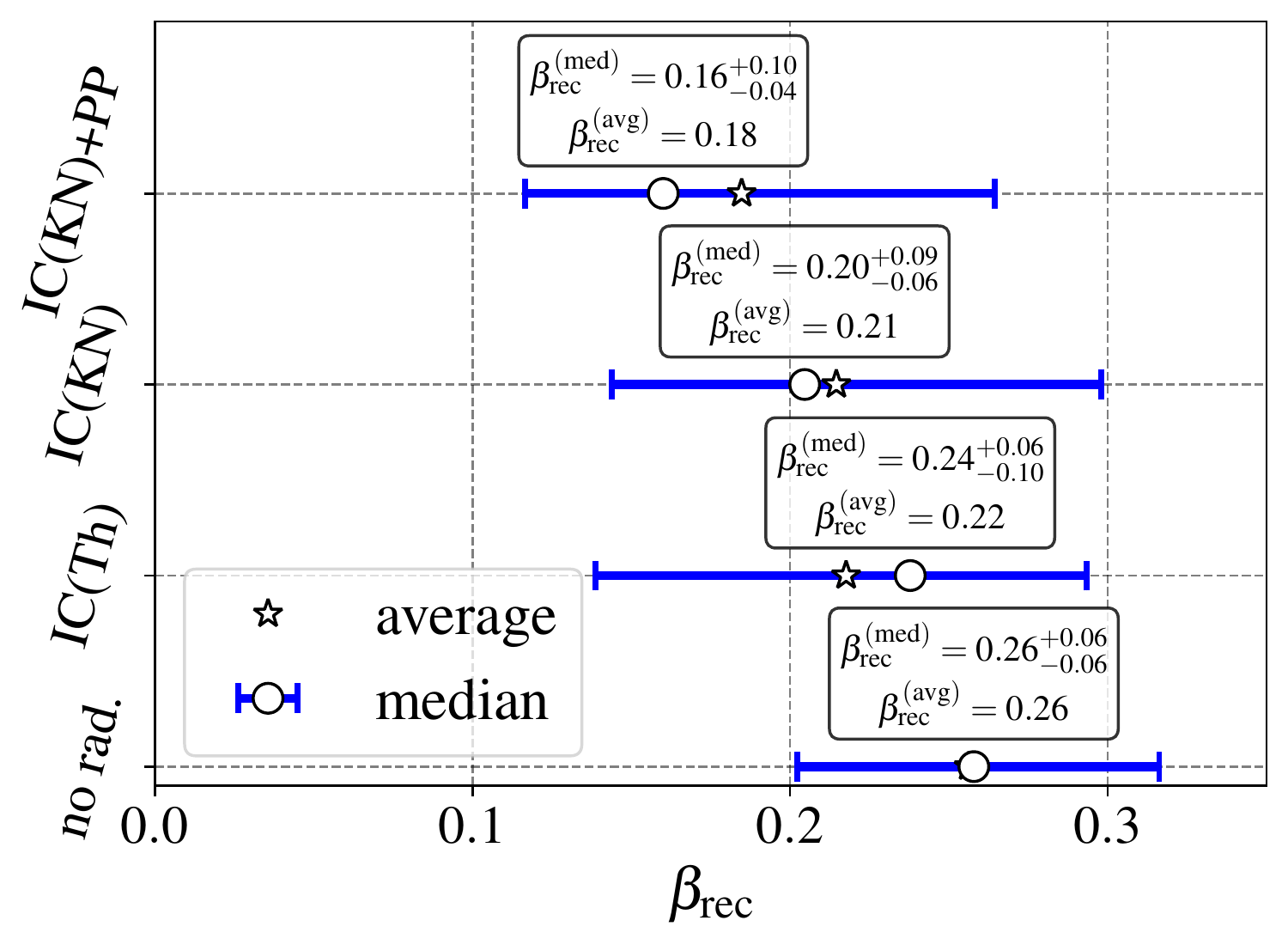}
    \caption{Per-simulation average, median, and one-sigma~($16$th and~$84$th) percentile reconnection rates. The reconnection rates between runs are statistically indistinguishable.}
    \label{fig:recrate}
\end{figure}

Consistent with their similar global evolution, all four of our principal runs exhibit statistically indistinguishable reconnection rates,~$\beta_{\rm rec}$. We measure these reconnection rates in Figs.~\ref{fig:recflux} and~\ref{fig:recrate}. Fig.~\ref{fig:recflux} shows the reconnected flux,~$\fluxsymb(t)$, in each simulation as a function of time,~$t$. To determine the characteristic reconnection rate, we consider the \textit{active phase} of each simulation, defined as the period,~$t \in [t_{\rm start}, t_{\rm end}]$, during which the middle 70 per cent of the change in reconnected flux (difference between initial and final points on the \mbox{Fig.-\ref{fig:recflux}} curves) occurs. This insulates the measurement from artificially slow values during reconnection onset while also reducing sensitivity to the late-time slowdown during reconnection saturation (see discussion in section~\ref{sec:globaldynamics}). We define the average reconnection rate as~$[\fluxsymb(t_{\rm end})-\fluxsymb(t_{\rm start})]/B_0 c (t_{\rm end} - t_{\rm start})$. Our measurements are s\ummariz[ed]in Fig.~\ref{fig:recrate}. 

Even in the active phase, however, reconnection does not proceed at a precisely uniform rate. To provide some idea of the temporal variation, we also compute instantaneous reconnection rates by differentiating the curves in Fig.~\ref{fig:recflux} with respect to time. We report the median and one standard-deviation~($16$th and~$84$th) percentiles of the resulting reconnection rate distributions in Fig.~\ref{fig:recrate}.

Concerning the differences between the simulations, the radiative runs generally reconnect about~$20$ per cent more magnetic flux than the n\onradiative[]simulation. This effect was previously noted by \citet{wpu19}, who interpreted it as resulting from radiative cooling tending to reduce plasma pressure support inside plasmoids, enhancing plasmoid compression \citep[cf.][]{sgu19, sgu23, hps19, hrp23}. As a result, more reconnected magnetic flux is needed to achieve the saturated geometry where the separatrix forms~$90^\circ$ angles with itself at the principal X-point (section~\ref{sec:globaldynamics}). This effect primarily alters the late-time flux saturation; it has little impact on the rate of reconnection during the active phases of our simulations.

Additionally, we observe some radiatively dependent skew in our measured reconnection rate distributions. Notably, the run IC(KN)+PP is almost bimodal in its reconnection rate distribution, reconnecting relatively quickly during the first half of its active period and slowing down during the second half. As a result, the median reconnection rate falls substantially below the mean. This could be a sign of the pair feedback anticipated by \citet{mwu21}, where the pairs produced in the upstream region load the upstream plasma, thereby reducing~$\sighgen$ and hence inhibiting reconnection, but it is not statistically significant. We examine the issue of pair feedback again in later sections.

\subsection{Non-thermal particle acceleration}
\label{sec:ptcldists}
\begin{figure*}
    \centering
    \begin{subfigure}{0.49\linewidth}
        \centering
        \includegraphics[width=\linewidth]{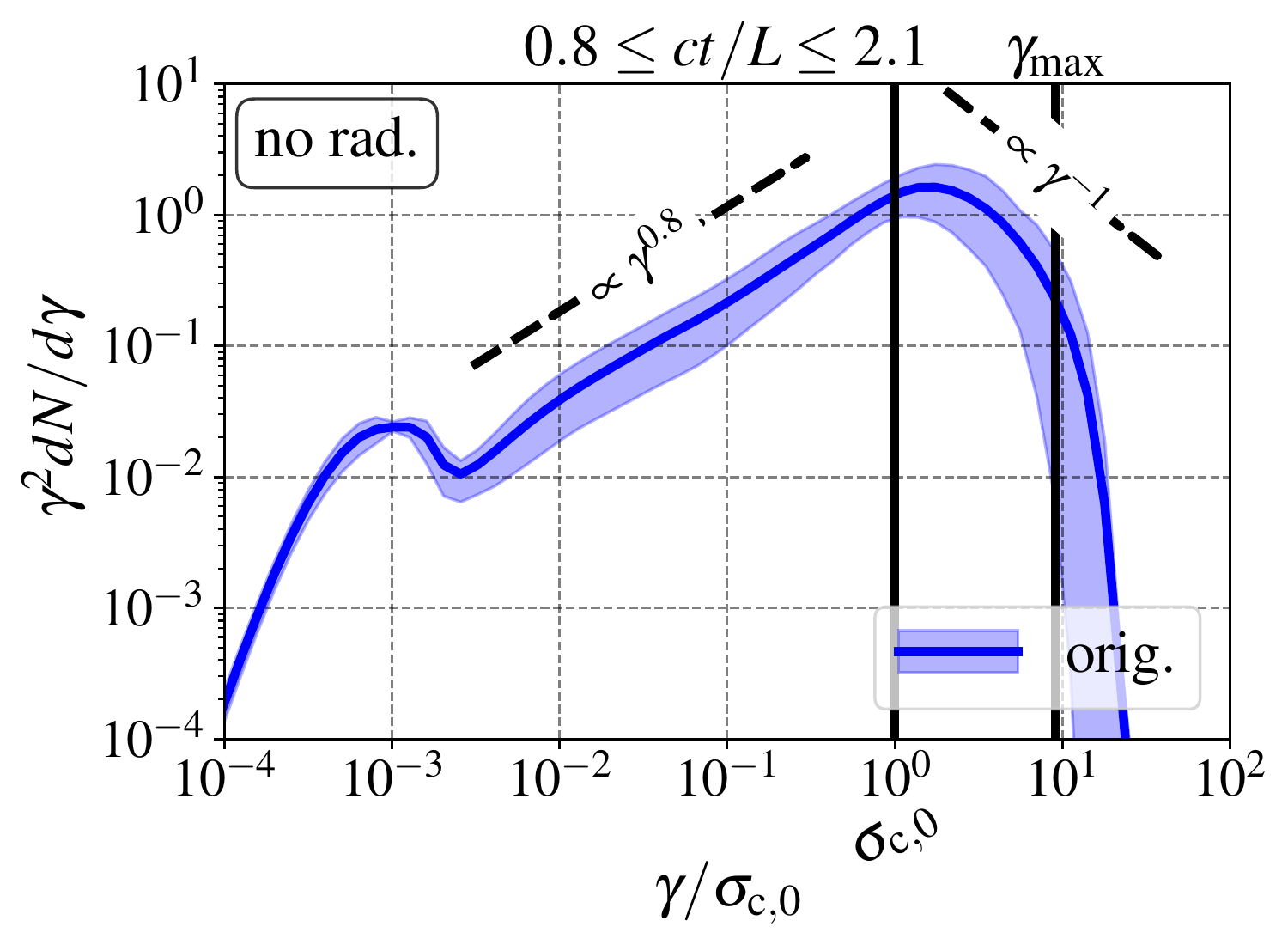}
    \end{subfigure}
    \begin{subfigure}{0.49\linewidth}
        \centering
        \includegraphics[width=\linewidth]{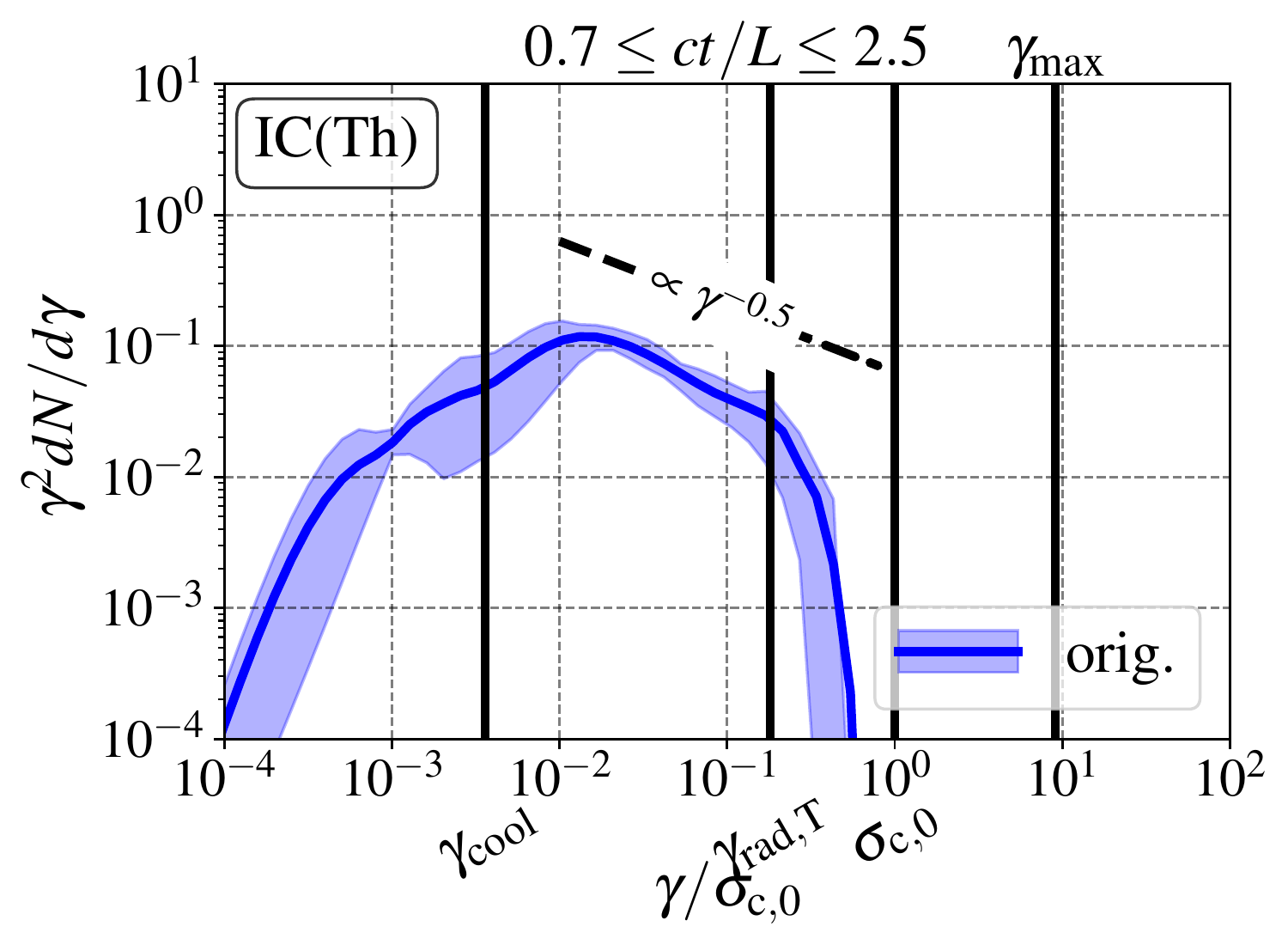}
    \end{subfigure}
    \begin{subfigure}{0.49\linewidth}
        \centering
        \includegraphics[width=\linewidth]{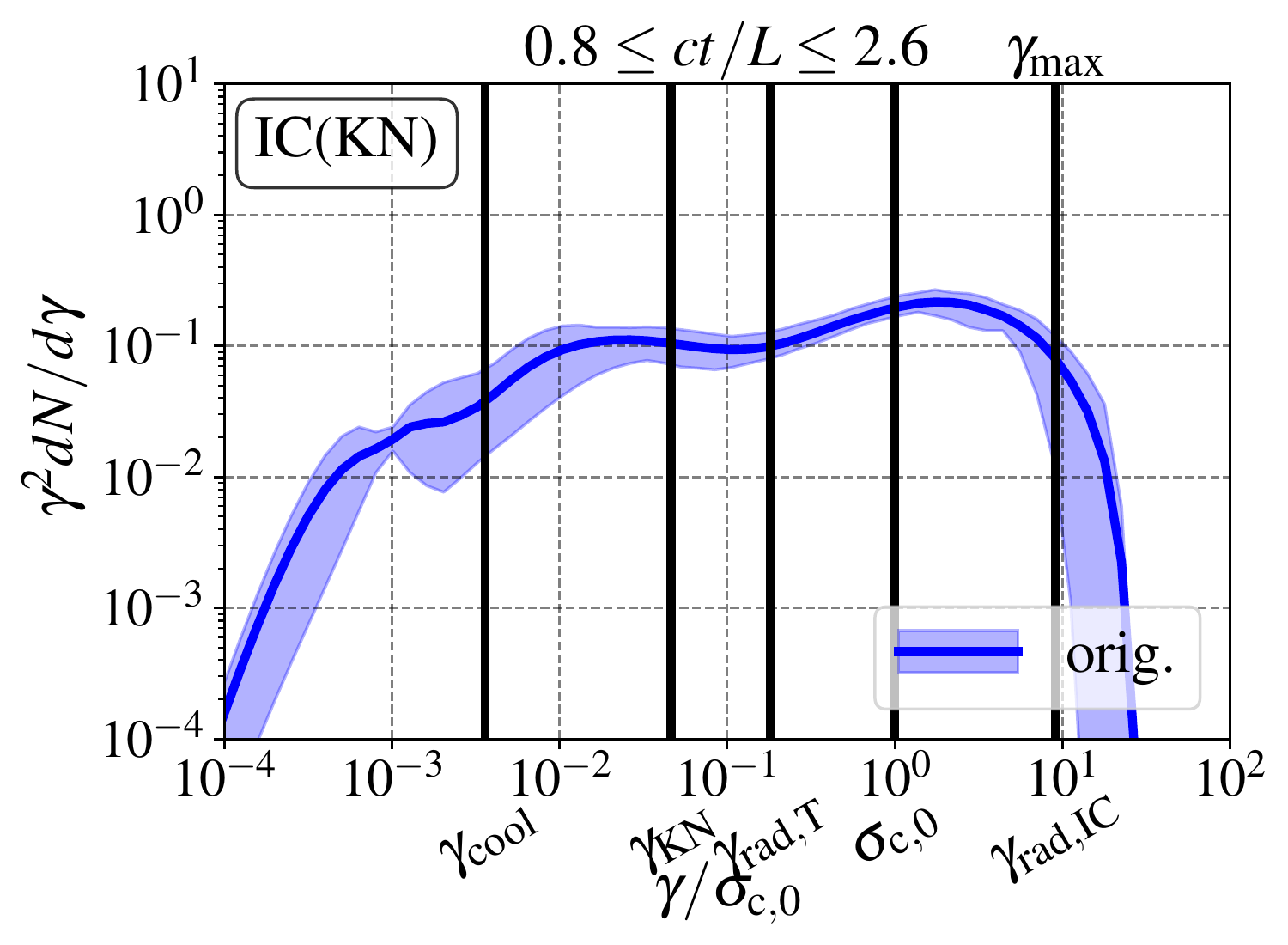}
    \end{subfigure}
    \begin{subfigure}{0.49\linewidth}
        \centering
        \includegraphics[width=\linewidth]{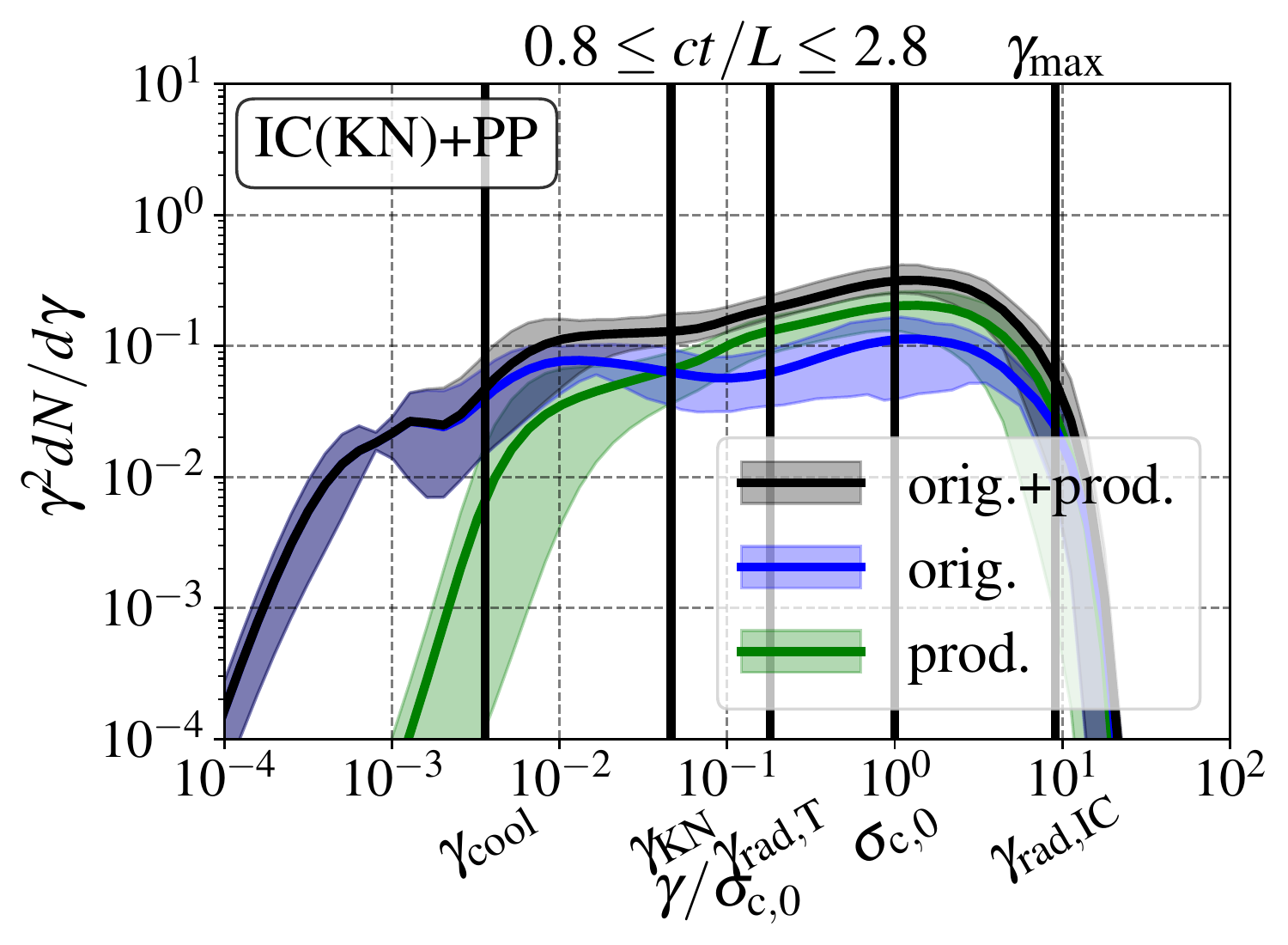}
    \end{subfigure}
    \caption{Time-averaged particle energy distributions for our four main runs. The averaging interval is the same as Fig.~\ref{fig:recflux}: the period during which the middle 70 per cent of the change in the reconnected flux occurs. Error envelopes display the one-sigma~($16$th and~$84$th) percentiles during the averaging interval in each energy bin. The n\ormaliz[ation]is arbitrary but equal on all panels. The IC(KN)+PP distribution is decomposed into contributions from particles that were originally present at time~$t=0$ (orig.) and those produced during the simulation (prod.). Klein-Nishina effects lead to NTPA that is intermediate between the n\onradiative[]and Thomson radiative regimes, with a high-energy \cutoff[]similar to the n\onradiative[]case but a power-law scaling intermediate between those of the n\onradiative[]and Thomson-cooled runs.}
    \label{fig:ptcldists}
\end{figure*}

In our simulations, magnetic reconnection results in efficient energy delivery to the plasma particles in the form of NTPA. 
To illustrate this, we present the particle energy distributions, time-averaged over each simulation's reconnection active phase (section~\ref{sec:recrate}), in Fig.~\ref{fig:ptcldists}. While reconnection-powered NTPA is efficient in all cases, differences between the runs now begin to emerge. In particular, the slope and extent of the high-energy power-law tail differ depending on the radiative physics involved.

In the n\onradiative[]run, reconnection promptly energizes particles up to Lorentz factors~$\gamma \sim \gx \equiv 4 \sigc$ with a hard power-law,~$\dif N / \dif \gamma \propto \gamma^{-p}$, of index~$p\simeq1.2$. This is expected in our weak guide-field, extremely highly magnetized~($\sigh\sim 10^3$) regime, in which numerous previous studies have found that~$p$ tends to decrease (hardening the power-law) with increasing~$\sigh$, asymptoting close to unity in the \mbox{large-$\sigh$} limit \citep[e.g.][]{zh01, zh08, ss14, sgp16, s22, mwf14b, gld14, gld15, gld19, gld21, wuc16, wu17, wub18, bso18, u22}. Because~$p<2$, most of the particle kinetic energy is stored in the high-energy tail of the distribution, forcing a departure from the~$p\simeq1.2$ scaling beyond Lorentz factors a few times~$\gx \equiv 4 \sigc$. Otherwise, the particles would carry more energy than initially available in the reconnecting magnetic field.

Beyond the steepening around~$\gamma = \gx$, the particle distribution declines and then eventually sharply cuts off near~$\gamma = \gmax$. Since the energies~$\gx \equiv 4 \sigc$ and~$\gmax$ are relatively close together by computational necessity in this study (section~\ref{sec:radscaleschoose}), it is difficult to determine whether the transition near~$\gamma = \gx$ leads to a steeper (softer) power-law, perhaps scaling roughly as~$p\simeq3$, or to an exponential \cutoff[]that later gives way to an even sharper \cutoff[]near~$\gamma\simeq\gmax$. The former case, if realized, would match the picture advanced by \citet{ps18} and \citet{hps21} where particles undergo additional slow acceleration after being processed across the reconnection separatrix and becoming trapped inside of adiabatically compressing plasmoids.

The addition of radiative cooling markedly changes the signatures of NTPA in our simulations. In the purely Thomson radiative run, radiative losses impose a decisive \cutoff[]on the maximum particle energy at the Lorentz factor~$\gradt$, as first studied by \citet{wpu19}. This \cutoff[]is well below not only the nominal system-size-limited Lorentz factor,~$\gmax$, but also the intrinsic maximum energy,~$\gx$, attainable by particles via non-ideal direct acceleration by the reconnection electric field near X-points. Thus, radiative losses compete with even very rapid particle acceleration. Slower secondary acceleration channels are suppressed altogether, and there is no evidence of a secondary power-law component associated with such channels (cf.\ \citealt{mwu20, mwu21, hrp23}; for a similar effect observed in the context of relativistic turbulence, see, e.g.\ \citealt{zuw20, zuk21, nb21, sns21, cs21}).

Besides the radiative \cutoff[]at~$\gradt$, the high-energy tail of the Thomson-cooled particle energy distribution features a steeper power-law scaling,~$\dif N / \dif \gamma \propto \gamma^{-p}$, with~$p \simeq 2.5$. This is also consistent with earlier work~\citep{wpu19, mwu20}, including the pile-up seen at intermediate energies (yielding a peak in~$\gamma^2 \dif N / \dif \gamma$) just beyond~$\gcool$. This pile-up results from intermittent episodes of explosive particle acceleration ignited at plasmoid mergers. In between such episodes, particles are rapidly cooled inside their host plasmoids, reaching a typical energy~$\gcool L / c \Delta t_{\rm coll}$ c\haracteriz[ed]by the time~$\Delta t_{\rm coll}$ between plasmoid collisions. While, in reality,~$\Delta t_{\rm coll}$ is different for different tiers in the plasmoid hierarchy, the relativistic plasmoid motion in the box of size~$L$ dictates that it should be~$<L/c$. This is consistent with the pile-up in the high-energy tail (peak in~$\gamma^2 \dif N / \dif \gamma$) occurring at a few-to-several times~$\gcool$ in the IC(Th) panel (top right) of Fig.~\ref{fig:ptcldists}.

Klein-Nishina effects lead to an NTPA regime that is largely intermediate between the n\onradiative[]and strongly Thomson-cooled cases. This is evidenced by both our runs with general Compton losses, IC(KN) and IC(KN)+PP. The Klein-Nishina reduction in radiative efficiency causes the particle energy distribution's tail to exhibit a flatter (harder) scaling,~$p\simeq2$, and to persist to higher energies (definitively cutting off by the time~$\gamma\simeq\gmax=\gradk$, though perhaps steepening sooner, near~$\gamma = \gx$) than when cooling proceeds purely in the Thomson limit. However, the tail is still steeper than that in the n\onradiative[]run. This intermediate behavi\spellor[]can be understood from the hierarchy of \ts[s,~$\tcoolt(\gamma) < \tcoolk(\gamma) < L/c$,]which holds at all Lorentz factors~$\gamma < \gmax$ in these simulations (Fig.~\ref{fig:tstop_kn}, middle panel).

The degree to which NTPA more resembles that in the Thomson or n\onradiative[]limits depends on the precise value of~$\gkn$. Increasing~$\gkn$ tends to bring the g\eneraliz[ed]cooling time,~$\tcoolk(\gamma)$, closer to its Thomson limit,~$\tcoolt(\gamma)$ (Fig.~\ref{fig:tstop_kn}, top panel). Once~$\gkn$ surpasses~$\gradt$, particles are forbidden from experiencing significant Klein-Nishina effects, and the system reverts to purely Thomson radiative reconnection. On the other hand, reducing~$\gkn$ lengthens the cooling time,~$\tcoolk(\gamma)$, and, in the extreme case of~$\gkn \lesssim \gcool$,~$\tcoolk(\gamma)$ exceeds~$L/c$ for all~$\gamma$: an effectively n\onradiative[]regime (Fig.~\ref{fig:tstop_kn}, bottom panel). We have verified these expectations by running simulations (not presented here) with differing~$\gkn$. These runs confirm that increasing~$\gkn$ moves the particle distribution \cutoff[]closer to~$\gradt$ while also steepening its power-law tail, causing NTPA to resemble the Thomson limit. Conversely, reducing~$\gkn$ maintains the sharp particle energy \cutoff[]at the system-size limit,~$\gmax$ (perhaps with an earlier \cutoff[]or power-law transition near~$\gx$), while simultaneously hardening the high-energy tail, transitioning the system towards n\onradiative[]NTPA. 

\begin{figure*}
    \centering
    \includegraphics[width=\linewidth]{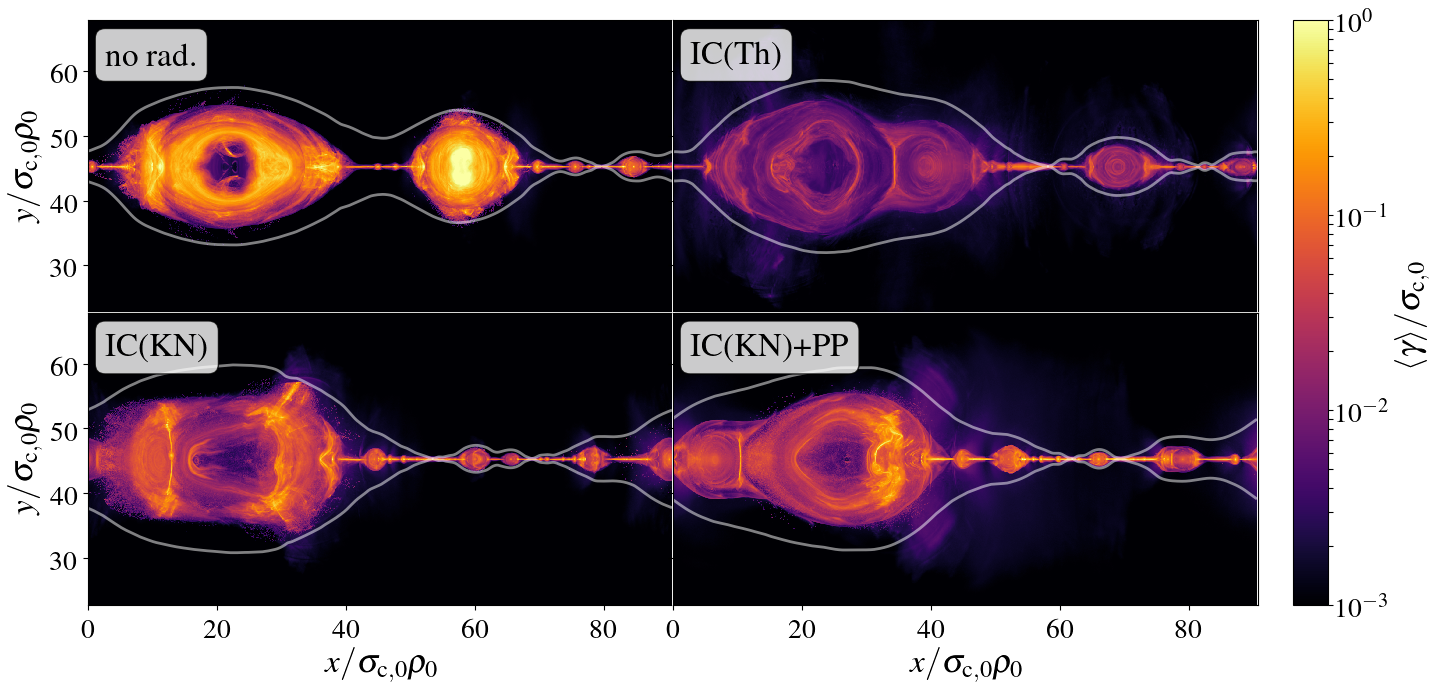}
    \caption{Local average Lorentz factor maps for our four main runs. Each run is pictured at the moment in time when half of its final change in reconnected magnetic flux has occurred. Commensurate with what happens in the particle energy distributions, Klein-Nishina effects result in maps that have properties that are intermediate between the n\onradiative[]and purely Thomson-cooled regimes. Namely, the IC(KN) and IC(KN)+PP runs both feature hot reconnection current sheets like the n\onradiative[]simulation, but plasmoids with temperatures between those of the plasmoids in the n\onradiative[]and Thomson-cooled cases.}
    \label{fig:spatialtemps}
\end{figure*}
In Fig.~\ref{fig:ptcldists}, for the run with pair production, IC(KN)+PP, we decompose the particle energy distribution into its separate contributions from particles that were present originally in the simulation and those produced on the fly. Strikingly, the produced particles dominate not only the high-energy tail, but also the total plasma energy contained in the simulation box. This is in spite of the fact that these particles are far less numerous than their originally present counterparts, which is only possible thanks to their extremely high average energy. That the produced particles should compete with the original particles for energetic dominance despite being fewer in number is in line with the basic predictions for this regime of reconnection advanced by \citet{mwu21}. We examine more thoroughly the differences between original and produced particles in section~\ref{sec:origprodcompare}.

To complement the distributions of Fig.~\ref{fig:ptcldists} with a spatial view into particle acceleration, we present maps of the local plasma average Lorentz factor for each of our four main runs in Fig.~\ref{fig:spatialtemps}. Here again, we see that the n\onradiative[]and Thomson-cooled cases represent two opposite extremes. On the one hand, the n\onradiative[]run contains hot reconnection current sheets and plasmoids (except the cold core of the large exhaust plasmoid cent\speller[d]at~$x \simeq 20 \sigc \, \rho_0$ -- it is composed of initially drifting plasma swept directly into this plasmoid near the beginning of the simulation without ever being energized by reconnection). On the other hand, the simulation with Thomson losses features cold plasmoid cores \citep[cf.][]{b17, sb20, ssb21, ssb23, ghb23} and merely warm current sheets.

Let us see how these differences arise. The plasmoids in the non-radiative run accumulate kinetic energy via the hot plasma that is exhausted away from reconnection X-points, thereby containing, collectively, a running tally of the dissipated magnetic energy. While plasmoids still collect particles in the IC(Th) case, they no longer amass liberated energy, which instead escapes as IC radiation. Then, instead of the area-filling, particle-accumulating plasmoids, it is the \mbox{quasi-1D} current-sheet singularities that host the energetic plasma -- i.e. where intense acceleration is actively taking place.\footnote{In~3D, the plasmoids would be volume-filling instead of area-filling, and the current sheets \mbox{quasi-2D} structures instead of \mbox{quasi-1D} ones.} And even in these special regions, the local mean energy is radiatively limited to~$\sim \gradt = 0.2 \sigc$, much lower than the intrinsic maximum X-point acceleration Lorentz factor,~$\gx = 4 \sigc$, reached in the (consequently much hotter) current sheets of the n\onradiative[]run. In the IC(Th) case, once particles vacate rapid acceleration zones near X-points to move into plasmoids, they quickly cool down, giving the plasmoid cores a characteristic mean energy of order the pile-up energy,~$\langle \gamma \rangle \sim 10^{-2} \sigc$, in the IC(Th) distribution of Fig.~\ref{fig:ptcldists}.

Moving now to the IC(KN) and IC(KN)+PP simulations in Fig.~\ref{fig:spatialtemps}, we see again that they are intermediate between the n\onradiative[]and Thomson-cooled extremes. Like the n\onradiative[]case, these two runs contain very hot reconnection current layers, with local Lorentz factors comparable to~$\sigc$ and far exceeding~$\gradt$. This reflects the fact that cooling losses do not substantially inhibit acceleration near reconnection X-points in these Klein-Nishina-regime runs. As for plasmoids, these are colder than in the n\onradiative[]simulation but warmer than in the IC(Th) case. This stems again from the cooling \ts[]hierarchy,~$\tcoolt(\gamma) < \tcoolk(\gamma) < L/c$: particles accelerated at current sheets are not efficiently cooled on the \ts[s]of their acceleration, but they are still efficiently cooled over one dynamical time, causing plasmoids to cool down -- just not as quickly as in the IC(Th) run.

To s\ummariz[e,]in this subsection, we have witnessed the first main differences emerge among our simulations, with the differing radiative physics leaving pronounced and distinguishing imprints on NTPA. Even though all runs exhibit n\onthermal[]power-law particle energy distributions,~$\dif N / \dif \gamma \propto \gamma^{-p}$, in their reconnection active phases, the slopes and extents of their power-laws differ dramatically. The non-radiative regime yields~$p$ close to unity, with a departure from this scaling near~$\gamma = \gx \equiv 4 \sigc$ and a subsequent sharp \cutoff[]at~$\gamma=\gmax$. The IC(Th) run represents an opposite regime, with~$p\simeq2.5$ followed by an abrupt \cutoff[]at~$\gamma\simeq\gradt \ll \gmax$. Klein-Nishina radiative cooling lies between these two extremes, and for the parameters of our IC(KN) and IC(KN)+PP runs, gives~$p\simeq2$ and a sharp \cutoff[]at~$\gamma\simeq\gmax$ (perhaps with an earlier steepening near~$\gamma=\gx$).

\subsection{Inverse Compton emission spectra}
\label{sec:icemitdists}
In this section, we connect the pronounced NTPA in our simulations to their IC emission spectra. Like the underlying particle energy distributions (section \ref{sec:ptcldists}), these spectra are highly extended and n\onthermal[,]and the distinctions among them reflect the differing radiative physics at play.

\begin{figure*}
    \centering
    \begin{subfigure}{0.49\linewidth}
        \centering
        \includegraphics[width=\linewidth]{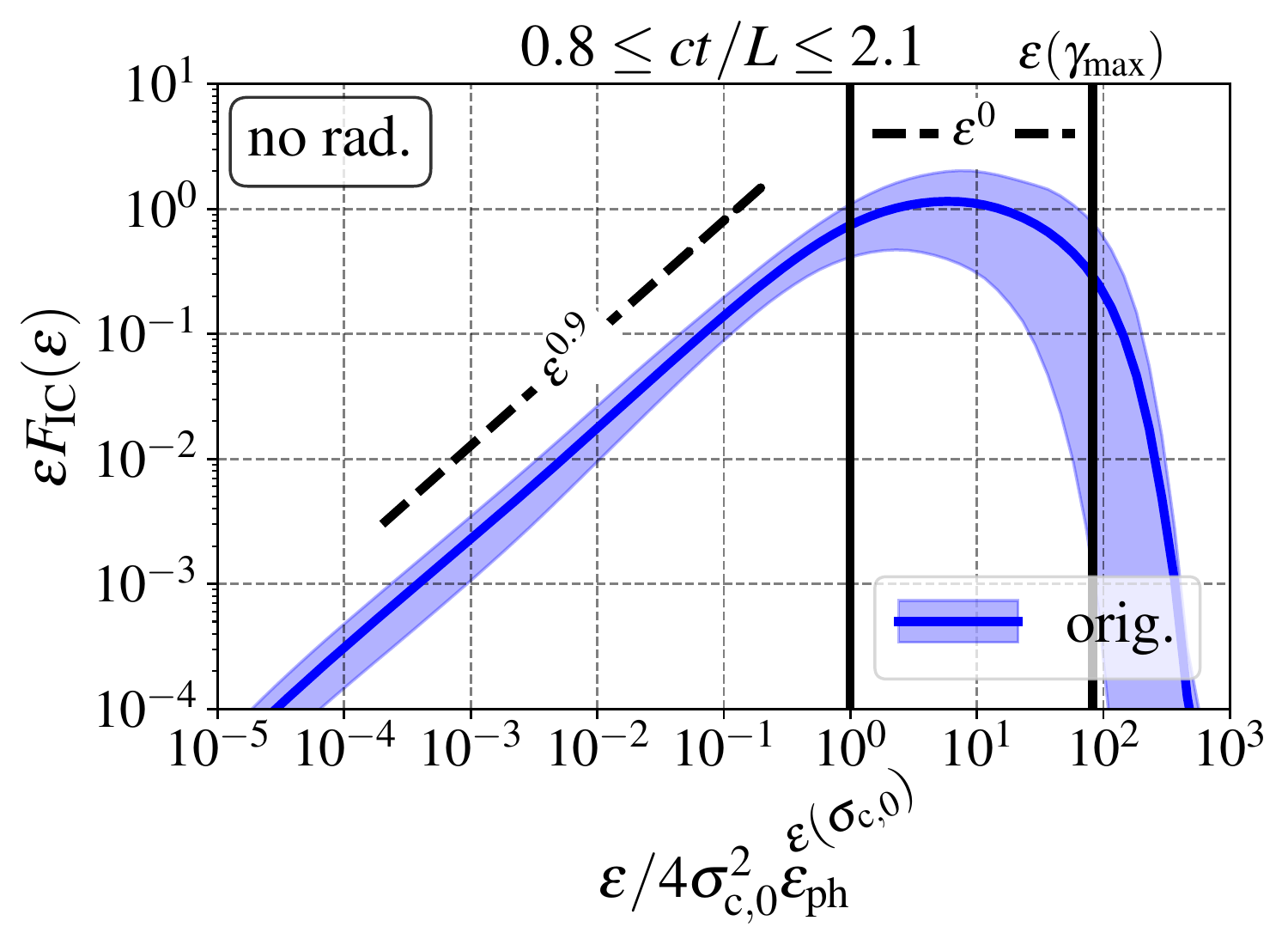}
    \end{subfigure}
    \begin{subfigure}{0.49\linewidth}
        \centering
        \includegraphics[width=\linewidth]{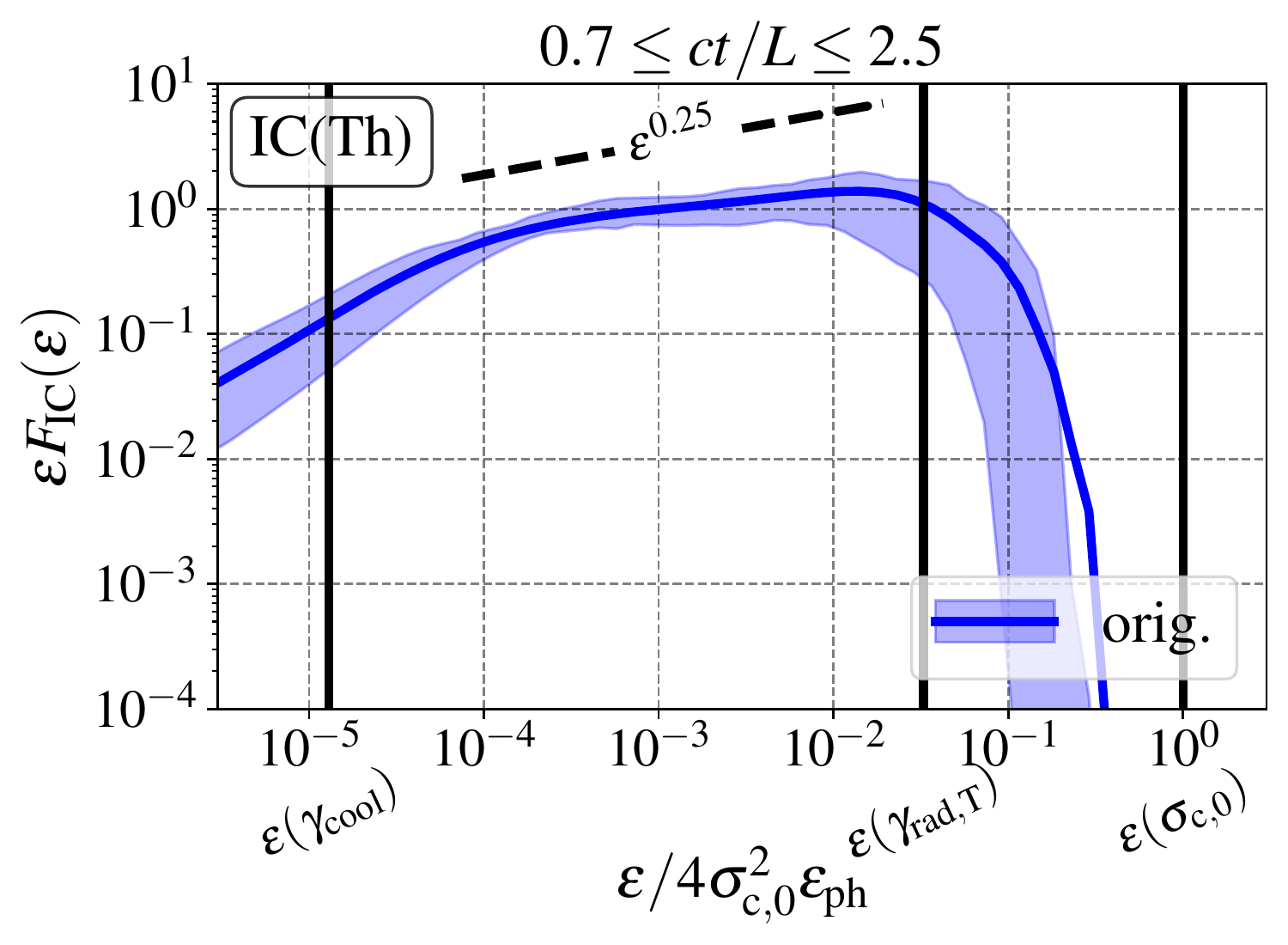}
    \end{subfigure}
    \begin{subfigure}{0.49\linewidth}
        \centering
        \includegraphics[width=\linewidth]{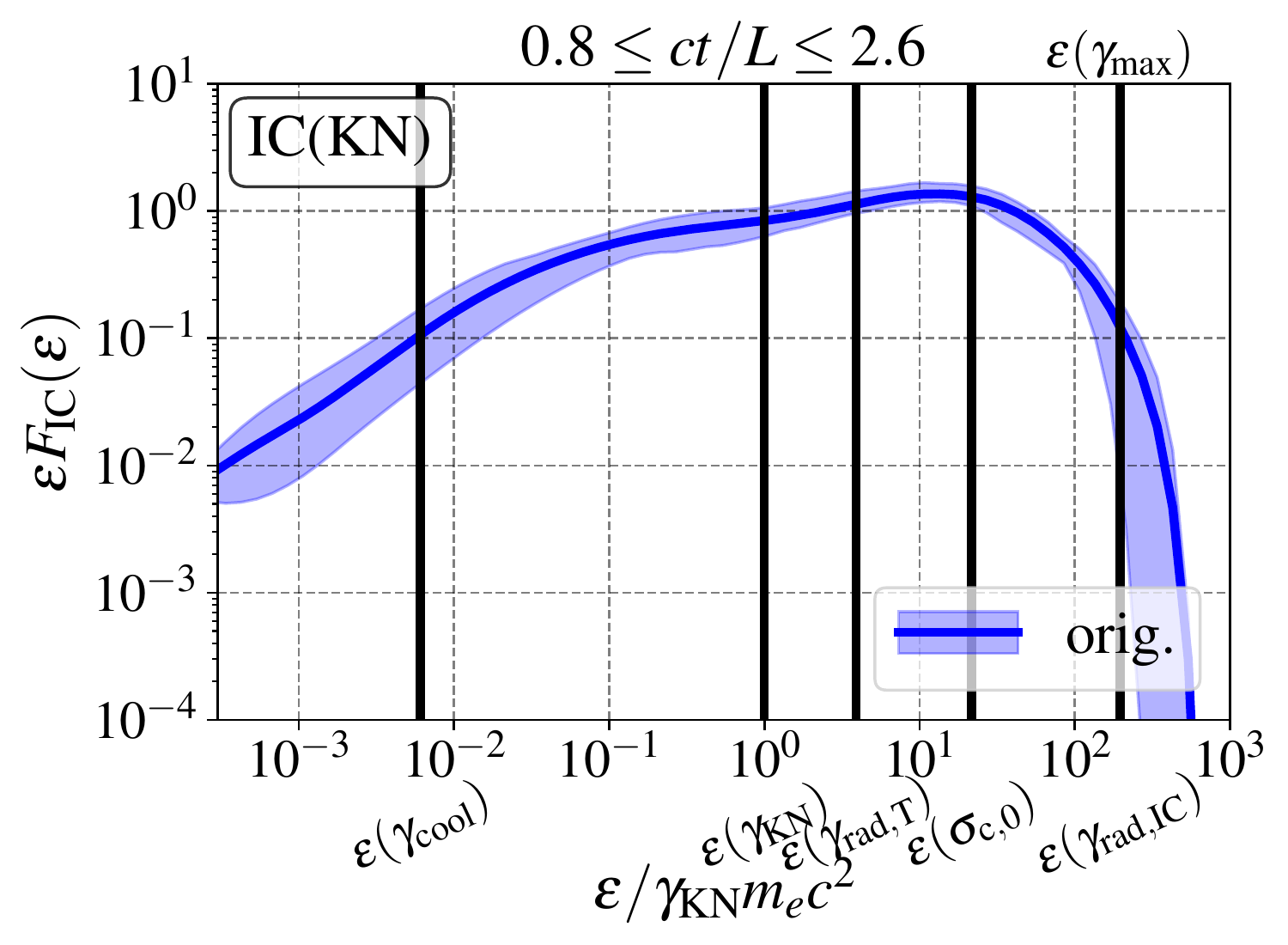}
    \end{subfigure}
    \begin{subfigure}{0.49\linewidth}
        \centering
        \includegraphics[width=\linewidth]{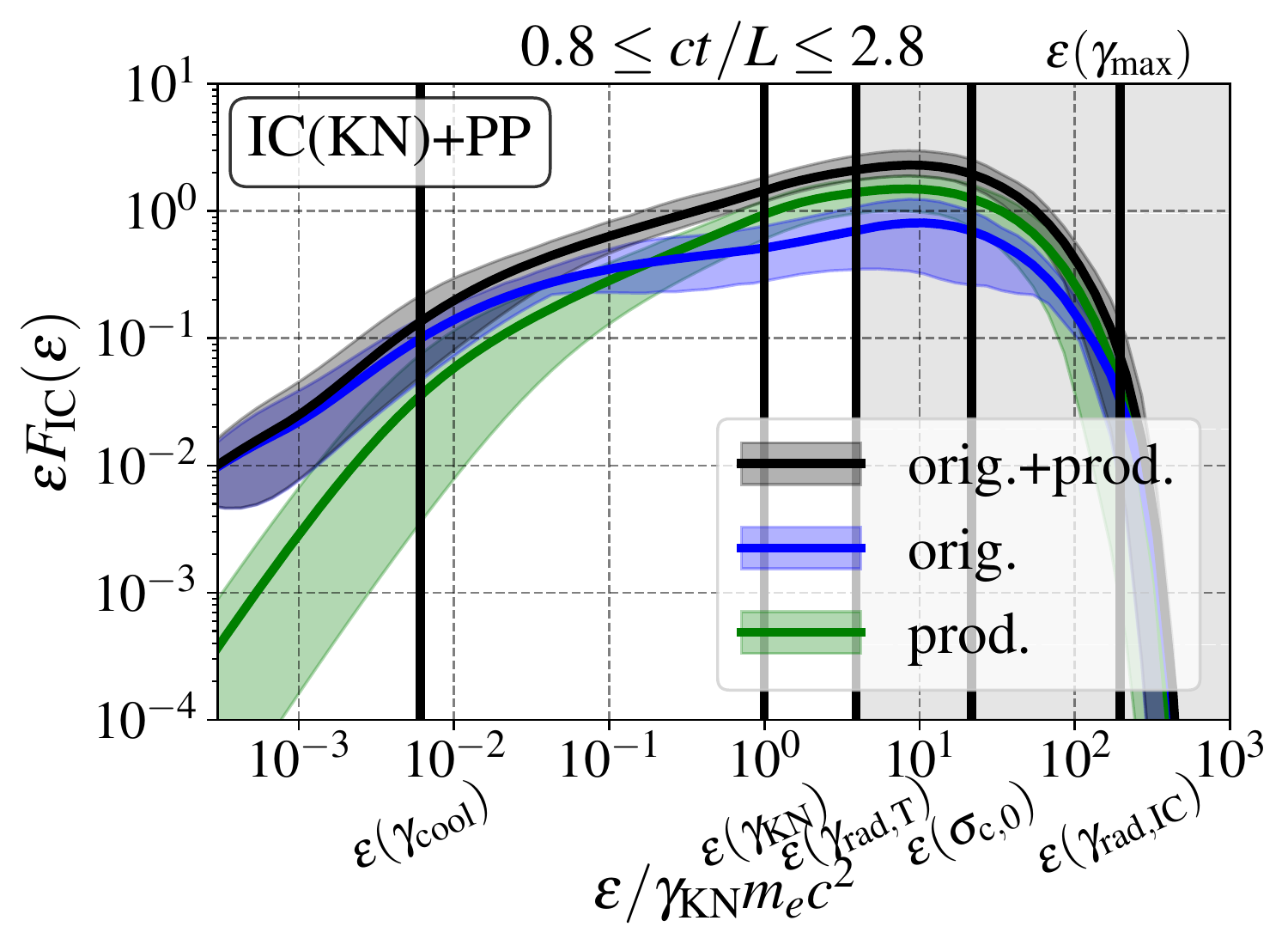}
    \end{subfigure}
    \caption{Time-averaged (over the same interval as in Figs.~\mbox{\ref{fig:recflux}-\ref{fig:ptcldists}}) IC emission spectra for our four main runs. Instantaneous spectra used for averaging are equal to the box-integrated isotropic IC emissivity,~$\epsilon F_{\rm IC}(\epsilon; t) \equiv \epsilon \int \dd[3] x \dd \Omega j_{\rm IC}(\epsilon, \myvec{x}, \Omega; t)$. Error envelopes display the one-sigma percentiles over the averaging interval in each energy bin. The n\ormaliz[ation]is arbitrary and equal only for the three radiative simulations (because the no rad.\ run is not dynamically cooled, its luminosity is not tied to the energetics of the simulation and so can be freely rescaled). The characteristic photon energy,~$\epsilon(\gamma)$, emitted by a particle of Lorentz factor~$\gamma$ is defined piecewise-continuously as~$4 \gamma^2 \eph$~($\gamma \me c^2$) for~$\gamma$ less (greater) than~$\gkn$. The IC(KN)+PP spectrum is decomposed into contributions from originally present (orig.) and produced (prod.) particles. Klein-Nishina effects suppress the IC \crosssection[,]hardening the particle energy distribution (Fig.~\ref{fig:ptcldists}) while softening the emitted spectrum. These competing effects result in IC(KN) spectral scalings that are not very different from that in the IC(Th) simulation. The gray region in the IC(KN)+PP panel indicates photon energies beyond pair-production threshold, $4 \gkn \me c^2$. Photons emitted at these energies do not make it to the observer; they are absorbed inside the system to produce new pairs.}
    \label{fig:icemitdists}
\end{figure*}
We present angle-integrated IC emission spectra for each of our four main runs in Fig.~\ref{fig:icemitdists}. These spectra are computed by summing together the individual spectra from every simulated particle, regardless of the particle's position or velocity direction. As a result, beaming and light-travel-time effects are neglected. In addition, for the IC(KN)+PP run, pair-production absorption of the emitted gamma-rays is temporarily ignored, but we discuss it briefly at the end of this subsection and in more detail in the next subsection.

In the n\onradiative[]and Thomson-cooled cases, the IC emission spectra are produced in the Thomson regime and are thus given straightforwardly in terms of the underlying particle energy distributions as follows. Particles of Lorentz factor~$\gamma$ upscatter photons to energies~$\epsilon(\gamma) \sim \gamma^2 \eph$. Because the scattering rate,~$\sigma_{\rm T} c \nph$ [cf.\ equation~(\ref{eq:icscatrate})], is independent of~$\gamma$, the number of photons emitted per unit time into a given energy interval,~$\dif \epsilon \dif N_{\rm ph}/\dif t \dif \epsilon$, is proportional to the number of particles at the corresponding scattering Lorentz factor,~$\dif \gamma \dif N / \dif \gamma |_{\gamma \propto \epsilon^{1/2}}$. Thus, if the particle energy distribution is a power-law,~$\dif N / \dif \gamma \propto \gamma^{-p}$, the emitted photon distribution is also a power-law:~$\dif N_{\rm ph}/ \dif t \dif \epsilon \propto (\dif N / \dif \gamma |_{\gamma \propto \epsilon^{1/2}}) (\dif \gamma / \dif \epsilon) \propto \epsilon^{-(p+1)/2}$ \citep[cf.][]{rl79}. In the~$\epsilon F_{\rm IC}(\epsilon)$ representation plotted in Fig.~\ref{fig:icemitdists}, this translates to~$\epsilon F_{\rm IC}(\epsilon) = \epsilon^2 \dif N_{\rm ph} / \dif t \dif \epsilon \propto \epsilon^{-(p-3)/2}$.

This result equips us to easily interpret the power-law components in the n\onradiative[]and Thomson-cooled IC emission spectra. The n\onradiative[]run's particle distribution power-law is approximately~$\dif N / \dif \gamma \propto \gamma^{-1.2}$ (Fig.~\ref{fig:ptcldists}), which yields the expected IC power-law,~$\epsilon F_{\rm IC}(\epsilon) \propto \epsilon^{0.9}$. This is in good agreement with the n\onradiative[]IC spectrum in the top left panel of Fig.~\ref{fig:icemitdists}. Additionally, if present, the second/steeper particle power-law component for this same simulation,~$\dif N / \dif \gamma \propto \gamma^{-3}$ -- putatively stemming from slower secondary acceleration channels -- should produce a flat spectrum,~$\epsilon F_{\rm IC}(\epsilon) \propto \epsilon^0 = \rm const$. While this is roughly consistent with the measured spectrum of Fig.~\ref{fig:icemitdists}, it is difficult to definitively say that such a component truly exists and is not just part of the spectral \cutoff[.]Regarding the Thomson-cooled run, the power law,~$\dif N / \dif \gamma \propto \gamma^{-2.5}$, should yield the gently increasing spectrum,~$\epsilon F_{\rm IC}(\epsilon) \propto \epsilon^{0.25}$, which agrees with Fig.~\ref{fig:icemitdists} (top right panel). Finally, all these ideas can be applied not just to connect spectral slopes between the particle and IC spectra, but also to relate their \cutoff[s.]In particular, the \cutoff[s]at~$\gmax$ and~$\gradt$ in the respective n\onradiative[]and Thomson-cooled particle distributions (Fig.~\ref{fig:ptcldists}) correspond to the observed \cutoff[s]at photon energies of order~$\gmax^2 \eph$ and~$\gradt^2 \eph$ in these simulations' IC emission.

This simple framework breaks down in the presence of Klein-Nishina effects. Then, the characteristic scattered photon energy becomes a \textit{broken} power-law function of the particle's energy:~$\epsilon(\gamma) \sim \gamma^2 \eph$ when~$\gamma < \gkn$ and~$\epsilon(\gamma) \sim \gamma \me c^2$ otherwise. Furthermore, the scattering rate,~$\sigma_{\rm T} c \nph g_{\rm KN}(\gamma / \gkn)$ [equation~(\ref{eq:icscatrate})], becomes a \nontrivial[,]decreasing function of~$\gamma$. This suppresses the emission efficiency and breaks the simple correspondence between the particle distribution power-law index and that of the IC emission spectrum. Let us examine how these effects manifest themselves in the IC(KN) and IC(KN)+PP panels (bottom left and bottom right, respectively) of Fig.~\ref{fig:icemitdists}. To begin with, the characteristic emission energies from particles at each of our Lorentz-factor scales are pushed closer together beyond the energy~$\epsilon(\gkn)$ (because, beyond~$\gkn$,~$\epsilon$ scales linearly with~$\gamma$ instead of quadratically). As an example, even though the scales~$\gmax$,~$\gradt$, and~$\gcool$ are all equally spaced on a logarithmic scale (because~$\gradt^2 = \gmax \gcool$), the corresponding photon energies,~$\epsilon(\gmax)$,~$\epsilon(\gradt)$, and~$\epsilon(\gcool)$, are not at all evenly spaced, with~$\epsilon(\gmax)$ and~$\epsilon(\gradt)$ closer together than~$\epsilon(\gradt)$ and~$\epsilon(\gcool)$ [because the former two lie above the break energy~$\epsilon(\gkn)$ while the latter lies below it]. Next, even though the power-law scalings of the IC(KN) and IC(KN)+PP particle distributions are both approximately~$\gamma^{-2}$ or shallower (i.e. harder), the corresponding~$\epsilon F_{\rm IC}(\epsilon)$ spectra are both steeper (i.e. softer) than the Thomson-limit prediction,~$\epsilon^{0.5}$, demonstrating the reduced radiative efficiency in the Klein-Nishina limit.

While the diminished Klein-Nishina \crosssection[]produces a softer emission spectrum for a given particle distribution, it also yields a particle distribution that is harder in the first place (section~\ref{sec:ptcldists}). These two effects somewhat cancel out, and, hence, not much change is observed in the spectral slope from the time-averaged IC(Th) spectrum to those yielded by the IC(KN) and IC(KN)+PP simulations \citep[cf.][]{msc05}. This is even despite the very different shape -- the result of different cooling physics -- in the particle distributions between these runs.

Finally, we note that the IC(KN) and IC(KN)+PP Compton emission spectra peak far above~$\epsilon(\gkn)$ -- deeply in the Klein-Nishina regime. As a result, most of the radiated energy in the IC(KN)+PP case (gray region in the lower right panel of Fig.~\ref{fig:icemitdists}) is emitted \textit{above pair-production threshold}, where [unlike the IC(KN) run] it will be recaptured by the system as hot newborn pairs. The peak in the intrinsic emitted IC spectrum is therefore \textit{invisible} to the observer, who sees only the indirect remnant of this radiation reprocessed to below-threshold energies. We elaborate the observable consequences of this effect in the next subsection.

\subsection{Lightcurves and spectral variability}
\label{sec:afterglow}
We now complement section~\ref{sec:icemitdists}'s energy-resolved view of the emission from our simulations by discussing the timing of the radiative signatures. When viewed through the lens of timing, the differences among the various radiative regimes are accentuated, resulting in highly distinct observable signatures.

\begin{figure}
    \centering
    \includegraphics[width=\linewidth]{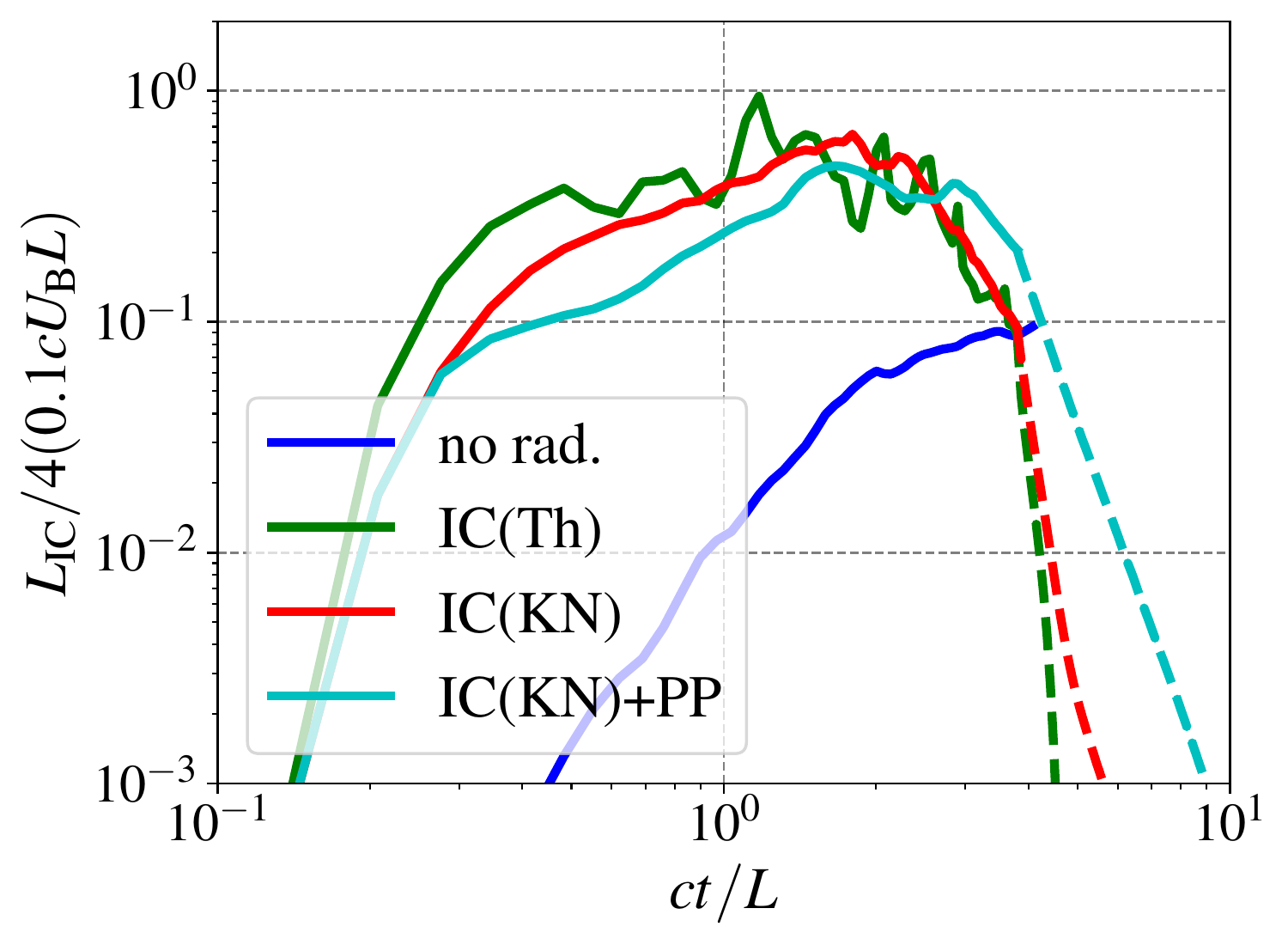}
    \caption{Lightcurves of the IC luminosity (instantaneous escaping radiated power) for each of our four main runs. The n\ormaliz[ation]for the n\onradiative[]simulation is not fixed by dynamical radiative cooling and so is rescaled to appear on the plot with the other simulations (cf.\ Fig.~\ref{fig:icemitdists}). Dashed portions of lightcurves are computed in postprocessing (see text).}
    \label{fig:lightcurves}
\end{figure}

We begin by presenting lightcurves of each run's bolometric luminosity (instantaneous total escaping emitted power) in Fig.~\ref{fig:lightcurves}. For consistency with the flow of energy in the simulations, we only include that part of the luminosity permanently lost by the simulation. This means that, for the lightcurves of the no rad.\ and IC(Th) runs, we report the frequency-integrated emitted power as a function of time. In contrast, for the IC(KN)+PP run, we report only the portion of the emission spectrum below pair-production threshold: at photon energies~$\epsilon < 4 \gkn \me c^2$. By the same reasoning, we are obligated to include all photon energies for the IC(KN) simulation, reporting the total emitted power in that case also -- otherwise, since pair production is artificially suppressed in that run, we would not count a fair fraction of the energy lost from the simulation. We have checked that this bookkeeping yields the same total energy radiated (integrals of the curves in Fig.~\ref{fig:lightcurves}) by each radiative simulation.

The lightcurves in Fig.~\ref{fig:lightcurves} are clearly separated into two main groups: the n\onradiative[]versus the radiative simulations. This dichotomy excellently illustrates a fundamental property of radiative reconnection: prompt emission. That is, in n\onradiative[]reconnection, particles are first accelerated and, then, over much longer \ts[s]than the duration of the reconnection process itself, radiate away their energy as potentially observable emission. In contrast, radiative reconnection features fundamentally prompt emission, where particles radiate their acquired energy on sub-dynamical \ts[s,]causing radiation to participate in the reconnection dynamics rather than, as in the n\onradiative[]case, passively trace energization that has already occurred. This is reflected in Fig.~\ref{fig:lightcurves} in that the three radiative lightcurves broadly track the instantaneous electromagnetic dissipation of their simulations, rising as reconnection gets going and falling again once the reconnected flux saturates (cf.\ Fig.~\ref{fig:recflux}). The lightcurve of the n\onradiative[]simulation, on the other hand, follows the \textit{time integral} of the electromagnetic dissipation, growing with the cumulative dissipated energy and reconnected flux.

\begin{figure}
    \centering
    \includegraphics[width=\linewidth]{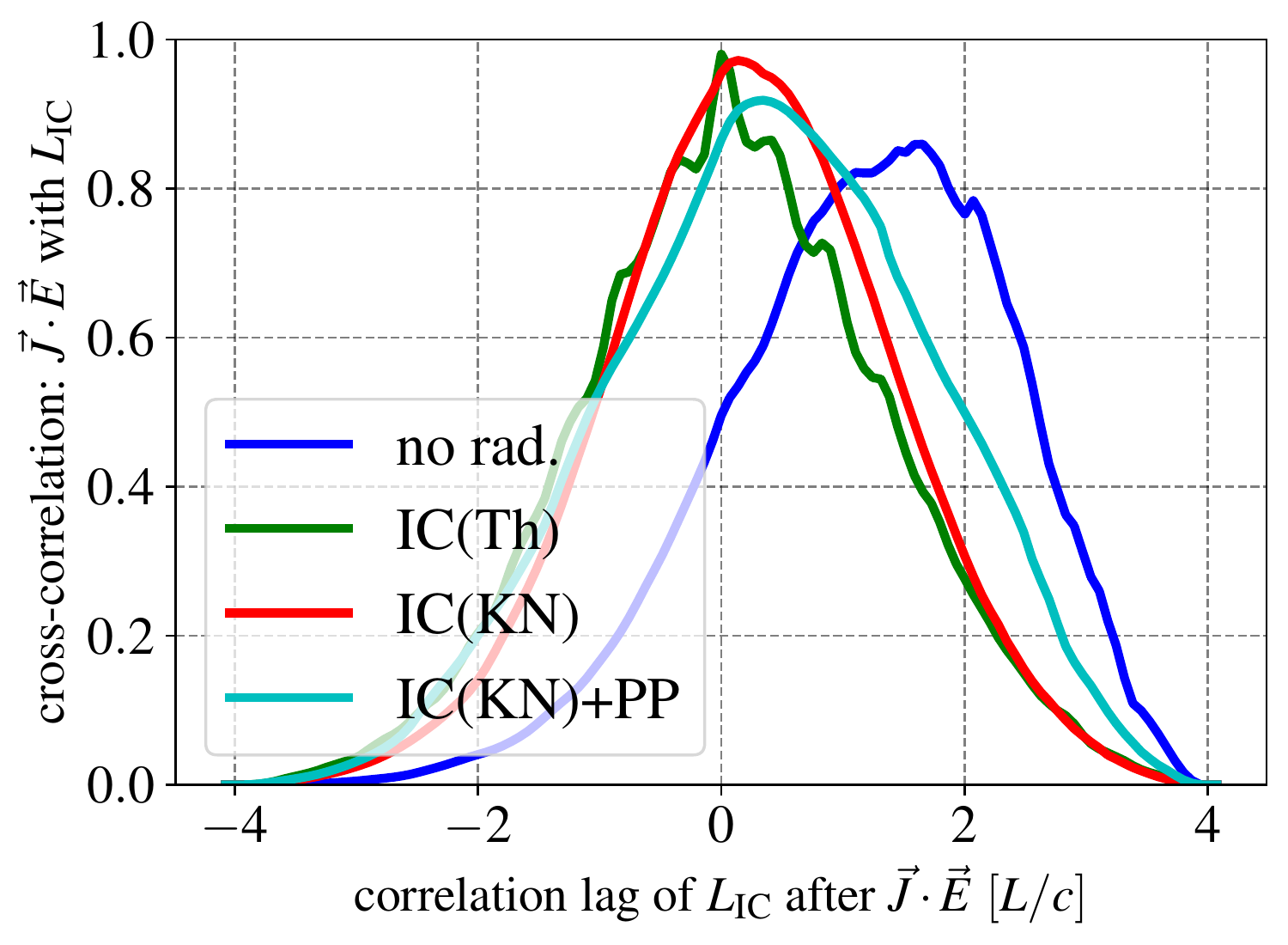}
    \caption{Cross-correlation of the instantaneous rate of electromagnetic dissipation with the escaping inverse Compton luminosity (lightcurves of Fig.~\ref{fig:lightcurves}). Emblematic of the distinction between radiative and n\onradiative[]reconnection, the n\onradiative[]simulation's luminosity lags its dissipation (proportional to the cumulative dissipated electromagnetic energy) whereas the radiative simulations' luminosities are s\ynchroniz[ed]with dissipation (proportional to its instantaneous rate). The IC(KN) run features a slightly longer lag than the IC(Th) run owing to its lower radiative efficiency. The IC(KN)+PP run features a still somewhat longer lag with a skewed distribution toward the high-lag end. This stems from the time it takes for radiation injected above pair-production threshold to become reprocessed to below-threshold (and, hence, escaping) energies.}
    \label{fig:iclag}
\end{figure}

To illustrate these remarks more thoroughly, we also supply Fig.~\ref{fig:iclag}, which shows the cross-correlation of each lightcurve from Fig.~\ref{fig:lightcurves} with the box-integrated electromagnetic dissipation,~$P_{\rm diss}(t) \equiv \int \dd[3] x\, \myvec{J}(t) \cdot \myvec{E}(t)$. As discussed above and confirmed by this figure, there is a significant lag of the n\onradiative[]IC luminosity behind~$P_{\rm diss}$. In contrast, all three radiative simulations lose their energy promptly, with peak lags close to zero.

There are, however, differences among the radiative runs. The IC(Th) luminosity exhibits the shortest variability \ts[s]in Fig.~\ref{fig:lightcurves}, with small bumps atop its overall envelope corresponding to bursts of particle acceleration at plasmoid mergers. In contrast, because the cooling \ts[s]are slightly longer for the IC(KN) and IC(KN)+PP runs, particles retain enough energy in between these episodic events to smooth out the variations in the resulting lightcurves. As one might expect from these remarks, the IC(Th) simulation exhibits the smallest (exactly zero) peak lag from its electromagnetic dissipation to its emitted luminosity. In contrast, the IC(KN) and IC(KN)+PP runs have small but finite peak lag. Of these, the IC(KN)+PP case has a slightly longer lag and a more skewed cross-correlation distribution, with more power concentrated at longer lags. This is the result of the reprocessing of gamma-rays (near the peak of the emission spectrum in Fig.~\ref{fig:icemitdists}) to lower energies through gamma-ray radiation and pair production. That is, it takes time for power injected at the high-energy, above-threshold peak of the IC emission spectrum to be processed down to lower energies where it can escape the system. 

\begin{figure*}
    \centering
    \includegraphics[width=\linewidth]{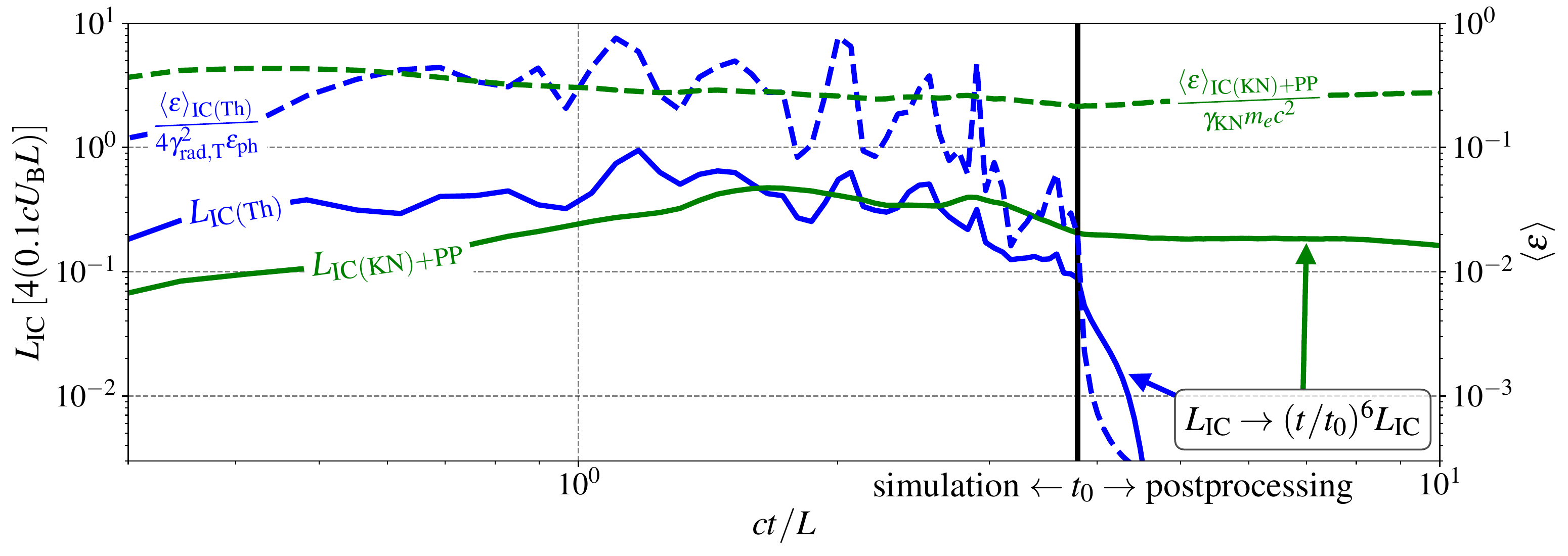}
    \caption{Lightcurves (reproduced from Fig.~\ref{fig:lightcurves}) of escaping IC luminosity and overall (computed using the escaping spectrum) average photon energy for the IC(Th) and IC(KN)+PP simulations. Both quantities are extrapolated beyond the simulation (from time~$t_0$ indicated by the vertical black bar) via postprocessing (see text). After~$t_0$ (but not before), the luminosities are compensated by~$(t/t_0)^6$, though the average photon energies are not modified. Several robust trends are evident that could distinguish these two types of reconnection in the context of reconnection-powered flares. Thomson radiative reconnection features a tight correlation between IC luminosity and mean photon energy. In contrast, the Klein-Nishina reconnection mean photon energy is~$\sim 0.3 \gkn \me c^2$ independently of the IC luminosity. These patterns hold both in the reconnection active phase and in the decaying phase when energization from the fields is either slowing down (as at late simulation times) or completely shut off (as in the postprocessing zone of the plot). In the limit of no further energy injection, Klein-Nishina reconnection exhibits a power-law~$t^{-6}$ brightness decay -- distinct from the much more precipitous decline in the Thomson run's luminosity.}
    \label{fig:afterglow}
\end{figure*}
When combined with spectral information, the lag induced by pair-production and gamma-ray absorption described above leaves a prominent imprint on the observable signatures of Klein-Nishina radiative reconnection -- one that could potentially distinguish it from other radiative regimes. To demonstrate this, we present a side-by-side comparison of the luminosities and average photon energies,~$\langle \epsilon \rangle \equiv \int \dif \epsilon\, \epsilon F_{\rm IC}(\epsilon) / \int \dif \epsilon\, F_{\rm IC}(\epsilon)$, viewed as a function of time for the IC(Th) and IC(KN)+PP runs in Fig.~\ref{fig:afterglow}.

To fully appreciate the observational differences between these runs, we extrapolate their spectra beyond the end of each simulation. This is done under the assumption that the simulations are evolved to the point where no further exchange of energy occurs between the particles and the fields, allowing the particle energy distributions to be passively Compton cooled in~1D energy space. Pair-production can be included in this~1D evolution thanks to the homogeneous, isotropic, and static nature of the seed photon bath, which introduces no spatial or velocity-direction dependence. Postprocessing the luminosity and average photon energy time series allows us to extend them in Fig.~\ref{fig:afterglow} from just shy of~$4L/c$ to more than~$10L/c$ -- a significant gain over the simulations themselves, which would otherwise need to be expensively evolved to more than double their actual duration to reach the same times.

Several facts, each of them accessible to observations, are immediately apparent from Fig.~\ref{fig:afterglow}. First, the overall luminosity and average photon energy emitted from the IC(Th) simulation are tightly correlated. During the short \ts[]variations of the lightcurve, sporadic magnetic reconnection acceleration yields an extended n\onthermal[]distribution of particles (Fig.~\ref{fig:ptcldists}) with the resulting IC emission spectrum peaking near the high-energy \cutoff[](Fig.~\ref{fig:icemitdists}): that is, the acceleration mechanism produces a correlation between~$\langle \epsilon \rangle$ and~$L_{\rm IC}$ on the rising side of each subpeak in their time series. In between these reconnection energization episodes, particles emitting at the highest energies -- the ones contro\ling[]both the overall luminosity and peak photon energy -- are also the most rapidly cooled. They thus suddenly plummet to lower energies, inducing simultaneous drops in both~$\langle \epsilon \rangle$ and~$L_{\rm IC}$: that is, the radiative cooling mediates the correlation between~$\langle \epsilon \rangle$ and~$L_{\rm IC}$ on the falling side of each peak in their time series. 

A second observationally pertinent property of the IC(Th) curves in Fig.~\ref{fig:afterglow} is that, once energization from the electromagnetic fields is shut off, both the IC photon energy and luminosity drop precipitously. In fact, even though the luminosity time series are compensated by~$(t/t_0)^6$ rightward of the transition time,~$t_0$, to the postprocessing stage, this does little to stem the fall of the IC(Th) luminosity. Flares from highly radiative Thomson reconnection are thus c\haracteriz[ed]by \textit{tight correlation between the observed luminosity and photon energy plus a rapid falling phase where both plummet together}.

Let us now examine how these qualities compare to those of the IC(KN)+PP run. Most importantly, the tight correlation between luminosity and photon energy is broken. Instead, irrespective of the instantaneous IC brightness, the average photon energy remains rock steady, persisting near~$0.3 \gkn \me c^2$ -- even in the decaying phase of the lightcurve after electromagnetic energization has ceased. In the following, we argue that this average photon energy is first set during active periods of reconnection-powered particle acceleration and subsequently reinforced, when such acceleration is inactive, by the Klein-Nishina radiative physics, explaining its persistence.

During active acceleration episodes, reconnection produces a hard distribution of radiating particles with corresponding upward-sloped~$\epsilon F_{\rm IC}(\epsilon)$ emission spectrum peaking well above pair-production threshold (Fig.~\ref{fig:icemitdists}). Most of the initially emitted energy is, thus, veiled by pair production, and the peak of the apparent/observed spectrum lies instead just before the absorption-induced \cutoff[:]i.e. at energies~$\sim \gkn \me c^2$, as seen in the IC(KN)+PP run. Subsequently, in between reconnection-powered acceleration events, the radiative physics takes over in determining the mean photon energy. As seen during these periods in the Thomson regime, the IC spectrum softens because rapidly cooling particles radiating at the spectral peak energy,~$\sim \gradt^2 \, \eph$ (Fig.~\ref{fig:icemitdists}), cannot be replenished by electromagnetic energization. The key difference in the IC(KN)+PP case is that particles emitting at the observed peak energy,~$\sim \gkn \me c^2$, can still be partially replenished by IC cooling and pair production, which actively reprocess radiation originally emitted at higher, absorbed photon energies down to the observed band. This s\tabiliz[es]the observed spectrum, even in the falling phase of the lightcurve when electromagnetic energization is completely absent. Hence, the photon energy~$0.3 \gkn \me c^2$ owes its luminosity-independent stability to the fact that the coupling of reconnection-powered NTPA to Klein-Nishina radiative physics results in the same natural photon energy scale as that produced by the radiative physics alone.

We note that radiative reprocessing of initially above-threshold photons also leads, after reconnection has concluded, to the self-similar~$t^{-6}$ power-law brightness decay shown in Fig.~\ref{fig:afterglow}. Although this is much slower than in the case of Thomson radiative cooling, it is still probably too abrupt for gamma-ray instruments to resolve. Thus, what we would like to stress as the main difference between gamma-ray flares powered by Thomson-radiative and Klein-Nishina reconnection is that the latter are c\haracteriz[ed]by \textit{a constant mean observed photon energy, irrespective of brightness}.

\begin{figure*}
    \centering
    \begin{subfigure}{0.49\linewidth}
        \centering
        \includegraphics[width=\linewidth]{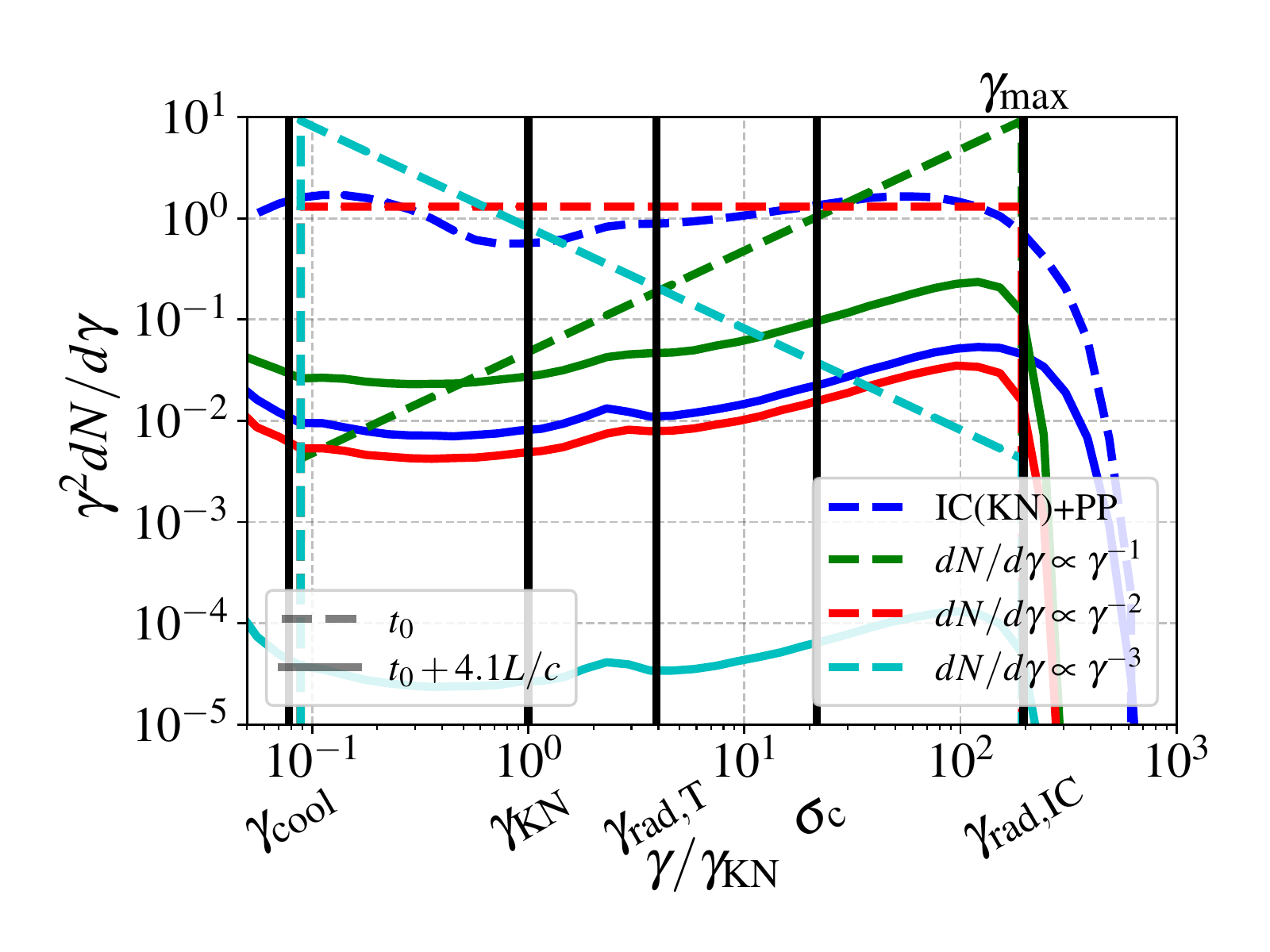}
    \end{subfigure}
    \begin{subfigure}{0.49\linewidth}
        \centering
        \includegraphics[width=\linewidth]{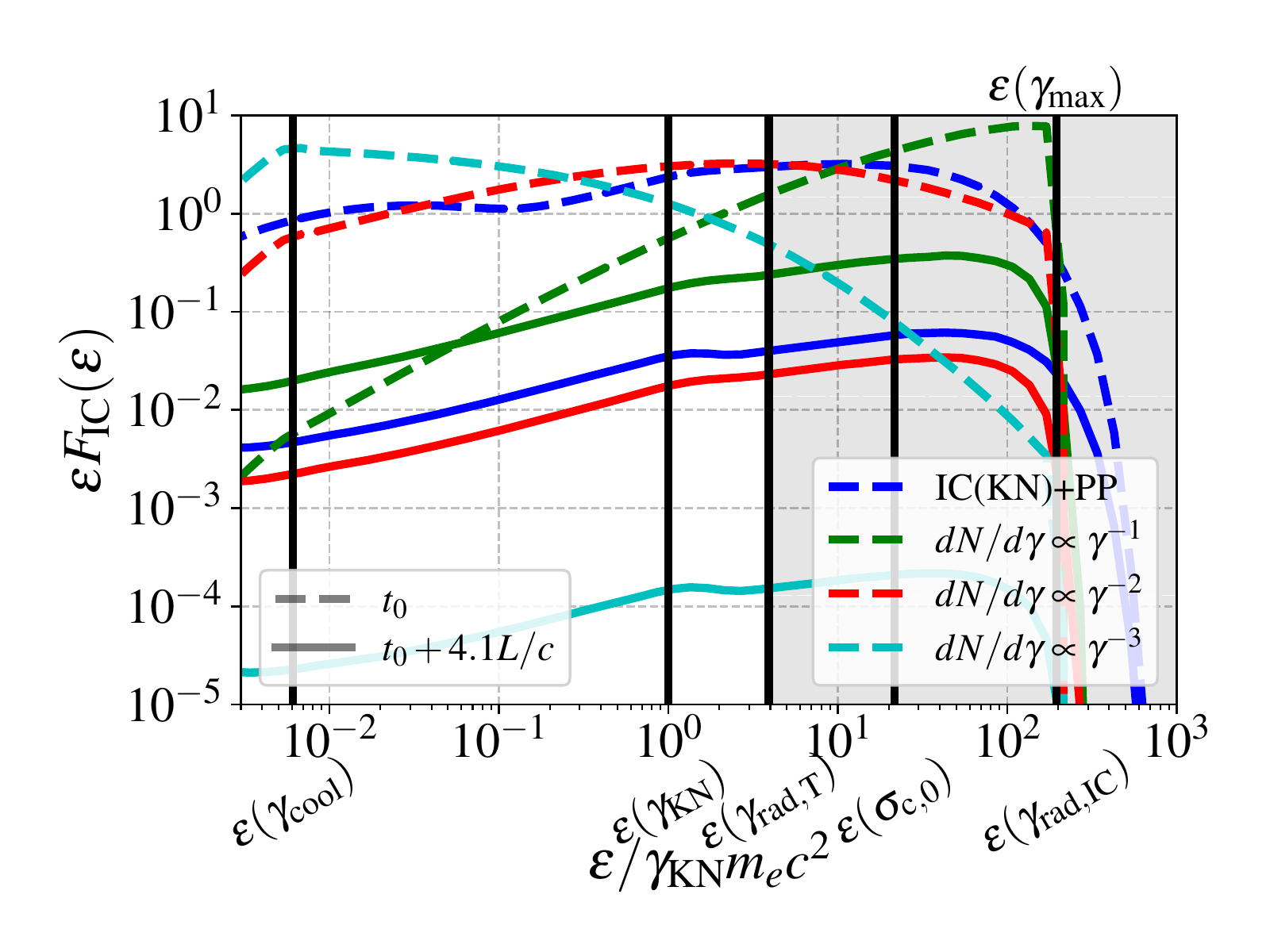}
    \end{subfigure}
    \caption{Left: Numerical experiment wherein different initial power-law particle distributions,~$\dif N / \dif \gamma \propto \gamma^{-p}$ for~$p=1,2,3$, are passively cooled under IC emission and pair production. All parameters are identical to the IC(KN)+PP simulation. Integration is done from time~$t_0$ (dashed curves) to~$t_0+4.1L/c$ (solid curves). For reference, we present, in blue, the passive cooling of the particle distribution yielded by the IC(KN)+PP simulation (as in Fig.~\ref{fig:afterglow}). Right: IC emission spectra corresponding to the particle distributions at left. A grey region indicates unobserved (above pair-production threshold) photon energies. After an initial-condition-\textit{dependent} relaxation period, all distributions converge to an initial-condition-\textit{independent} universal shape. At observable energies -- i.e.\ in the Thomson regime,~$\epsilon \lesssim \gkn \me c^2$~($\gamma \lesssim \gkn$) -- this universal shape features an approximate~$\gamma^{-2}$ power-law, corresponding to~$\epsilon F_{\rm IC}(\epsilon) \propto \epsilon^{0.5}$. (The apparent upturn in the particle distributions at low energies is an artifact of short integration times -- it results from a bulk population of cooled particles that would gradually become still colder with time, extending the~$\gamma^{-2}$ scaling to even lower energies.)}
    \label{fig:distspostproc}
\end{figure*}
We conduct postprocessing experiments in Fig.~\ref{fig:distspostproc} that suggest that the presented properties of the IC(KN)+PP lightcurve and spectrum in the absence of particle acceleration are \textit{universal}. In these experiments, we evolve different initial power-law distributions of particles --~$\dif N / \dif \gamma \propto \gamma^{-p}$ for~$p = 1,2,3$ and~$\gcool < \gamma < \gmax$ (dashed lines in Fig.~\ref{fig:distspostproc}, left panel) -- solely under the influence of Klein-Nishina emission and pair-production (as in the postprocessing phase of~Fig.~\ref{fig:afterglow} except that, there, the initial particle distribution is taken from time~$t_0$ of our PIC simulation). We find that, irrespective of the initial power-law slope, the particle distribution always relaxes, in the Thomson regime,~$\gamma \lesssim \gkn$, to a~$\gamma^{-2}$ scaling (solid lines in Fig.~\ref{fig:distspostproc}, left panel).\footnote{The slope,~$\gamma^{-2}$, can be calculated by considering the trickle of particles from~$\gamma>\gkn$ as monochromatic particle injection at~$\gamma=\gkn$ \citep{mwu21}.} This corresponds, for the Thomson part,~$\epsilon \lesssim \gkn \me c^2$, of the emission spectrum, to~$\epsilon F_{\rm IC}(\epsilon) \propto \epsilon^{-(2-3)/2} = \epsilon^{1/2}$ (see section~\ref{sec:icemitdists}) -- a rising spectrum that continues almost up to the pair-production threshold energy~$\epsilon = 4 \gkn \me c^2$ (solid lines in Fig.~\ref{fig:distspostproc}, right panel). These universal shapes, once reached, are maintained by the particle distribution and emission spectrum as they fall off, resulting in a self-similar~$t^{-6}$ luminosity decay law and constant mean photon energy,~$\langle \epsilon \rangle \sim 0.3 \gkn \me c^2$: the same as Fig.~\ref{fig:afterglow}.

The above exercise enables us to reason about the observable signatures of Klein-Nishina reconnection in regimes not probed by our simulations where reconnection-powered NTPA is known to yield a steeper particle energy distribution. This occurs, for example, in the presence of a strong guide field \citep{wu17} or in the transrelativistic regime of electron-proton plasmas \citep{wub18}. In such cases, we speculate that one would observe an initially steep emission spectrum corresponding to intrinsic reconnection-powered NTPA, followed, in the passive cooling phase, by a transition -- in fact, a hardening! -- to the identified universal shape as the flux decays (similar to the~$\dif N / \dif \gamma \propto \gamma^{-3}$ initial condition in Fig.~\ref{fig:distspostproc}). This might be difficult to observe, however, as the spectrum may dim too much before relaxing to the expected shape, starving gamma-ray instruments of a sufficient number of photons to reconstruct it (cf.\ the large gap between the initial and final~$\dif N / \dif \gamma \propto \gamma^{-3}$ curves in Fig.~\ref{fig:distspostproc}).

In this section, we have seen how the different regimes of radiative cooling treated by this study -- including their influence on particle acceleration and the resulting IC emission spectra -- give rise to highly distinct temporal radiative signatures. At the most coarse-grained level, emission from radiative reconnection tracks the electromagnetic dissipation in real time, whereas emission from n\onradiative[]reconnection traces only energization that has occurred in the past. Focusing on more specific observable differences in the context of reconnection-powered flares, the lightcurve from Thomson radiative reconnection is highly correlated with the observed average photon energy (provided one observes near the spectral peak at~$\sim \gradt^2 \eph$) and features an abrupt decay phase where both drop simultaneously. Klein-Nishina radiative reconnection with pair production, however, yields an exactly opposite trend, with no correlation between the average photon energy,~$\sim 0.3 \gkn \me c^2$, and the overall brightness, including in the (slower than in the Thomson regime, but still relatively fast) decay phase. These findings can be directly compared with, and tested by, observations of gamma-ray flares from suitable astrophysical systems (see section~\ref{sec:discussion}).

\subsection{Newborn pair energy budget and particle count}
\label{sec:origprodcompare}
\begin{figure*}
    \centering
    \includegraphics[width=\linewidth]{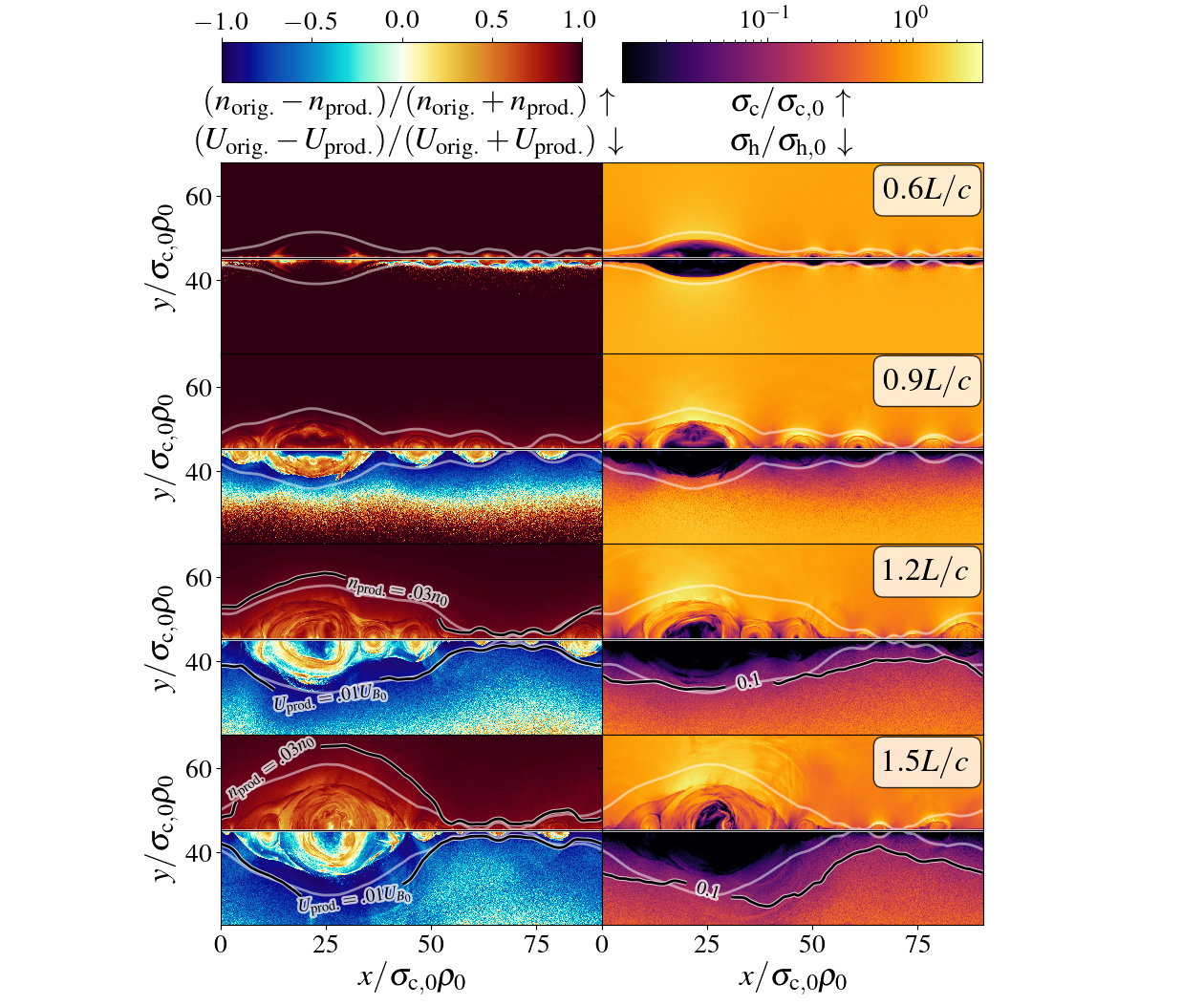}
    \caption{Left column: Snapshots of the relative contributions of original (orig.) and produced (prod.) pairs to (on the top half of each panel) the local particle number density,~$(n_{\rm orig.} - n_{\rm prod.})/(n_{\rm orig.} + n_{\rm prod.})$, and (on the bottom half of each panel) the local particle energy density,~$(U_{\rm orig.} - U_{\rm prod.})/(U_{\rm orig.} + U_{\rm prod.})$. Right column: Snapshots of the local cold,~$\sigcgen / \sigc$ (top halves of panels), and hot,~$\sighgen / \sigh$ (bottom halves of panels), magnetizations. Contours display values of key quantities along, or just upstream of, the reconnection separatrix. For the~$U_{\rm prod.}$ contour,~$U_{B_0} \equiv B_0^2 / 8 \pi$. In agreement with \citet{mwu21}, newborn pairs are everywhere subdominant in terms of their number density, but compete with the original particles for energetic dominance. Moreover, the newborn particles create an energetically dense coat around the reconnection layer where they load the local hot magnetization but not the cold magnetization.}
    \label{fig:spatialpairdom}
\end{figure*}

In the preceding parts of section~\ref{sec:results}, we explored consequences of Klein-Nishina and pair-production physics on magnetic reconnection, using as control cases the Thomson-radiative and n\onradiative[]regimes. We presented first what is similar to the latter two cases -- e.g.\ the overall spatial dynamics and the reconnection rate -- and then discussed the main distinctions, culminating with the very different observable signatures of the various radiative regimes. We now go one step farther, leaving behind our control cases in order to address issues that only exist in the context of Klein-Nishina reconnection with pair production. In particular, we comment on the newborn pairs' contribution to the reconnection system's energy and particle number budgets, which is presented graphically in Fig.~\ref{fig:spatialpairdom}. 

The left column of Fig.~\ref{fig:spatialpairdom} illustrates the local differences in number densities,~$(n_{\rm orig.} - n_{\rm prod.})/(n_{\rm orig.} + n_{\rm prod.})$, and energy densities,~$(U_{\rm orig.} - U_{\rm prod.})/(U_{\rm orig.} + U_{\rm prod.})$, between originally present (orig.) and produced (prod.) particles. In the number density panels, the newborn particles remain subdominant across time and space. However, once reconnection ignites above-threshold gamma-ray emission and pair production, the produced particles build up an energetically dense coat around the reconnection layer. Inside this coat, the newborn pairs completely dominate the energy density of the reconnection upstream region; in the reconnected flux region, while not as clearly dominant, they still vie for control of the energy budget. This demonstrates the main peculiarity of Klein-Nishina reconnection -- previously anticipated by \citet{mwu21} -- that, for a broad range of parameters (made more precise below), the produced particles are \textit{hot but tenuous}, competing with the original particles for energetic dominance of the system despite being much fewer in number. A parallel view of these effects is presented in the right-hand column of Fig.~\ref{fig:spatialpairdom}, where we plot the local cold magnetization~$\sigcgen \equiv (B_x^2+B_y^2)/[4 \pi (n_{\rm orig.}+n_{\rm prod.}) \me c^2]$ and the hot magnetization~$\sighgen \equiv (B_x^2 + B_y^2)/4 \pi w$ [note: the enthalpy density, for our ultrarelativistic particle distributions, is~$w=4(U_{\rm orig.}+U_{\rm prod.})/3$; see equation~(\ref{eq:sigh}) and surrounding discussion]. Because the newborn pairs load the upstream plasma energy density without appreciably changing the number density, they reduce~$\sighgen$ from its native/initial value,~$\sigh$, while leaving~$\sigcgen\simeq\sigc$ essentially untouched.

Even though the newborn pairs are everywhere less numerous than those originally present, they discernibly contribute to the plasma density [e.g. by changing the hue of the~$(n_{\rm orig.} - n_{\rm prod.})/(n_{\rm orig.} + n_{\rm prod.})$ spatial maps] close to and inside of the reconnection separatrix. \citet{mwu21} forecasted that the newborn pairs should begin to contribute non-negligibly to the plasma number density passing across the separatrix when~$\sigc$ exceeds~$\gkn$ by more than a factor of~$10$ or so. This is consistent with our findings on the Klein-Nishina reconnection pair yield presented in the next section, which suggest that the IC(KN)+PP run~($\sigc \simeq 20 \gkn$) is indeed beginning to border on a regime where the \textit{in-situ} produced pairs contribute more significantly to the particle count. For the rest of this section, however, we focus on the present regime where the newborn particles are energetically dense but numerically few.

As discussed by \citet{mwu21}, the energy density injected into the upstream region via the hot newborn pairs leads to a \textit{pair-loaded magnetization},~$\sighgen^*$, to which the system tends to regulate itself (provided the initial magnetization,~$\sigh$, exceeds~$\sighgen^*$). However, if the pair loading is efficient enough, the system may never actually attain a steady state with~$\sighgen = \sighgen^*$. It will instead flood the upstream energy density via pair production, overshooting to a lower magnetization,~$\sighgen < \sighgen^*$, and quenching subsequent NTPA until the upstream plasma is vacated and a high~$\sighgen > \sighgen^*$ is restored. This would restart the process, leading to a limit cycle: the system would indefinitely ricochet between a high and a low magnetization on either side of~$\sighgen^*$ throughout the duration of reconnection.

\citet{mwu21} quantified the dependence of the long-term fate of the system -- whether it smoothly regulates to, or violently oscillates about,~$\sighgen^*$ -- on the efficiency of upstream pair loading. We measure this efficiency from our simulations in Appendix~\ref{sec:pairreg}, finding that it is too low, based on the analysis of \citet{mwu21}, to trigger~$\sighgen$-mediated limit cycles. We therefore conclude that \textit{limit cycles mediated by pair-loading of~$\sighgen$ are unlikely in astrophysical Klein-Nishina reconnection}.

In this section, we have shown that our simulations probe a Klein-Nishina reconnection regime where, as previously predicted by \citet{mwu21}, the newborn upstream pairs are \textit{hot but tenuous}, loading the upstream plasma pressure, energy density, and hot magnetization, but not the upstream plasma density or cold magnetization. The simulations further provide measurements (Appendix~\ref{sec:pairreg}) that disfav\spellor[~$\sighgen$-mediated]limit cycles. However, if a regime existed featuring a large multiplicity of newborn pairs, the possibility of~$\sigcgen$-mediated limit cycles would remain an open question. It is to the overall pair yield of Klein-Nishina reconnection, including the potential existence of such a regime, that we now turn. 

\section{Pair yield}
\label{sec:pairyield}
We have already seen how the pairs produced in Klein-Nishina reconnection contribute unique aspects to its observable signatures and self-consistent internal dynamics. In addition to these intrinsic features, another important implication of Klein-Nishina reconnection is its interaction with its environment. Here, pair production opens up a coupling channel that is unique to QED reconnection: the possibility to change the ambient plasma composition (e.g.\ the positron-to-proton ratio). Thus, in this section, we c\haracteriz[e]the total pair yield from a Klein-Nishina reconnection event.

To map the dependence of the pair yield on reconnection parameters, we introduce two auxiliary simulation campaigns. Each one explores the pair yield's dependence on one principal variable. The first campaign explores the main new quantity introduced by Klein-Nishina physics,~$\gkn$. The second campaign is a system-size~($L$) scan. It doubles as an opportunity to diagnose how well our results may g\eneraliz[e]to astrophysical Klein-Nishina reconnection, where the layer lengths,~$L$, are expected to be much larger compared to the plasma microscales (e.g.~$\sigc \, \rho_0$) than is possible to simulate. The full sets of parameters used for the respective~$\gkn$- and~$L$-scans are s\ummariz[ed]in Tables~\ref{table:gknscan} and~\ref{table:lscan}.
\begin{table}
\centering
\begin{threeparttable}
    \caption{Parameters for the~$6$-simulation scan over~$\gkn$. Values that are the same as those of Table~\ref{table:params} are omitted.}
\label{table:gknscan}
\begin{tabular}{l r l r l l}
  \hline
  Symbol & Value & (=equivalent) \\
  \hline \hline
  $\sigc$ & $10^4$ & \\
  $\sigh$ & $10^2$ & \\
  $\itemp$ & $2.5 \times 10^{-4} \sigc$ & $=25$ \\
  $L$ & $210 \sigc \rho_0$ \\
  $\gmax$ & $21 \sigc$ & $= 2.1 \times 10^5$ \\
  $B_g$ & $0.1 B_0$ & \\
  $\Delta x , \, \Delta y$ & $\sigc \rho_0 / 24$ & \\
  $\Delta t$ & $0.99 \Delta x / \sqrt{2} c$ & \\
  $N$ & $5120$ & \\
  \hline \hline
  $\gcool$ & $2.3 \times 10^{-2} \sigc$ & $= 230$ \\
  $\gradt$ & $0.70 \sigc$ & $= 7.0 \times 10^3$ \\
  $\gkn$ & $ (0.091, 0.12, 0.15) \sigc$ & $= (\,\,\,910, 1200, 1500)$ \\
  \quad \ldots & $ (\,\,\, 0.19, 0.27, 0.34)  \sigc$ & $= (1900, 2700, 3400)$ \\
  $\tau_{\gamma\gamma}$ & $(2.4, 3.1, 4.0, 5.2, 7.2, 9.1)$ & \\
  \hline
\end{tabular}
\end{threeparttable}
\end{table}
\begin{table}
\centering
\begin{threeparttable}
    \caption{Parameters for the~$3$-simulation scan over~$L$. Values that are the same as those of Table~\ref{table:params} are omitted. The value of~$\gkn$ is adjusted slightly between runs to keep~$\gradk = \gmax$.}
\label{table:lscan}
\begin{tabular}{l r l r l l}
  \hline
  Symbol & Value & (=equivalent) \\
  \hline \hline
  $\sigc$ & $400$ & \\
  $\sigh$ & $400$\tnote{*} & \\
  $\itemp$ & $0.1$ \\
  $L$ & $(34, 67, 100) \sigc \rho_0$ \\
  $\gmax$ & $(3.4, 6.7, 10.) \sigc$ & $= (1300, 2700, 4000)$ \\
  $B_g$ & $0.1 B_0$ & \\
  $\Delta x , \, \Delta y$ & $\sigc \rho_0 / 76$ & \\
  $N$ & $(2560, 5120, 7680)$ & \\
  \hline \hline
  $\gcool$ & $(12, 5.9, 4.0) \times 10^{-3} \sigc$ & $= (4.7, 2.4, 1.6)$ \\
  $\gradt$ & $0.2 \sigc$ & $= 80$ \\
  $\gkn$ & $(0.065, 0.055, 0.051) \sigc$ & $= (26, 22, 20.)$ \\
  $\tau_{\gamma\gamma}$ & $(3.3, 5.6, 7.8)$ & \\
  \hline
\end{tabular}
\begin{tablenotes}
\item[*]Because the initial plasma temperature,~$\itemp=0.1$, is n\onrelativistic[ally]cold,~$\sigh=\sigc$.
\end{tablenotes}
\end{threeparttable}
\end{table}

\begin{figure}
    \centering
    \includegraphics[width=\columnwidth]{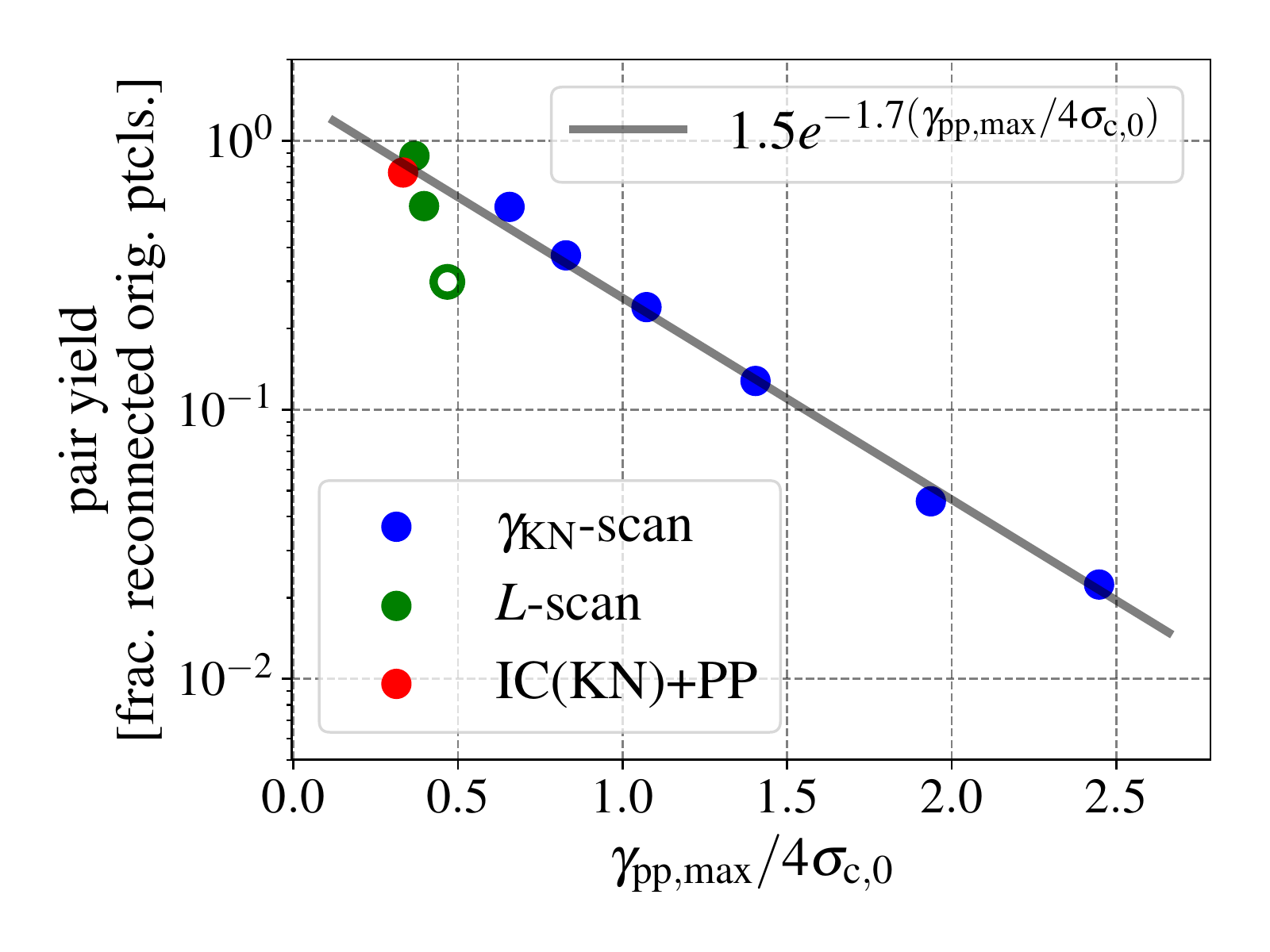}
    \caption{Pair yield for the simulations of the~$\gkn$ parameter scan (blue; Table~\ref{table:gknscan}), the~$L$ parameter scan (green; Table~\ref{table:lscan}), and the IC(KN)+PP run (red; Table~\ref{table:params}). The smallest-$L$ simulation (open circle) is omitted when fitting for the exponential scaling law. Despite the large differences in their fiducial and scanned parameters, all simulations' pair yields (with the exception of the smallest one in the~$L$-scan; see text) fall onto a scaling law with a single control parameter. This parameter is the ratio of two Lorentz factors: that,~$\gamppmax \simeq 30 \gkn$, of a particle that scatters photons up to the peak energy of the pair-production \crosssection[;]and the maximum typical Lorentz factor,~$\gx \equiv 4 \sigc$, that particles attain at reconnection X-points. This suggests that the pair yield is contro\led[]mainly by the efficiency of particle acceleration up to and beyond the optimal pair-producing particle energy.}
    \label{fig:pairyield}
\end{figure}

Besides examining the impact of their respective variables, these additional sets of simulations are separately calibrated to different fiducial parameters. The~$\gkn$-scan has a larger n\ormaliz[ed]system size,~$L=210 \sigc \rho_0$, and lower magnetizations,~$\sigc=10^4$ and~$\sigh=10^2$, than our base run, IC(KN)+PP (for which~$L \simeq 91 \sigc \rho_0$,~$\sigc=1.2\times10^5$, and~$\sigh=1250$; Table~\ref{table:params}). The~$L$-scan has still different initial magnetizations,~$\sigc=\sigh=400$, and features a n\onrelativistic[]initial upstream plasma,~$\itemp=0.1$. Spreading out, in this way, our auxiliary campaigns around the Klein-Nishina reconnection parameter space helps us identify a reduced set of control parameters (in fact, one single parameter; Fig.~\ref{fig:pairyield}) that decides the pair yield. These control parameters, in turn, shed light on the main physical mechanisms responsible for the pair-production efficiency while also providing a potential method for estimating this efficiency in astrophysical systems.

We define the pair yield in our simulations as the ratio of the total number,~$N_{\rm prod.}$, of leptons (electrons and positrons) produced on the fly to the cumulative count,~$N_{\rm rec.}$, of originally present leptons processed by reconnection (i.e.\ swept across the separatrix). Generally, about~$60$ per cent of the initial upstream particles cross the separatrix before reconnection saturates, the same as the percentage of the initial magnetic flux that is reconnected (see section~\ref{sec:recrate} and Fig.~\ref{fig:recflux}). Though the processing of upstream magnetic field and original particles essentially finishes by the end of our simulations, pair production does not; there remains a prominent population of high-energy particles that have yet to cool down by emitting pair-producing gamma-rays. In section~\ref{sec:afterglow}, we postprocessed the passive cooling associated with these particles to diagnose the decaying phase of the reconnection-powered lightcurve. Here, we use the same technique to continue evolving the total newborn pair count past the end of each simulation. Once the count saturates (typically by~$6L/c$ or so), we record it as~$N_{\rm prod.}$. As long as we begin the postprocessing after the energy transfer from fields to particles is mostly complete, the final~$N_{\rm prod.}$ does not depend much (less than~$10$ per cent) on the exact moment in time when the postprocessing starts.

The above method for calculating the total pair yield, like our periodic simulation boundaries, ignores the potential for particles and photons to escape the ambient radiation field before pair production is complete. This issue is less important when the pair-production optical depth,~$\tau_{\gamma\gamma}$, is large, corresponding to shorter mean-free-paths of photons and more rapid particle cooling times (e.g.\ Fig.~\ref{fig:tstop_kn}). When~$\tau_{\gamma\gamma}$ becomes small, the pair yield calculated in this way still has meaning as long as the extent of the ambient radiation field is much larger than the size of the reconnection system (and provided, once particles exit the reconnection system, they are no longer significantly energized).

We present the pair yield calculated for our auxiliary~$\gkn$ and~$L$ simulation campaigns, as well as for the single IC(KN)+PP run discussed earlier (section~\ref{sec:results}), in Fig.~\ref{fig:pairyield}. Remarkably, when plotted as a function of just the single control parameter,~$\gamppmax / 4 \sigc$, the pair yields from all simulations -- despite the very different fiducial and scanned parameters -- collapse onto the same exponential scaling law,
\begin{align}
    N_{\rm prod.} / N_{\rm rec.} = 1.5 \exp \left(-1.7 \gamppmax / 4 \sigc\right) \, .
    \label{eq:pairyield}
\end{align}
Here,~$\gamppmax \equiv 3.6 \times 8 \times \gkn \simeq 30 \gkn$ is the characteristic Lorentz factor of particles that scatter background photons to energies at peak pair-production \crosssection[,~$\escat \sim \gamppmax / 2 \sim 3.6 (\me c^2)^2 / \eph$.] Meanwhile,~$4 \sigc$ is the characteristic maximum energy,~$\gx \equiv 4 \sigc$, that particles acquire near reconnection X-points. Equation~(\ref{eq:pairyield}) suggests that what controls the final pair yield of reconnection is how broad a distribution of high-energy particles can be energized near reconnection X-points to radiate photons close to or above the peak pair-production \crosssection[:]i.e.\ by how much~$\gx$ exceeds~$\gamppmax$.

\citet{mwu21} predicted that the ratio of the newborn-to-original upstream pair densities flowing across the reconnection separatrix should be proportional to~$\sigc / \gkn$ (times a \nontrivial[]function depending on NTPA in the reconnection layer). Although this number density ratio is not the same as the global ratio of newborn-to-reconnection-processed particles measured here -- the latter also includes the n\onnegligible[]number of pairs born on the exhaust side of the reconnection separatrix -- both results share the same main contro\ling[]parameter,~$\sigc / \gkn \propto \gx / \gamppmax$.

Now let us discuss where the scaling~(\ref{eq:pairyield}) may break down. First, all of the simulations in our campaigns have an effective radiative \cutoff[]Lorentz factor,~$\gradk \geq 4 \sigc$. This means that we need to be vigilant as~$\gkn$ increases (coinciding with larger~$\gamppmax/4 \sigc$), because~$\gradk$ may then fall below~$4 \sigc$. In that case, our present interpretation of equation~(\ref{eq:pairyield}) leads us to expect another \cutoff[]in the pair yield, e.g.\ of the form,~$\exp\,(-\gamppmax/\gradk)$, to kick in, signa\ling[]that X-point particle acceleration has become limited by~$\gradk$ instead of by~$\sigc$. Second, in the opposite limit, where~$\gamppmax / 4\sigc$ becomes small, we are likely to transition to a regime in which extremely efficient particle acceleration (giving a particle distribution power-law index approaching the n\onradiative[]limit,~$p\simeq1$) causes the pair yield to continue to grow with~$4 \sigc / \gamppmax$. Though this is not what a naive extrapolation of~(\ref{eq:pairyield}) produces, it is more coherent with our present interpretation of that formula, in which more particles being accelerated into the optimal pair-producing energy range enhances the pair yield.

Let us examine, for the sake of argument, what might happen if pushing~$\gamppmax / 4 \sigc$ to smaller values than studied here indeed led to a copious pair yield. Then Klein-Nishina reconnection would begin to move out of the regime of our simulations (discussed in section~\ref{sec:origprodcompare} and Appendix~\ref{sec:pairreg}) where the produced particles are \textit{hot but tenuous}. Instead, the newborn particles would become \textit{hot and abundant}, potentially even loading the upstream cold magnetization to a reduced value,~$\sigcgen^* < \sigc$, whereas in our simulations they modify only the hot magnetization (section~\ref{sec:origprodcompare}). If a steady state were to exist in such a regime, the cold magnetization would need to self-adjust until the pair yield as defined in equation~(\ref{eq:pairyield}) attains unity (times some efficiency factor to account for the fact that not all of the produced particles would be born into the upstream region; cf.\ Appendix~\ref{sec:pairreg}). According to Fig.~\ref{fig:pairyield}, this occurs when~$\gamppmax / 4 \sigcgen^* \simeq 0.25$, or, equivalently,~$\sigcgen^* \simeq \gamppmax \simeq 30 \gkn$. Such a result would be highly attractive, for it would open up the possibility of measuring the lepton material composition (via~$\sigcgen^*$) in terms of the seed photon energy, a much easier quantity to infer observationally.

Lastly, we note that the trend identified in Fig.~\ref{fig:pairyield} is broken at smaller system sizes. Indeed, we excluded the pair yield measurement from our~$L=34 \sigc \, \rho_0$ simulation when fitting for equation~(\ref{eq:pairyield}). The break from the formula in smaller systems reflects findings of earlier simulation studies that collisionless reconnection transitions to an asymptotically large-system limit, corresponding to the multiple X-point, plasmoid-mediated regime, only once~$L \gtrsim 40 \sigc \rho_0$ \citep{wuc16}. When respecting this limit, our simulations overlap the identified pair yield trend. However, we cannot rule out larger systems yielding even more efficient pair production than~(\ref{eq:pairyield}).

In this section, we have c\haracteriz[ed]the pair yield of Klein-Nishina reconnection in terms of a one-dimensional exponential scaling law, equation~(\ref{eq:pairyield}) -- despite the high-dimensional parameter space of this problem. This scaling law appears to be robust across an order of magnitude or more in~$\gkn$,~$\sigh$,~$\sigc$, and~$\itemp$, while being respected across a factor of~$3$ or~$4$ in system size (the most that we can afford to probe in the large-system,~$L > 40 \sigc \rho_0$, regime given the stringent parameter constraints of the problem; cf.\ section~\ref{sec:setup}). Equation~(\ref{eq:pairyield}) may need to be modified in the transition region between Klein-Nishina~($\gkn < \gradt$) and Thomson~($\gkn > \gradt$) radiative reconnection, and it may also give way to a new regime of efficient (much greater than order-unity) pair yield when~$\gamppmax / 4 \sigc$ becomes much smaller than the values we test. Nevertheless, we are able to capture an order of magnitude in the control parameter,~$\gamppmax / 4 \sigc$, including near the point,~$\gamppmax / 4 \sigc \simeq 0.25$, where the pair yield reaches~$1$.

\section{Discussion}
\label{sec:discussion}
In this section, we discuss the relevance of our findings to gamma-ray observations of selected astrophysical systems: flat-spectrum radio quasars, black hole accretion d\isk[]coronae, the M87$^*$ magnetosphere, and gamma-ray binaries. In each case, we argue why the operation of Klein-Nishina reconnection in these systems is expected on theoretical grounds. We further discuss consequences, for each system, of the results of sections~\ref{sec:afterglow} and~\ref{sec:pairyield}. For reference, we briefly recapitulate those results here in the context of potential links to observations.

The main finding of section~\ref{sec:afterglow} is the marked departure of the observable signatures of radiative reconnection in the Klein-Nishina regime (with pair production) from the Thomson-cooled regime (without pair production). Namely, while Thomson-cooled reconnection features a tight correlation between the mean observed photon energy and the system's total luminosity (i.e.\ \quoted[),]{harder-when-brighter}Klein-Nishina reconnection breaks this correlation, featuring a steady mean photon energy irrespective of the luminosity. This includes the decaying phase of a flaring event, wherein the average photon energy is preserved even as the luminosity drops. These findings serve therefore as an \textit{observational diagnostic}. They can be directly compared to gamma-ray observations to build a case (or not) for the operation of Klein-Nishina reconnection in a given object. In contrast, the results of section~\ref{sec:pairyield} function as an \textit{inference criterion}, providing a method to estimate a quantity that is difficult to constrain from observations -- the emitting region's material composition -- using quantities that may be more readily measured or estimated (specifically, the magnetization,~$\sigc$, and the Lorentz factor,~$\gamppmax$, of particles whose photons are at maximum pair-production \crosssection[]with the radiation bath).

Following our system-by-system discussion, we s\ummariz[e]our broad conclusions across all systems in section~\ref{sec:discsummary} and Table~\ref{table:discussion}. While readers interested in a particular object may skip directly to the corresponding subsection, those seeking a more general overview may wish to skip first to the summary material.

\subsection{Flat-spectrum radio quasars and other blazars}
\label{sec:fsrqs}
Blazars are AGNs that launch bipolar relativistic jets, one of which (hereafter, the singular \quoted[)]{jet}travels toward the Earth. The jet's relativistic motion Doppler boosts its emission, leading to dramatic observable consequences. For example, blazars dominate the discrete sources on the extragalactic gamma-ray sky \citep[e.g.][]{tevcat, fermi20a} and, in the optical band, they routinely outshine the cumulative starlight of their host galaxies \citep{olk16}. Blazar jet emission is also exceptionally broad, extending from radio frequencies up to gamma-rays in a characteristic n\onthermal[]double-humped spectrum \citep{fmc98, g11, grc17}. The lower-energy spectral hump originates from synchrotron radiation by relativistic electrons and positrons (henceforth \quoted[)]{leptons}spiraling around magnetic field lines in the jet. The higher-energy peak is frequently attributed to IC radiation also by relativistic jet leptons \citep[e.g.][]{pg22}.

Blazars are phenomenologically decomposed into two main subdivisions: flat-spectrum radio quasars (FSRQs) and BL Lacs.\footnote{For our purposes, we fold the extreme/ultra-high-frequency-peaked BL Lacs (EHBLs/UHBLs) high-frequency-peaked BL Lacs (HBLs), intermediate-frequency-peaked BL Lacs (IBLs), and low-frequency-peaked BL Lacs (LBLs) into the BL Lac class.} \mbox{FSRQ}s exhibit lower-energy spectra, with the synchrotron component peaking at IR energies and the higher-energy IC component peaking in the MeV-to-GeV gamma-rays. In contrast, the maxima of the synchrotron and IC spectra in BL Lacs typically lie in the UV/X-ray and GeV-to-TeV bands, respectively. Despite their lower photon energies, \mbox{FSRQs} are more luminous and exhibit much larger ratios of IC-to-synchrotron power. Finally, where the norm for BL Lacs is featureless n\onthermal[]spectra, FSRQs usually exhibit prominent broad emission lines or quasi-thermal radiation at lower energies. These are usually attributed to emission by the underlying AGN accretion d\isk[]and to reprocessing of the accretion d\isk[]light by circumnuclear material. \citep[Illustrative references pertaining to this entire paragraph include:][]{fmc98, g11, ms16, grc17, bmr19, pg22}.

The circumnuclear regions that are observed at lower energies in FSRQs can provide intense sources of seed photons for IC emission in the jet \citep{bs87, mk89, sbr94}. Particularly bright are the broad emission line region (BLR) and the hot dust region (HDR). Of these two, the BLR is smaller, occupying an inner zone (up to roughly~$0.1 \, \rm pc$ from the nucleus) where irradiation from the accretion d\isk[]ionizes the ambient gas, and subsequent recombination emits line emission, broadened by rapid orbital motion, of characteristic UV energy
\begin{align}
    \eblr = 10 \, \rm eV \, 
    \label{eq:eblr}
\end{align}
onto the jet \citep{tg08, ssm09, nbc12, mwu21}. The outer circumnuclear reprocessing region (up to roughly~$4 \, \rm pc$ from the nucleus) is the HDR, which comprises dust radiatively heated by the accretion d\isk[]light up to a temperature of about~$T_{\rm HDR} \sim 1200 \, \rm K$. The hot dust shines a quasithermal spectrum onto the jet of characteristic energy \citep{nsi08a, nsn08b, ssm09, nbc12, mwu21}
\begin{align}
    \ehdr \sim 3 k_{\rm B} T_{\rm HDR} = 0.3 \, \rm eV \, .
    \label{eq:ehdr}
\end{align}

The radiation fields from the BLR and the HDR furnish excellent conditions for comparing with our simulations. First, they are energetically dense at distances far from the central engine such that the magnetic field energy density is small compared to that of the seed photons,~$B_0^2/8 \pi \ll \uph$, a necessary condition for neglecting synchrotron losses, as we do in our simulations. Second, the resulting radiation field is expected to be homogeneous not just across the reconnection region, but also across the whole jet width. This creates a direct opportunity for applying our pair yield law found in section~\ref{sec:pairyield}, which ignores the possibility of above-threshold photons escaping the ambient radiation field before being absorbed to produce electron-positron pairs.

We conduct a detailed analysis of scenarios where reconnection powers high-energy IC emission in FSRQ jets in our previous analytic work, \citet{mwu21}. There, we estimate the Lorentz factor energy scales~$\gcool$,~$\gkn$,~$\gradt$, and~$\gmax$ either for the case where the reconnection region lies within the more energetically dense BLR, and thus leptons scatter primarily BLR photons, or for the case where the reconnection region is outside the BLR but inside the HDR such that the BLR radiation field is diluted and the HDR supplies the dominant seed photons. Our estimates in both scenarios yield fiducial energy scales that are in the required order,~$\gcool < \gkn < \gradt < \gmax$ [equation~(\ref{eq:knregime})], to r\ealiz[e]Klein-Nishina reconnection.

In \citet{mwu21}, we also pointed out that the BLR and HDR are optically thick to gamma-rays above the pair-production threshold energies 
\begin{align}
    \epsilon_{\rm th,BLR} = \me^2 c^4 / \eblr = 30 \, \rm GeV
    \label{eq:eblrth}
\end{align}
and
\begin{align}
    \epsilon_{\rm th,HDR} = \me^2 c^4 / \ehdr = 0.9 \, \rm TeV \, ,
    \label{eq:ehdrth}
\end{align}
respectively. The corresponding characteristic energies radiated by a particle of energy~$\gkn$ in each case are~$\gamma_{\rm KN,BLR} \me c^2 / 2 = \epsilon_{\rm th,BLR} / 8 = 3 \, \rm GeV$ and~$\gamma_{\rm KN,HDR} \me c^2 / 2 = \epsilon_{\rm th,HDR} / 8 = 0.1 \, \rm TeV$. This means that observations by the \textit{Fermi} LAT, which is sensitive roughly to energies in the~$0.1 - 100 \, \rm GeV$ range \citep[][]{fermi09_techspecs}, are able to probe emission by particles at~$\gamma_{\rm KN,BLR}$ up through the BLR gamma-ray absorption \cutoff[.]At the same time, Imaging Atmospheric Cherenkov Telescopes (IACTs), typically sensitive in the~$0.1 - 10 \, \rm TeV$ band \citep{cta19}, stand best to capture the analogous physics for the IC(HDR) scenario. This is fortuitous because particles with energies near~$\gkn$ are precisely those responsible for the characteristic spectral and timing signatures of Klein-Nishina reconnection uncovered in section~\ref{sec:afterglow}. These particles radiate just below pair-production threshold, producing the highest-energy observable photons, and they are actively replenished by radiative reprocessing from higher (above-threshold and, hence, invisible) photon energies, which s\tabiliz[es]the observed spectral energy density. Thus, FSRQ flares in the GeV and TeV bands are ideally suited to probe the expected observational signatures of reconnection in the regime studied in this work.

\begin{figure*}
    \centering
    \includegraphics[width=\linewidth]{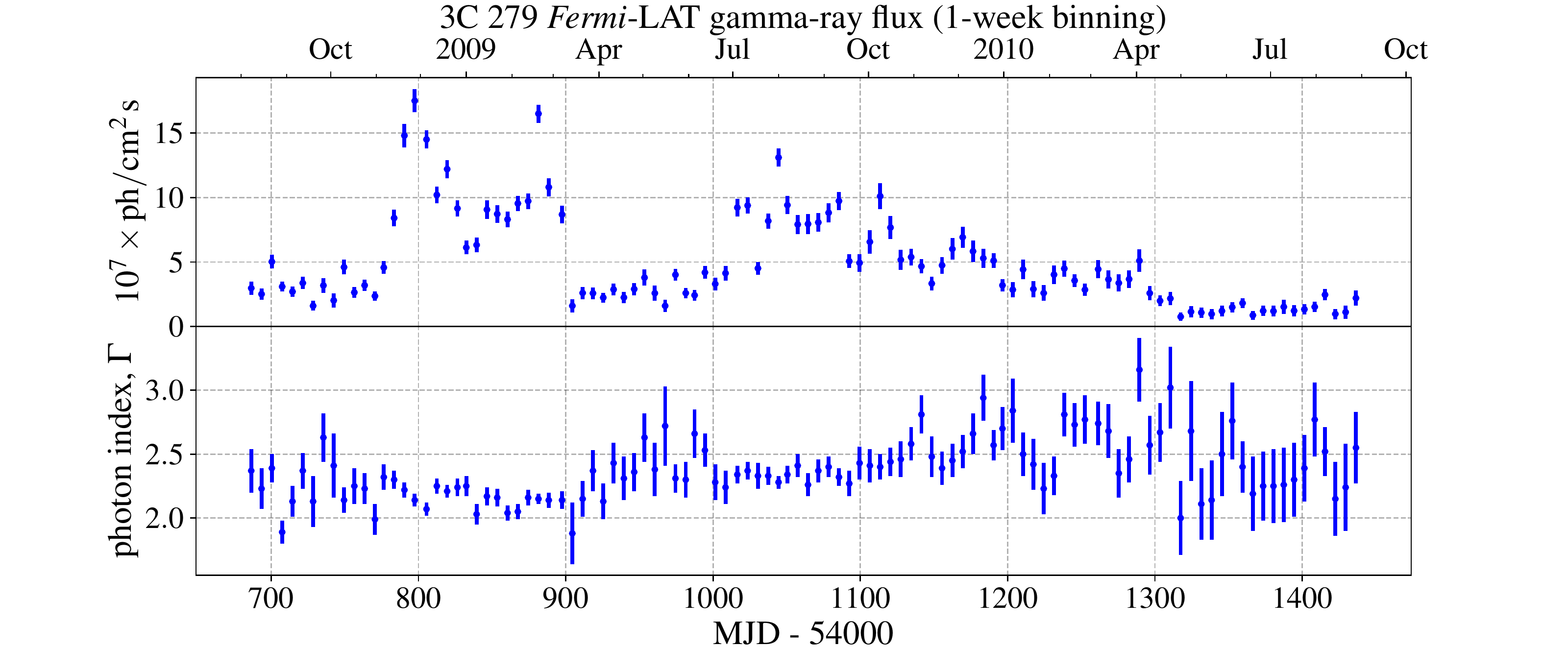}
    \caption{\textit{Fermi} LAT photon flux (top) and spectral index (bottom), both binned into one-week time intervals, detected from 3C 279 during the period presented by \citet{hmn12}. Error bars indicate symmetric Gaussian equivalent~$1$-sigma error. Top and bottom panels, correspond, respectively, to panels~(c) and~(f) of those authors' fig.~$1$. The data were retrieved from the Fermi LAT Light Curve Repository \citep{fermi23}. The photon index,~$\Gamma$, is defined such that the flux of photons between energies~$\epsilon$ and~$\epsilon + \dif \epsilon$ is proportional to~$\epsilon^{-\Gamma}$. This is connected to the~$\epsilon F(\epsilon)$ representation, e.g.\ of Fig.~\ref{fig:icemitdists}, in that~$\epsilon F(\epsilon) \propto \epsilon^{-\Gamma+2}$.}
    \label{fig:lc3c279}
\end{figure*}
\begin{figure}
    \centering
    \includegraphics[width=\linewidth]{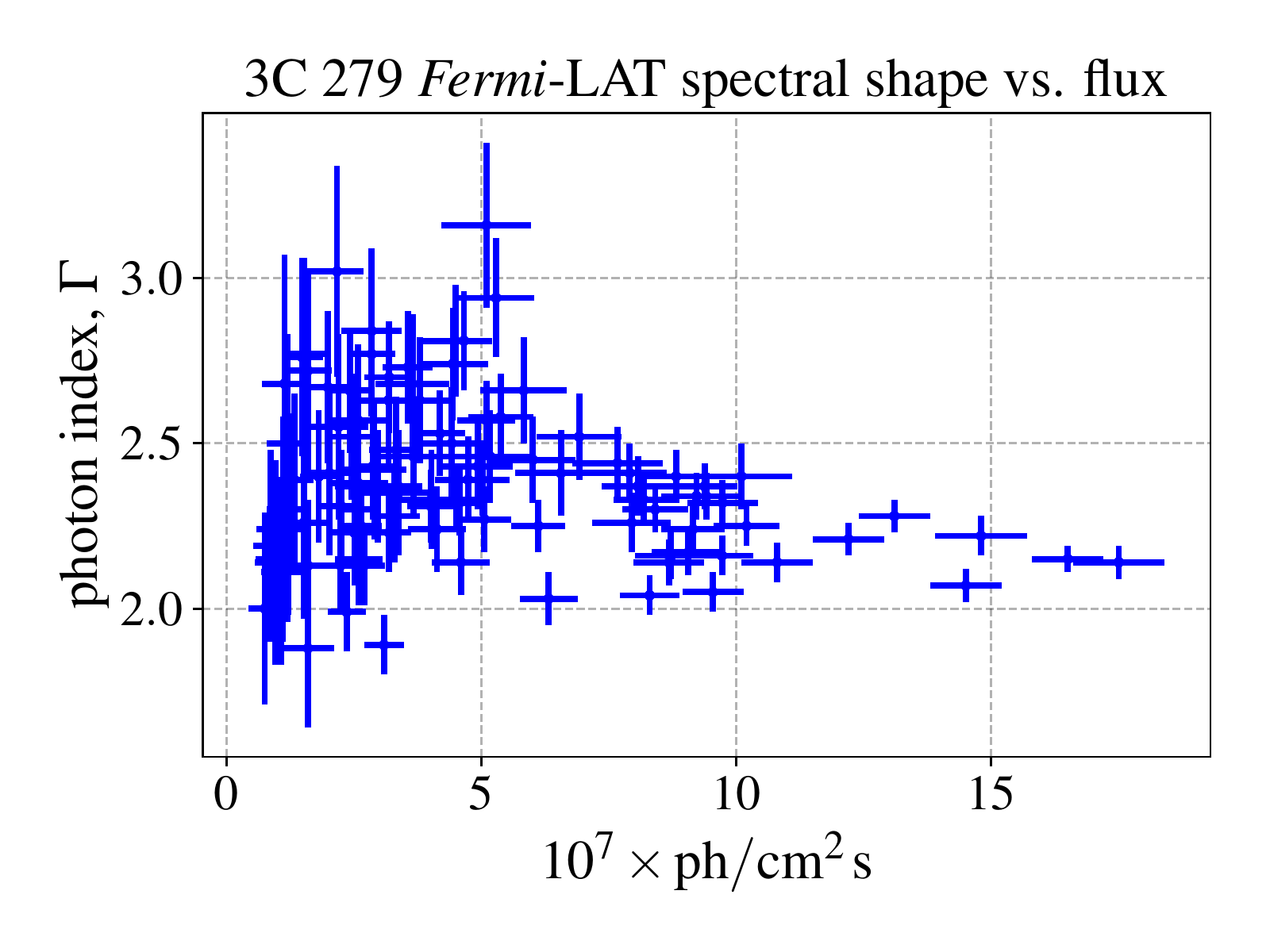}
    \caption{The data from Fig.~\ref{fig:lc3c279} rearranged into a scatter plot of gamma-ray spectral index vs.\ photon flux. This figure roughly corresponds to fig.~$3$, panel~(A) of \citet{hmn12}. The s\tabiliz[ation]of the spectrum during high-flux periods, notably from MJD~$54800$ to~$54900$ and from MJD~$55000$ to~$55100$ in Fig.~\ref{fig:lc3c279}, is quite reminiscent of that found in section~\ref{sec:afterglow} (cf.\ Fig.~\ref{fig:afterglow}).}
    \label{fig:hi3c279}
\end{figure}
Let us examine a few instructive flares observed by the \textit{Fermi} LAT [which, again, probes the IC(BLR) scenario] from the FSRQ,~3C~279. \citet{hmn12} report~3C~279 flaring periods over the first two years of \textit{Fermi} operations. For reference, we reproduce the lightcurve and photon index time series presented by those authors in their fig.~$1$, as well as the correlation between the total gamma-ray flux and spectral shape shown in their fig.~$3$, in our respective Figs.~\ref{fig:lc3c279} and~\ref{fig:hi3c279} using data retrieved from the Fermi LAT Light Curve Repository \citep{fermi23}. \citet{hmn12} note mild \quoted{harder-when-brighter}behavi\spellor[]over the entire observation period, but this is somewhat quenched during the brightest periods (Fig.~\ref{fig:hi3c279}), during which the \textit{Fermi}-measured spectral index becomes remarkably flux-independent (e.g.\ between MJD~$54800$ and~$54900$ as well as between MJD~$55000$ and~$55100$ in Fig.~\ref{fig:lc3c279}). Roughly similar behavi\spellor[]is often, but not uniformly, seen in later observations of the same object. In two even brighter outbursts from~3C~279 reported by \citet{hnm15} and \citet{fermi16a}, the gamma-ray flux reached high-enough levels to reconstruct spectra for individual orbits of the \textit{Fermi} satellite. In the latter event, the flaring individual-orbit spectra reveal a photon energy index that remains between about~$1.9$ and~$2.1$ while the flux varies across about a factor of~$3$ (table~$1$ of \citealt{fermi16a}). The pre- and post-outburst phases of the event also feature rather stable photon indices (though this appears more statistically significant for the pre-outburst phase; see fig.~$1$ of \citealt{fermi16a}). As an example of when such spectral stability is not seen, the first flare a\nalyz[ed]by \citet{hnm15} exhibits an extreme hardening of the photon index that then softens on the trailing edge of the flare.

\begin{figure}
    \centering
    \includegraphics[width=\linewidth]{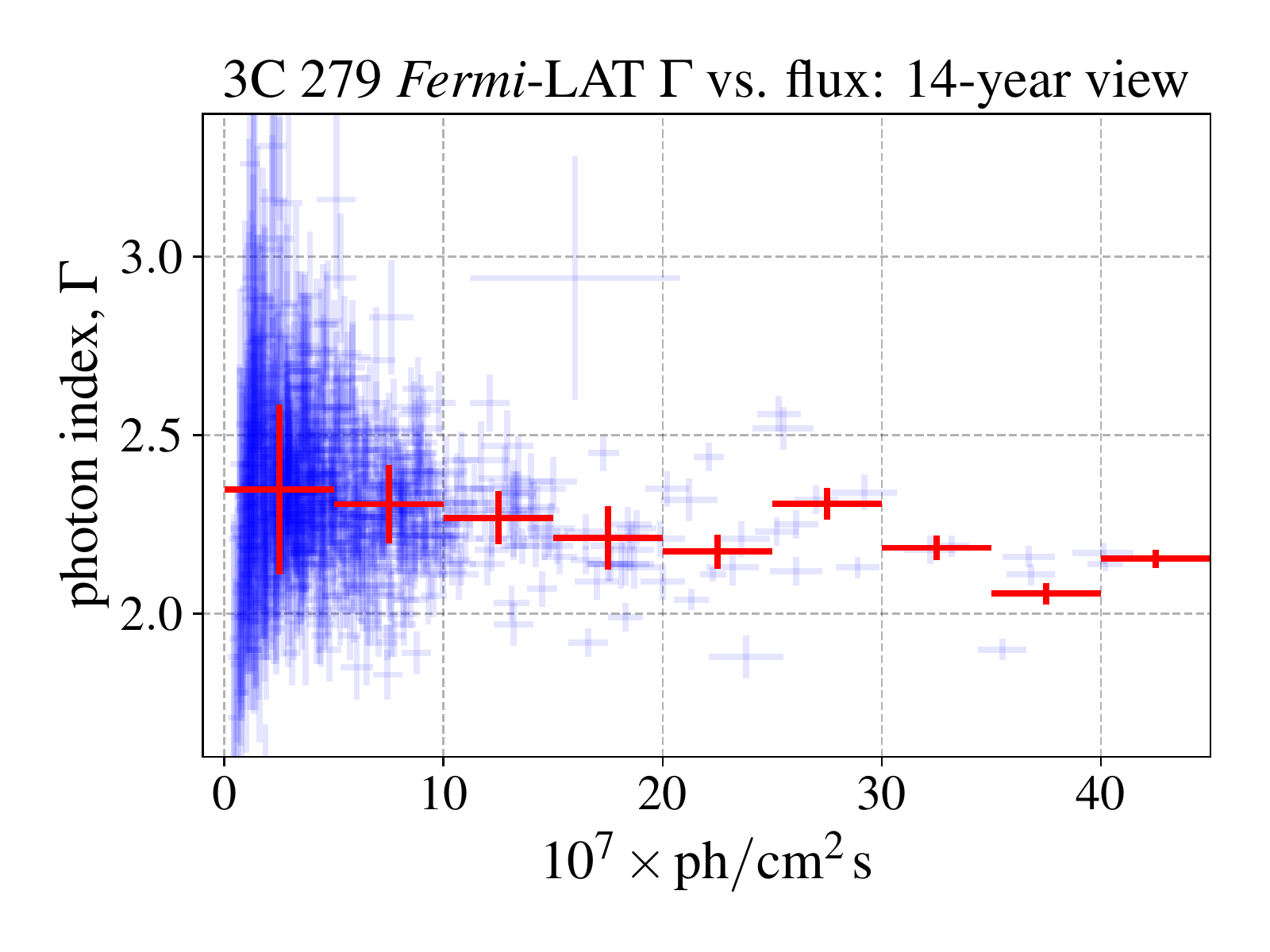}
    \caption{The same as Fig.~\ref{fig:hi3c279} but plotting all (blue) of the one-week-binned LAT flux levels against corresponding spectral indices for over~$14$ years of~3C~279 observations in the Fermi LAT Light Curve Repository \citep{fermi23}. Additionally, we average the spectral indices (red) in flux bins of width~$5 \times 10^7 \, \rm ph / cm^2 \, s$. This bin width is indicated by horizontal red error bars. Vertical red error bars show~$1$-sigma uncertainty levels after bin averaging.}
    \label{fig:hi3c279all}
\end{figure}
To provide a complete view of the correlation between spectral hardness and luminosity of~3C~279 in the GeV band, we also supply Fig.~\ref{fig:hi3c279all}. This is similar to Fig.~\ref{fig:hi3c279} except that the \textit{Fermi} LAT spectral index is plotted against the gamma-ray flux level for all~$14+$ years of archived data. The full set of \textit{Fermi} observations demonstrates broad consistency with, for example, the individual flaring period presented by \citet{hmn12} and displayed in Figs.~\ref{fig:lc3c279} and~\ref{fig:hi3c279}: as the object brightens, the variation in photon index appears to decrease, suggesting a stabler spectrum during flaring periods.

At an even more general level, \citet{msb19} conducted a statistical analysis of the brightest flares from~$6$ of the most luminous FSRQs detected by \textit{Fermi}. They find hints of \quoted{harder-when-brighter}behavi\spellor[]in some flares from some objects, but no statistical significance. At the same time, they report that higher flaring flux tends to coincide with reduced spectral variability. These remarks seem to be representative of other individual \textit{Fermi}-detected FSRQ outbursts that are (quasi-)contemporaneous with TeV flares seen by one or more IACTs, including from PKS~1222+216 \citep{magic11}, PKS~1441+25 \citep{veritas15}, PKS~1510-089 \citep{magic17}, PKS~0736+017 \citep{hess20}, and QSO~B1420+326 \citep{magic21}. In these events, the GeV \mbox{(sub-)flare} often coincides with a relative s\tabiliz[ation]of the GeV spectral index: a seeming decoupling between the flux and the spectral shape -- including on the falling part of the flare (albeit the spectral index error bars are larger there). Nevertheless, one does sees hints of \quoted{harder-when-brighter}trends in some of the GeV \mbox{(sub-)flares.}

Thus, FSRQ flares in the \textit{Fermi} LAT sensitivity band exhibit broad, though perhaps imperfect, consistency with Klein-Nishina reconnection coupled to BLR Compton seed photons. The trends seem to persist even into the flare decay: if radiative losses reverted entirely to the Thomson regime, this would induce a strong (and unobserved) correlation between the gamma-ray spectral index and the flux in the decaying part of the flares. This suggests the role of Klein-Nishina effects, as unveiled in this work, in s\tabiliz[ing]the spectrum even as the lightcurve declines. For the cases where a harder-when-brighter trend appears more evident, this could be induced by weak synchrotron losses perturbing the dominant temporal-spectral behavi\spellor[]induced by Klein-Nishina and pair-production physics.

One caveat to our association of GeV FSRQ flares with IC(BLR) Klein-Nishina reconnection is the occasional \mbox{(quasi-)c}ontemporaneous detection of TeV gamma-rays (a few examples of which are listed two paragraphs prior). For these cases, a simple one-zone emission model for both the GeV and TeV outburst places the emission region outside the BLR, which would otherwise absorb the TeV photons [equation~(\ref{eq:eblrth})]. This caveat is made more severe by population studies of \textit{Fermi}-detected FSRQs purely in the GeV band, which find no evidence for gamma-ray absorption of the BLR seed photons in the vast majority of objects \citep{cct18, msb19}. As noted by \citet{cct18}, however, even within a single zone framework, these constraints become less severe if one attributes the flaring emission to a structure (in our case, a reconnection layer) that comoves with the jet rather than a stationary feature (e.g.\ a standing shock). Then, over an observing period,~$\Delta t_{\rm obs}$, of just~$1$ day (typically comparable to or shorter than variability \ts[s] identified by VHE FSRQ observations, with two notable exceptions: \citealt{magic11} and \citealt{hessmagic21}), Doppler time-compression of the lightcurve allows the emitting zone to travel a distance,~$d \sim \Gamma_{\rm j}^2 c \Delta t_{\rm obs} \sim 10^{17}\, \rm cm$, from the central engine for a fiducial jet Lorentz factor,~$\Gamma_{\rm j} = 10$. This is at the edge of the BLR, which only extends up to roughly~$0.1 \, \rm pc \sim 10^{17} \, \rm cm$ from the nucleus \citep{tg08, ssm09, nbc12, mwu21}, reducing the importance of absorption for the (potentially up to TeV) part of the emission produced at larger distances. In this view, attributing emission from longer flaring periods (e.g.\ as in Fig.~\ref{fig:lc3c279}) to IC(BLR)-coupled reconnection demands positing the ejection of multiple reconnecting structures from the central engine, as, for example, in striped-jet models \citep{gu19}.

We next consider FSRQ flares at TeV energies observed by IACTs. Such events probe the the IC(HDR) scenario wherein reconnection couples to Klein-Nishina and pair-production physics mediated by the HDR seed photons. Suitable observations are much more difficult to obtain here. Not only do FSRQs shine intrinsically weakly in the TeV band (recall that their quiescent IC spectral hump peaks already at MeV or, sometimes, GeV energies), but they suffer both intrinsic and external absorption at these energies. Intrinsically, TeV emission produced inside the BLR will be absorbed, rendering invisible TeV flaring regions too close to the central engine. Furthermore, because FSRQs (unlike BL Lacs) are distributed in the Universe preferentially at higher redshift \citep{fermi15, fermi20b}, their TeV gamma-rays may be absorbed by the extragalactic background light while en route toward Earth. Owing to these combined effects, only a handful of FSRQs have even been detected at TeV energies \citep[9 at the time of writing,][]{tevcat}. Of these, quiescent emission has only been seen from one object, PKS 1510-089 \citep{magic18}. The rest are detected exclusively in high or flaring states (e.g.~3C~279, \citealt{magic08}; PKS 1222+216, \citealt{magic11}; PKS 1441+25, \citealt{veritas15}; QSO B0218+257, \citealt{sbb15, magic16}; PKS 0736+017, \citealt{hess20}; QSO B1420+326, \citealt{magic21}). Even during flares, the gamma-ray flux is rarely sufficient to provide detailed temporal information for the total luminosity itself, much less for the spectrum.

Considering mostly time-integrated spectra, what is generally seen for FSRQs in the TeV band is a much steeper spectrum, even after correcting for absorption by the extragalactic background light, than in the \textit{Fermi} range. TeV spectral indices are almost always greater than~$2.5$ and commonly larger than~$3$: for~3C~279,~$\Gamma\simeq4.2$ \citep{hess19}; for PKS 1510-089,~$\Gamma\simeq2.5$ \citep{magic14a},~$\Gamma\simeq3.2,4.3$ \citep{magic17},~$\Gamma\simeq2.9,3.4$ \citep{zbc17}, and~$\Gamma\simeq3.3$ \citep{magic18}; for PKS 1222+216,~$\Gamma\simeq2.7$ \citep{magic11}; for PKS 1441+25,~$\Gamma\simeq3.4$ \citep{veritas15}; for QSO B0218+257,~$\Gamma\simeq2.4$ \citep{magic16}; and for QSO B1420+326,~$\Gamma\simeq2.9$ \citep{magic21}. In the IC(HDR) scenario, these steep spectral indices suggest that reconnection proceeds in a regime, unlike that probed in detail in this study, where its intrinsic particle acceleration index is steeper \citep[e.g.\ with a strong guide field:][]{wu17}. Even in this case, Klein-Nishina radiative physics should still harden the spectrum in the decaying part of a flare (section~\ref{sec:afterglow}). However, such detailed temporal behavi\spellor[]has not yet been accessible to TeV instruments. From this point of view, the results of our model remain, for the moment, predictions. The coming online of the Cherenkov Telescope Array (CTA) over the next few years will provide increased sensitivity in the TeV band \citep{cta19}, enabling enhanced temporal resolution and, hence, a more thorough probing of the IC(HDR) scenario.

We note that BL Lacs are typically much brighter than FSRQs at TeV energies, and thus many more IACT observations of BL Lacs exist -- and typically with much greater temporal resolution -- than of FSRQs \citep{tevcat}. However, without strong external photon fields supplied by the BLR or the HDR, the source of seed photons for reconnection in BL Lacs is far less constrained. If these are the synchrotron photons produced during reconnection \citep[a single-zone synchrotron self-Compton, or SSC, setup, cf.][]{mgc92, bm96}, a much more detailed seed photon model -- taking into account multichromaticity, time-dependence, anisotropy, and spatial inhomogeneity -- needs to be folded in to the simulations in order to make robust predictions. However, if the seed photons impinge on the putative reconnection region from elsewhere in the jet (a multi-zone SSC paradigm), then our static seed photon model is more appropriate, but it would still potentially need to be g\eneraliz[ed]to the non-monochromatic case. We note that in the famous minute-scale flare of PKS 2155-304 presented by \citet{hess07}, no evidence of spectral variability was found. This is consistent with our reconnection model, however the photon indices, even after correcting for the background absorption \citep[e.g.][]{hess09} are steeper than one would expect on the decaying side of the flare, wherein Klein-Nishina and pair-production physics should harden the spectrum closer to the universal shape of section~\ref{sec:afterglow}. Ultimately, not enough is known about the seed photons to draw firm conclusions, however.

We close this discussion by examining the potential pair yield of Klein-Nishina reconnection in the IC(BLR) and IC(HDR) scenarios. Suppose that the jet is launched with an electron-proton composition (no positrons). Let us also assume that the jet evolves so as to be moderately magnetized in its rest frame,~$B_0^2 / 4 \pi n_0 \massp c^2 \sim 1 - 10$, at the parsec scale \citep[cf.][]{g13, gu19, mwu21}, which is near the transition point where the HDR overtakes the BLR as the dominant seed photon source \citep{nbc12, mwu21}. The cold \textit{electron} magnetization is then~$\sigc \sim (\massp / \me)(B_0^2 / 4 \pi n_0 \massp c^2) \sim 2 \times 10^{3-4}$. In our earlier work \citep{mwu21}, we estimate (in the jet's rest frame)~$\gamma_{\rm KN,BLR} \sim 300$ and~$\gamma_{\rm KN,HDR} \sim 1 \times 10^4$. This means that the pair yield control parameter (section~\ref{sec:pairyield}) is~$\gamppmax / 4 \sigc \simeq 30 \gkn / 4 \sigc \sim 0.1-1$ for reconnection illuminated by the BLR and~$4-40$ for HDR irradiation. This control parameter only becomes small (the regime of high pair yield) in the IC(BLR) scenario, which is, incidentally, also the scenario for which available flaring observations, as discussed in this section, best corroborate a Klein-Nishina reconnection model. If pair production is predominantly decided by leptonic physics as unveiled by our pair-plasma simulations, then the present estimates suggest that reconnection functions as an \textit{in situ} source of antimatter in FSRQ jets strongly illuminated by the BLR, possibly creating more than 1 positron per electron and effectively transforming any initially electron-proton plasma into a plasma with a prominent positron component. This is an important result in blazar studies because the jet composition is notoriously difficult to ascertain observationally \citep{ms16}; if it is true, it could mean that BLR-illuminated FSRQ jets generically carry a strong pair-plasma component downstream of the GeV emission zone.

In summary, the bright Doppler-boosted emission from blazars highlights the physics of particle acceleration in their jets. In particular, for the \mbox{FSRQ} sources, GeV and TeV observations function as respective probes of scenarios where magnetic reconnection is coupled, through Klein-Nishina and pair-production physics, to soft seed photons produced by the BLR and the HDR. Observations of FSRQ flares in the GeV band are in broad agreement with the spectral-temporal signatures of Klein-Nishina reconnection expected from this study (section~\ref{sec:afterglow}), constituting potential evidence that this type of reconnection occurs within the BLR of FSRQs. The expected pair yield (section~\ref{sec:pairyield}) of IC(BLR)-coupled reconnection could also contribute significantly to the antimatter content of FSRQ jets far away from their central engines. The IC(HDR) scenario cannot be probed at the same level of detail as the IC(BLR) case with the current generation of IACTs, a limitation which the CTA will help to overcome. Finally, while Klein-Nishina reconnection could also operate in BL Lacs, there are a lot more uncertainties concerning the seed photons in those sources, pushing detailed observational comparisons into the domain of future work.

\subsection{Black hole accretion d\isk[]coronae}
In our previous work, \citet{mwu21}, we made a case for Klein-Nishina reconnection operating in the coronae of accreting black hole X-ray binaries (BHXRBs) in their high/soft states. Conducting basic estimates, we showed that, for the case where an underlying optically thick, geometrically thin accretion d\isk[]\citep{ss73} supplies
\begin{align}
    \epsilon_{\rm d\isk[]} \sim 1 \, \rm keV   
    \label{eq:edisk}
\end{align}
seed photons to reconnection in a highly magnetized coronal plasma, the radiative scale hierarchy is~$\gcool < \gkn < \gradt < \gmax$ [equation~(\ref{eq:knregime})], pushing reconnection into the Klein-Nishina regime. Due to the intense accretion d\isk[]radiation bath, absorption fiducially kicks in for gamma-rays above the threshold energy,
\begin{align}
    \epsilon_{\rm th,d\isk[]} = \me^2 c^4 / \epsilon_{\rm d\isk[]} = 260 \, \rm MeV \, .
    \label{eq:ediskth}
\end{align}
Meanwhile, the typical photon energy emitted by a~$\gkn$-particle is~$\gamma_{\rm KN,d\isk[]} \me c^2 / 2 = \epsilon_{\rm th,d\isk[]} / 8 = 30 \, \rm MeV$, where~$\gamma_{\rm KN,d\isk[]} \sim 100$.

To our knowledge, such energies have only been detected in the high/soft state of an accreting BHXRB -- for which their origin in an ejected jet is not expected -- in one object: Cyg~X-1. This was during an approximately~$100 \, \rm Ms$-exposure by the \textit{Fermi} LAT, presented by \citet{zmc17}, who report the detection of gamma-rays up to a \cutoff[]energy of about~$20-40 \, \rm MeV$, somewhat below~(\ref{eq:ediskth}). As pointed out in \citet{mwu21}, this \cutoff[]may be consistent with gamma-ray absorption because the coronal region is likely highly radiatively compact (which translates into a high fiducial pair-production optical depth,~$\tau_{\gamma\gamma} \gg 1$), meaning that absorption is still prominent at energies below~$\epsilon_{\rm th,d\isk[]} \sim 300 \, \rm MeV$ by seed photons in the exponential tail of the d\isk[]spectrum. 

Given the long exposure time necessary for the gamma-ray detection \citep{zmc17}, the prospects for examining temporal signatures of Klein-Nishina reconnection in high/soft BHXRBs at the energy scales near~$\gamma_{\rm KN,d\isk[]} \me c^2$, where Klein-Nishina effects are likely most prominent, are not promising. At lower, X-ray energies, millisecond variability was detected from the high/soft state of Cyg X-1 by \citet{gz03}. Unlike the case of the IC(BLR) and IC(HDR) scenarios discussed in the preceding section, these X-ray observations exhibit a strong harder-when-brighter correlation between the flaring flux and the spectral shape. This could hint at the importance of synchrotron cooling in this system. Alternatively, due again to the high radiative compactness, the radiative cooling \ts[]may be so short that, even in the presence of Klein-Nishina and pair-production effects, the reconnection-energized particles cool down instantaneously on the observationally resolved \ts[s,]masking the radiatively s\tabiliz[ed](on presumably faster \ts[s;]section~\ref{sec:afterglow}) spectrum.

Using our previous estimate of the coronal magnetization \citep{mwu21},~$\sigc \sim 10^{2-4}$, we can infer a pair yield control parameter of~$\gamppmax / 4 \sigc \sim 30 \gamma_{\rm KN,d\isk[]} / 4 \sigc \sim 0.08-8$, indicating Klein-Nishina reconnection as a potentially important source of electron-positron pairs in the coronae of BHXRBs in their high/soft states. However, due to the high radiative compactness, another likely source of pair production is the collision of d\isk[-Comptonized] photons with each other \citep{b17}. This interaction occurs predominantly among photons both with energies~$< \epsilon_{\rm th,d\isk[]} = \me^2 c^4 / \epsilon_{\rm d\isk[]}$, since those with energies beyond~$\epsilon_{\rm th,d\isk[]}$ are most likely to be absorbed by the much denser d\isk[-supplied]radiation field. The total pair yield is then the sum of the contributions from both channels.

To s\ummariz[e,] the case for coronal Klein-Nishina reconnection in the high/soft states of BHXRBs is excellent on theoretical grounds \citep{mwu21}. It is likely, for example, that reconnection in this regime contributes to the pair content in these objects (section~\ref{sec:pairyield}). However, given that even in an exceptionally bright source, Cyg~X-1, the gamma-ray flux is far too low to probe corona-scale reconnection-powered flaring variability, firm connections with the characteristic temporal signatures of Klein-Nishina reconnection (section~\ref{sec:afterglow}) will likely remain out of reach for the foreseeable future.

\subsection{The M87$^*$ magnetosphere}
The M87 galaxy has been monitored in TeV gamma-rays for nearly 20 years \citep[e.g.][]{hess06, magic20, ehtmwl21}.
This includes 3 major flaring periods, one in 2006 \citep{hess06}, one in 2008 \citep{magic08, veritas09}, and one in 2010 \citep{veritas12a, hess12}, that exhibited rapid, one-day variability \ts[s]-- of order the \lc[]time of the galaxy's central supermassive black hole,~M87$^*$.
For two of these TeV-loud periods, observations at longer, more spatially resolved wavelengths revealed (nearly) contemporaneous flux enhancements from the galactic core \citep{veritas09, hess12}. Thus, variability considerations combined with the multi-wavelength context suggest~M87$^*$ itself (more precisely, its immediate plasma environment) as a viable site of TeV emission.

Direct polarized imaging of~M87$^*$ suggests that accretion proceeds in a so-called magnetically arrested (MAD) state \citep{eht21_viii}, c\haracteriz[ed,]as revealed in large part by magnetohydrodynamics (MHD) simulations \citep[e.g.][]{i08, tnm11, rbp20, rlc22, pmy21, cbl21, sdb22}, by quasi-periodic cycles of gradual accumulation of magnetic flux onto the black hole, eventual flux saturation, and finally abrupt and violent flux expulsion. These expulsion events are mediated in the black-hole magnetosphere by large-scale reconnecting current sheets \citep{rlc22}. The current sheets are irradiated by the larger-scale accretion flow, which provides a target photon bath for reconnection-accelerated particles to Comptonize up to the observed TeV energies \citep{hrp23}. If particle acceleration is efficient enough, the tail of the Comptonized radiation spectrum falls above pair-production threshold with the accretion flow seed photons, triggering potentially copious pair creation \citep{ccd21, ccd22}. Thus, reconnection in the M87$^*$ magnetosphere is: (1) a potential source of the observed TeV emission, and (2) likely coupled to the same radiative physics as treated in this study. 

However, in addition to IC radiation, high-energy particles in this context also suffer strong synchrotron losses. In fact, synchrotron cooling is expected to be much more efficient than IC cooling \citep{rlc22, hrp23}.\footnote{We refer only to IC cooling resulting from particles scattering photons impinging from the accretion d\isk[.]We do not consider Comptonization of the layer-emitted synchrotron photons (synchrotron self-Compton) or of synchrotron emission from pairs born upstream of the reconnection layer, both of which are expected to be subdominant IC channels \citep{hrp23}.} Because we ignore synchrotron losses in this study, we cannot directly apply our findings to reconnection in the M87$^*$ magnetosphere. In what follows, we instead present arguments to sketch how our results might g\eneraliz[e]in the presence of strong synchrotron cooling.

Let us first anticipate what may change in the observable signatures of reconnection uncovered in this work. We note that synchrotron cooling is quantitatively similar to Thomson IC cooling: the power radiated per particle,
\begin{align}
    P_{\rm syn}(\gamma) = 2 \sigma_{\rm T} c \beta^2 \gamma^2 U_{\rm B} \sin^2 \alpha \, ,
    \label{eq:psyn}
\end{align}
is identical to the Thomson IC power formula~(\ref{eq:pthom}) but with~$\uph$ replaced with~$3 U_{\rm B} \sin^2\alpha/2$, where~$U_{\rm B} \equiv \myvec{B}^2/8\pi$ is the local magnetic field energy density and~$\alpha$ is the pitch angle between the radiating particle's velocity and the local magnetic field. Thus, modulo special regions such as reconnection X-points where either the magnetic field or the particle pitch angle become small, strong synchrotron radiation is expected to play a dynamically similar role to strong Thomson IC cooling. Hence, similar to our findings for Thomson-cooled reconnection, the radiative signatures of the M87$^*$ magnetosphere are likely to exhibit a much tighter coupling between spectral shape and total luminosity -- i.e. \quoted{harder-when-brighter}-- than when Klein-Nishina IC cooling and pair production dominate the radiative physics (cf.\ section~\ref{sec:afterglow}). We note that this holds whether one observes the synchrotron or the IC radiation, as both probe the same underlying distribution of reconnection-energized particles. The argument for a prominent harder-when-brighter trend is, however, in mild tension with the general picture of the few M87 TeV flares, which indicate this trend only mildly \citep{veritas12a} and not in every event \citep{veritas10}.

Let us now consider the potential ramifications of pair production between IC photons and the seed radiation field from the accretion flow. We set this discussion in the context of recent semianalytical works, \citet{hrp23} and \citet{cud23}, on the matter content of the putative magnetospheric~M87$^*$ reconnection layer. \citet{hrp23} conduct a detailed analysis of the various radiation and pair-production channels that may operate in this context. They predict that the brightest radiation emerges through the synchrotron band and peaks between roughly~$1$ and~$20$ MeV. Though the synchrotron photons are not above pair-production threshold with the ambient accretion flow radiation, they are above threshold with themselves. There is, however, little chance for an individual synchrotron photon to be absorbed by another since the optical depth,~$\taus \sim 10^{-4}$, presented by the synchrotron radiation field is small. Thus, whereas the high optical thickness furnished by the ambient radiation causes nearly \textit{all} of the above-threshold IC photons to be absorbed close to the reconnection current sheet, covering it with a thin pair coat, only a small fraction of the synchrotron radiation is absorbed, leading to diffuse pair production throughout the magnetosphere. The resulting pairs feed the reconnection layer with a highly magnetized,~$\sigc \sim 5 \times 10^7$, plasma. The picture presented by \citet{cud23} is similar, but those authors predict more copious synchrotron-synchrotron pair production, yielding a reduced magnetization,~$\sigc \sim 6 \times 10^4$.

In the context of our reconnection model, the diffuse synchrotron-synchrotron pair production predicted by \citet{hrp23} and \citet{cud23} supplies the background plasma on top of which pair production between IC photons and the radiation impinging from the accretion flow (hereafter, \textit{IC pair production}) may add supplementary pairs localized near the current sheet. If we naively apply our findings for the reconnection-powered IC pair yield (temporarily ignoring potential changes due to synchrotron cooling), we infer that a substantial amplification of the background (synchrotron-synchrotron) pair density requires~$4 \sigc \geq \gamppmax \simeq 30 \gkn$. To estimate~$\gkn$, we adopt a seed photon energy of~$\eph \sim 10^{-2} \, \rm eV$, corresponding to the radiation field calculated by \citet{ydc21} in the inner magnetosphere for MAD accretion (cf. their fig.~6), which yields~$\gkn \equiv \me c^2 / 4 \eph \simeq 1 \times 10^7$. Comparing to~$\sigc = 5 \times 10^7$ from \citet{hrp23}, we estimate~$\gamppmax / 4 \sigc \simeq 2$, which is slightly too high to attain an order-unity IC pair yield according to this study (section~\ref{sec:pairyield}). One should keep in mind, however, that the uncertainties on these order-of-magnitude estimates are high, while even a reduction in the ratio~$\gamppmax / 4 \sigc$ by a factor of~$4$ is enough to bring the IC pair yield up to unity. However, if synchrotron-synchrotron pair production is more efficient, as predicted by \citet{cud23}, then the inflowing plasma is probably not sufficiently magnetized to lead to appreciable IC pair production. In the event that a substantial IC pair yield is r\ealiz[ed,]it induces a transverse density gradient on the magnetic field lines forming the jet funnel, with a higher plasma load on the field lines that participate in reconnection near the jet walls. This would then be potentially important for jet-boundary interactions \citep[e.g.][]{rbp20, srn21, cdr22, crs23, ecc22, ecc23}, such as may power observed limb-brightening \citep{lwj07, whd18, kkl18}.

Finally, whereas in the preceding discussion we simply applied the forecasted pair yield obtained from the present study to the M87$^*$ case, we now discuss how strong synchrotron losses in this context may alter the picture of pair production as revealed in this work. First, the IC pairs born into the upstream region are likely to cool down much more before entering the reconnection layer, changing them from a hot population to a merely warm or even cold one (and thus reducing their contribution to the upstream plasma energy density; cf.\ section~\ref{sec:origprodcompare}). Second, synchrotron losses could restrict the IC pair yield since high-energy particles are likely to emit fewer pair-producing IC photons before radiating away their energy as synchrotron light. Interestingly, however, the global magnetospheric simulations of \citet{ccd22}, which include the same processes of IC emission and pair creation studied here, find that IC pair production is efficient enough to fuel the reconnection layer with plasma even when synchrotron losses are made as strong as numerically possible. This is likely connected to the fact that TeV emitting particles are accelerated near reconnection X-points \citep{hrp23} where their synchrotron losses are suppressed. Previous numerical work thus hints that synchrotron radiation does not quench IC pair production.

To s\ummariz[e,]because we neglect synchrotron cooling in this work, we cannot directly apply our results to reconnection in the M87$^*$ magnetosphere. The discussion in this section is therefore mostly speculative. When formulating expectations for observable signatures and the IC pair yield, we find that, on both counts, the simplest arguments are not convincingly supported by recent studies. For example, one expects synchrotron losses to induce a tighter correlation between spectral shape and total luminosity, making the radiative signatures of reconnection more Thomson-like (with a more prominent harder-when-brighter trend). However, observational evidence for this argument is somewhat ambiguous, providing, at best, limited support \citep{veritas12a} and, at worst, mild tension \citep{veritas10, veritas12b}. Furthermore, the naive expectation that synchrotron cooling shuts down the IC pair yield seems to be in conflict with first-principles global simulations \citep{ccd22}. The fact that straightforward physical arguments do not satisfactorily fill the gap between this study and the~M87$^*$ case creates fertile ground for future work to self-consistently incorporate synchrotron losses and, thereby, to shed light on the perplexing issues raised here.

\subsection{Gamma-ray binaries}
\label{sec:gammaraybins}
Gamma-ray binaries consist of a relativistic compact object (neutron star or black hole) and a massive stellar companion (generally of type O or Be) and are defined by a spectral energy density peaking, in the~$\nu F(\nu)$ representation, above~$1 \, \rm MeV$ \citep[e.g.][]{d13, dgp17}. Of the handful (less than a dozen) of known gamma-ray binaries, only two are directly observed to host pulsars \citep{d13, thp18, cmp19, cm20}. Nevertheless, several general observed features suggest that the compact object in these systems is generically a rotation-powered pulsar \citep{d06, d13}.

Adopting this view, two often-invoked gamma-ray emission sites are the pre- and post-shocked pulsar wind, where the shock in question interfaces between the winds of the pulsar and the massive companion \citep[e.g.][]{kbs99, bk00, bd01, sb05, sb08, cdh08, kab12} and not, as would be the case in isolated pulsars, between the pulsar wind and the interstellar medium. As argued in the review by \citet{d13}, the fact that the high-energy~($0.1-10 \, \rm GeV$) spectra of gamma-ray binaries are often similar in terms of slope and \cutoff[]to isolated pulsars could hint at a similar emission mechanism between the two object classes, motivating an investigation of the unshocked pulsar wind, and perhaps even of the pulsar magnetosphere, as gamma-ray production sites in gamma-ray binaries. However, the GeV spectra also exhibit modulations on the binary orbital period, which would seem to disfav\spellor[]the magnetosphere as the dominant emission zone, since it is insensitive to the orbit of the binary \citep{d13}. Therefore, in this section, we consider the possibility that the unshocked pulsar wind significantly contributes to the observed high-energy gamma-rays in gamma-ray binaries \citep[cf.][]{bk00, bd01, cdh08, kab12}.

The unshocked pulsar wind behaves exactly as that of an isolated pulsar except for one key difference: it is illuminated from beyond by the hot massive companion star. The temperature,~$T_\star \sim 40000 \, \rm K$, of the companion's surface produces a characteristic blackbody photon energy,
\begin{align}
    \estar \sim 3 k_{\rm B} T_\star \sim 10 \, \rm eV \, ,
    \label{eq:estar}
\end{align}
and radiation energy density,
\begin{align}
    \ustar = \frac{\sigma_{\rm SB} T_\star^4}{c} \left( \frac{R_\star}{d} \right)^2 \sim 1 \times 10^3 \, \rm erg \, cm^{-3} \, ,
    \label{eq:ustar}
\end{align}
where~$d=0.1 \, \rm AU$ is the typical separation at periastron and~$R_\star = 10 R_\odot$ the stellar radius \citep{d13}. Equation~(\ref{eq:estar}) implies a critical Klein-Nishina Lorentz factor in the pulsar wind of
\begin{align}
    \gknstar \equiv \frac{\me c^2}{4 \estar} \sim 1 \times 10^4
    \label{eq:gknstar}
\end{align}
and that pair production becomes possible above the threshold
\begin{align}
    \epsilon_{\rm th,w} = \me^2 c^4 / \estar = 30 \, \rm GeV \, \rm .
    \label{eq:ethstar}
\end{align}
The typical photon energy emitted by particles with~$\gamma = \gknstar$ is then~$\gknstar \me c^2 / 2 = \epsilon_{\rm th,w} / 8 \sim 3 \, \rm GeV$.

Let us consider the effect that illumination by the companion may have on the pulsar wind. Here, we adopt the theoretical picture \citep{c90, m94, b99, lk01, l03a, ks03, klp09}, brought into sharper focus by recent first-principles kinetic simulations \citep{cp17, ps18_philippov, cpd20}, that this wind is not purely cold, but is instead \textit{striped} -- laced with a large-scale reconnecting current sheet that expands radially outward while undulating about the pulsar's rotational equator. The angular excursions about the equator approximately equal the obliquity angle between the magnetic and spin pulsar axes. Reconnection converts the outgoing wind Poynting flux into a combination of bulk acceleration and n\onthermal[]particle acceleration.

An isolated pulsar is not bathed in the intense light of a companion star, and so cooling of accelerated wind particles remains dominated by synchrotron losses. In the present case, however, cooling via IC scattering of the intense radiation bath~(\ref{eq:ustar}) likely outpaces synchrotron cooling once the local magnetic field energy density falls below~$\ustar$. This occurs at a critical magnetic field strength,
\begin{align}
    B_{\rm IC} \equiv \sqrt{8 \pi \ustar} \sim 200 \, \rm G \, .
    \label{eq:bic}
\end{align}
For a pulsar rotational period~$P \sim 100 \, \rm ms$ (characteristic of the two confirmed pulsars in gamma-ray binaries: PSR J2032+4127, \citealt{fermi09_psrdetections}; and PSR B1259-63, \citealt{jml92}) and a surface magnetic field~$B_{\rm surf}=10^{12} \, \rm G$, the magnetic field is diluted to~$B_{\rm LC} \simeq B_{\rm surf} (R_{\rm psr} / R_{\rm LC})^3 \sim 9 \times 10^3 \, \rm G$, at the light cylinder,~$R_{\rm LC} = c P / 2 \pi$, where~$R_{\rm psr} = 10 \, \rm km$ is the assumed pulsar radius. Beyond the light cylinder, the magnetic field falls off slower, as~$R_{\rm LC}/R$ where~$R$ is the cylindrical radius measured from the pulsar's spin axis, and, hence, even for the strong surface field~$B_{\rm psr} = 10^{12} \, \rm G$, the striped wind's radiative losses become IC dominated at~$R/R_{\rm LC}\simeq (B_{\rm surf}/B_{\rm IC}) (R_{\rm psr} / R_{\rm LC})^3 \sim \ricval$. This is far before the shock with the companion's wind, which is expected to occur on scales~$R \sim d \sim 10^3 R_{\rm LC}$ \citep{d13}. Moreover,~$R\sim \ricval R_{\rm LC}$ is also before the point where the pulsar wind's electromagnetic flux is expected to be fully dissipated, which kinetic simulations anticipate at~$R/R_{\rm LC}$ roughly between~$10^2$ and~$10^4$ \citep{cpd20}. Taken together, these estimates suggest that \textit{most of the pulsar wind's dissipation takes place before the shock with the companion's wind, but after the critical radius where IC losses surpass synchrotron losses}.

Not only does most of the pulsar wind dissipate through reconnection subject to strong IC cooling, but, as we show now, reconnection may occur in the Klein-Nishina regime of this study, attaining the critical scale hierarchy~(\ref{eq:knregime}),~$\gcool < \gkn < \gradt < \gmax$. Assuming~$B_{\rm LC} = 9 \times 10^3 \, \rm G$,~$P = 100 \, \rm ms$, and that the reconnecting magnetic field strength,~$B_0$, is~$B_0 = B_{\rm LC} R_{\rm LC} / R$, as well as identifying the length,~$L$, of the reconnection layer with the local radius,~$R$, in the pulsar wind, we have, by equation~(\ref{eq:gmax}),
\begin{align}
    \gmaxstar \equiv \frac{0.1 e B_0 R}{\me c^2} \sim 3 \times 10^8 \, ,
    \label{eq:gmaxstar}
\end{align}
which is independent of~$R$. Similarly, we have, by equation~(\ref{eq:gradt}),
\begin{align}
    \gradtstar \equiv \sqrt{\frac{0.3 e B_0}{4 \sigma_{\rm T} \ustar}} \sim 3 \times 10^6 \left( \frac{R}{\ricval R_{\rm LC}} \right)^{-1/2} \, .
    \label{eq:gradtstar}
\end{align}
We then can use~$\gradt^2 = \gcool \gmax$ [equation~(\ref{eq:gradgeomean})], to estimate
\begin{align}
    \gcoolstar \sim 3 \times 10^4 \left( \frac{R}{\ricval R_{\rm LC}} \right)^{-1} \, .
    \label{eq:gcoolstar}
\end{align}
Finally, using equation~(\ref{eq:taugg}), we estimate the pair-production optical depth as
\begin{align}
    \tauggstar \equiv \frac{3 \gknstar}{5 \gcoolstar} \sim 0.3 \left( \frac{R}{\ricval R_{\rm LC}} \right) \, .
    \label{eq:tauggstar}
\end{align}
Note that we have normalized~$R$ to the critical radius where IC losses overtake synchrotron losses. The pulsar wind extends much farther than this, which pushes~$\tauggstar$ above unity at the largest radii. The geometry of the striped wind is somewhat peculiar, however, in that the spacing between stripes is~$\sim R_{\rm LC} \ll R$ and, hence, even if the overall current sheet length is optically thick to pair production, the spacing between current sheets is thin, such that photons emitted in one stripe may be absorbed inside another.

Let us now examine what the implications of Klein-Nishina reconnection are on the pre-shocked pulsar wind of gamma-ray binaries. We discuss first the potential effect of pair production on the wind. Following \citet{cpd20}, the cold magnetization at the light cylinder is
\begin{align}
    \sigcstar = \frac{B_{\rm LC}^2}{4 \pi \kappa n_{\rm GJ} \me c^2} = \frac{e P B_{\rm LC}}{4 \pi \kappa \me c} = 1 \times 10^5 \left( \frac{B_{\rm LC}}{9 \times 10^3 \, \rm G} \right) \left( \frac{\kappa}{10^4} \right)^{-1} \, ,
    \label{eq:sigcstar}
\end{align}
where~$n_{\rm GJ} \equiv B_{\rm LC} / e c P$ is the Goldreich-Julian number density and~$\kappa$ is the multiplicity. This magnetization is \quoted{frozen-in}at the light cylinder in the sense that, because~$B_0 \propto 1/R$ and~$n_0 \propto 1 / R^2$ beyond the light cylinder,~$\sigcstar$ remains constant (modulo pair production) in the unreconnected plasma of the wind throughout its expansion. The pair yield control parameter in the pre-shocked wind is then~$\gamppmax / 4 \sigcstar \simeq 30 \gknstar / 4 \sigcstar\sim 0.7$, which is in the regime of order-unity pair yield. Furthermore, if the pulsar magnetosphere fails to launch the pulsar wind with~$\kappa \sim 10^4$, but instead with a lower multiplicity, the magnetization~$\sigcstar$ increases, leading to potentially copious \textit{in situ} pair production in the expanding wind. Using our results from section~\ref{sec:pairyield}, we infer a critical self-regulated magnetization of~$\sigcgg \simeq \gamppmax \simeq 30 \gknstar \sim 4 \times 10^5$. If the plasma is injected with a higher magnetization than this (e.g.\ by virtue of an underdense multiplicity), Klein-Nishina pair production fills in the plasma deficit, pulling up the multiplicity toward~$\kappa = e P B_{\rm LC} / 4 \pi \me c \sigcgg \sim 3 \times 10^3$. This critical multiplicity depends solely on the pulsar properties and those of the ambient radiation field.

Let us close by examining prospects for observing signatures of Klein-Nishina reconnection in gamma-ray binaries. The temporal observables uncovered in section~\ref{sec:afterglow} are most likely to manifest themselves during transient flares. Of the known gamma-ray binaries, PSR~B1259-63 is known to flare once per~$3.4$-year orbit \citep[e.g.][]{tht11, fermi11b, ccl15, thp18, czc21}. However, this binary is far less compact than others, with~$d \sim 0.9 \, \rm AU \gg 0.1 \, \rm AU$, even at periastron. Moreover, the flares occur significantly later in the orbit than the time of periastron. Such wide separations dilute the radiation field from the stellar companion, tending to move reconnection out of the Klein-Nishina regime. The binaries LS~5039, LS~I~+61$^\circ$303, and 1FGL~J1018.6–5856 are all much more compact~($d \sim 0.1 \, \rm AU$ at periastron; \citealt{d13}), but, instead of sudden flares, GeV gamma-ray observations reveal smooth modulations all throughout each binary's orbit (for LS~5039, \citealt{fermi09_ls5039}; for LS~I~+61$^\circ$303, \citealt{fermi09_lsi61303}; for 1FGL~J1018.6–5856, \citealt{fermi12}). Such modulations likely probe quasistatic changes to the unshocked pulsar wind, including its orbit-dependent illumination by the companion. However, in order to interpret them, one needs to go beyond the flaring-centric treatment of observable signatures adopted in this work and conduct explicit global modeling of the steady-state wind.

To summarize, we have shown that Klein-Nishina reconnection likely takes place in short-period gamma-ray binaries if the compact object in the binary is a pulsar. In that case, reconnection occurs in the pre-shocked striped pulsar wind and is immersed in a bright seed photon bath supplied by the companion star. As a result, the reconnecting stripes produce a minimum pair-plasma density, setting a multiplicity/density floor in the wind (even if it is launched underdense from the pulsar). Observational support for this scenario is somewhat limited, as the few observed bright flares occur in configurations where the stellar companion is too far removed to supply a sufficiently dense radiation bath for Klein-Nishina reconnection. More compact systems, on the other hand, do not exhibit rapid flares, but rather smooth modulations to their GeV signal over the entire orbit. Detailed predictions of this signal require global modeling, which we leave to a future study. 

\subsection{Summary}
\label{sec:discsummary}
\begin{table*}
\centering
\begin{threeparttable}
    \caption{A graphic summary of section~\ref{sec:discussion}. Column~(1) indicates the object class (roughly one per subsection~\ref{sec:fsrqs} through~\ref{sec:gammaraybins}). Column~(2) indicates whether basic theoretical estimates suggest that the Klein-Nishina reconnection scale hierarchy~(\ref{eq:knregime}),~$\gcool < \gkn < \gradt < \gmax$, is achieved in the given system. Column~(3) indicates our judgment of how firmly the results of this study -- in particular the observable signatures of Klein-Nishina reconnection discussed in section~\ref{sec:afterglow} -- can be connected to presently available observational data. If strong observational connections cannot be made, column~(4) indicates what, in our view, is the primary reason for this. Finally, column~(5) gives a short explanation of the judgments in columns~(3) and~(4).}
\label{table:discussion}
\begin{tabulary}{\textwidth}{lcccL}
    \toprule
    1) Object class & 2) KN hierarchy & 3) Observational & 4) Limiting & 5) Explanation \\
    & r\ealiz[ed?] & connection & factor & \\
    \midrule
    Flat-spectrum radio quasars & & & & \\
    \quad IC(BLR) scenario & Yes & Strong & -- & \textit{Fermi}-LAT observations are broadly consistent with anticipated temporal-spectral signatures of KN reconnection (section~\ref{sec:afterglow}; Figs.~\ref{fig:lc3c279}-\ref{fig:hi3c279all}). \\
    \quad IC(HDR) scenario & Yes & Limited & Instrumental & The upcoming CTA will provide enhanced temporal resolution at the relevant TeV energies, making possible more explicit comparisons with this work. \\
    Black hole accretion d\isk[]coronae & Yes & Limited & Sources & Sources are not bright enough in the relevant $10$+ MeV range to probe temporal variability. However, the spectral \cutoff[]in Cyg~X-1 is potentially consistent with a KN reconnection model. \\
    M87$^*$ magnetosphere & Yes\tnote{*} & -- & Mode\ling & Need to account for synchrotron cooling in order to make relevant predictions. \\
    Gamma-ray binaries & Yes & Limited & Mode\ling & Orbital modulations of lightcurves at the relevant GeV energies necessitate global mode\ling[.]\\
    \bottomrule
\end{tabulary}
\begin{tablenotes}
\item[*]We have checked that the KN hierarchy is likely r\ealiz[ed]in the M87$^*$ magnetosphere, but do not detail those estimates in this manuscript.
\end{tablenotes}
\end{threeparttable}
\end{table*}
We supply a graphic recapitulation of the discussion in this section in Table~\ref{table:discussion}. In all four types of systems -- FSRQs, (high/soft-states of) black hole accretion d\isk[]coronae, the M87$^*$ magnetosphere, and gamma-ray binaries -- a strong case can be made that the basic Klein-Nishina reconnection scale hierarchy~(\ref{eq:knregime}),~$\gcool < \gkn < \gradt < \gmax$, is r\ealiz[ed.]However, only for the FSRQs [particularly the IC(BLR) scenario] do suitable observations exist for comparing with the expected temporal signatures of Klein-Nishina reconnection. For these objects, we find that \textit{Fermi}-LAT data are, on the whole, consistent with the picture of a reconnection-powered flaring spectral energy density s\tabiliz[ed]by pair-production (section~\ref{sec:afterglow}). For the other object classes, more detailed observational connections are inhibited either by current instrumental sensitivity (that will soon be alleviated), intrinsically dim sources (that probably cannot be overcome), or mode\ling details (upon which future work will be able to improve). 

\section{Conclusions}
\label{sec:conclusions}

In this work, we construct (section~\ref{sec:setup}) a numerical model of the \textit{Klein-Nishina reconnection} regime. We consider a reconnection system immersed in a background radiation bath that is static, homogeneous, isotropic, and monochromatic, c\haracteriz[ed]solely by its total energy density~$\uph$ and individual photon energy~$\eph$ [equation~(\ref{eq:umono})]. We also choose a two-dimensional reconnection setup, a consequence of the large box sizes needed to maintain an adequate separation between the many scales in the problem, several of which are introduced by the QED physics. In particular, we need to satisfy the critical hierarchy of Lorentz factor energy scales~(\ref{eq:knregime}),~$\gcool \ll \gkn \ll \gradt \ll \gmax$. This hierarchy, which is apparently satisfied in a variety of astrophysical systems (section~\ref{sec:discussion}; \citealt{mwu21}), permits: (1) efficient IC radiative losses on \ts[s]much shorter than the \lc[]time,~$L/c$; (2) copious particle acceleration above the energy,~$\sim \gkn$, where particles emit photons above pair-production threshold; and (3) a high optical depth,~$\tau_{\gamma\gamma} \sim \gkn / \gcool \gg 1$, such that nearly all of the above-threshold radiation is absorbed inside the system.

We present simulations (sections~\ref{sec:results}-\ref{sec:pairyield}) of Klein-Nishina reconnection performed with a QED-enabled version (section~\ref{sec:arch}) of the \zeltron[]PIC code. These include various control runs -- one non-radiative, one Thomson-radiative, and one Klein-Nishina radiative with pair-production artificially suppressed -- to elicit the unique properties induced by the novel QED physics. We draw the following main conclusions for Klein-Nishina reconnection:

\begin{itemize}
    \item From sections~\ref{sec:globaldynamics}-\ref{sec:recrate}:

        The added Klein-Nishina and pair-production physics does not substantially change the large-scale qualitative aspects of reconnection: the reconnection rate and hierarchical plasmoid chain remain essentially unchanged.
    \item From section~\ref{sec:ptcldists}:

        Reconnection-powered NTPA remains efficient but is somewhat inhibited by radiative cooling. On time-average, the reconnection-energized plasma is colder, and the particle energy distribution's extended n\onthermal[]tail steeper, compared to the n\onradiative[]regime, but the plasma is not as cold, nor the distribution as steep, as in the Thomson radiative regime.
    \item From section~\ref{sec:icemitdists}:

        The time-averaged IC emission spectrum is steeper than in the n\onradiative[]regime but exhibits little change in slope from the Thomson-cooled case. This owes to Klein-Nishina effects suppressing the IC \crosssection[,]and, hence, largely cance\ling[]the hardening of the underlying particle energy distribution with respect to the Thomson case \citep[cf.][]{msc05}.
    \item From section~\ref{sec:afterglow}:

        The intrinsic IC emission peaks above pair-production threshold, meaning that most of the initially radiated energy is reprocessed (by radiative cooling and pair production) to longer wavelengths before escaping the system. This leads, despite similarities in time-\textit{averaged} spectra, to profound differences in the time-\textit{dependent} signatures of reconnection between the Klein-Nishina and Thomson radiative regimes. In the latter case, the observed spectrum exhibits a tight correlation between total luminosity and prominence of the high-energy n\onthermal[]tail (i.e.\ harder-when-brighter). In the Klein-Nishina case, however, the spectral shape becomes virtually independent of total luminosity, including in the decaying phase after reconnection-powered particle energization has ceased. 
    \item From section~\ref{sec:afterglow}:

        When particle energization is shut off, Klein-Nishina IC cooling and pair production conspire to produce a universal spectral shape, with the particle energy distribution scaling as~$\dif N / \dif \gamma \propto \gamma^{-2}$ for~$\gamma < \gkn$ and IC emission spectrum as~$\epsilon F_{\rm IC}(\epsilon) \propto \epsilon^{1/2}$ for~$\epsilon < \gkn \me c^2$.
    \item From section~\ref{sec:origprodcompare}:

        The upstream plasma enthalpy -- and, hence, the upstream hot magnetization -- is not sufficiently loaded by newborn pairs to trigger the limit cycles sketched by \citet{mwu21}, according to the pair-loading efficiency requirement derived by those authors. This conclusion, however, applies only when the newborn upstream pairs are hot and tenuous, contributing significantly to the pressure of the reconnection inflow plasma but not to its number density. In regimes (suggested to exist by our findings concerning the pair yield below) where pair production loads the upstream plasma number density -- and, thus, the \textit{cold} magnetization in addition to the hot magnetization -- the possibility of limit cycles remains an open issue.
    \item From section~\ref{sec:pairyield}:

        The pair yield (per reconnection-processed lepton) follows an empirical exponential decay law, equation~(\ref{eq:pairyield}), in the parameter,~$\gamppmax / 4 \sigc \simeq 30 \gkn / 4 \sigc$. When~$\gamppmax \lesssim 4 \sigc$, reconnection-powered NTPA cuts off at~$\gamma \sim \gx \equiv 4 \sigc$, far beyond the energy,~$\gamppmax$, where typical Comptonized photons lie at peak pair-production \crosssection[]with the seed photons, enabling an order-unity pair yield. We speculate that the exponential formula breaks down in the limit,~$4 \sigc \gg \gamppmax$ (not probed by our simulations), where reconnection would instead produce copious pairs. In such a case, the population of newborn pairs would no longer be hot and tenuous as in section~\ref{sec:origprodcompare}, but rather hot and \textit{abundant}.
    \item From section~\ref{sec:discussion}:

        The Klein-Nishina reconnection scale hierarchy (\ref{eq:knregime}) is likely satisfied in at least four classes of astrophysical systems: \mbox{FSRQs}, high/soft states of \mbox{BHXRBs}, the magnetosphere of~M87$^*$, and gamma-ray binaries. Of note are \mbox{FSRQs}, where the often-observed spectral stability during GeV flaring states could be due to Klein-Nishina effects as discussed (and s\ummariz[ed]above) in section~\ref{sec:afterglow}. For the other object classes, further mode\ling[]or instrumental development will enable firmer observational connections. A more detailed summary of section~\ref{sec:discussion} can be found in section~\ref{sec:discsummary} and~Table~\ref{table:discussion}.
\end{itemize}

A few broad implications of the section-\ref{sec:afterglow} results merit additional discussion. The spectral stability properties in that section stem from the similar shape of the Klein-Nishina reconnection IC spectrum to that produced when an initial population of particles is allowed to passively cool through IC emission and pair production. This introduces a fundamental degree of degeneracy in efforts to leverage astrophysical observations to learn about plasma physical particle acceleration processes. Namely, any Klein-Nishina-coupled particle accelerator with an IC spectrum resembling that to which the radiative physics inevitably relaxes (i.e.\ after the accelerator is turned off) may yield the same spectral stability as reconnection. However, this degeneracy is, thankfully, incomplete, because acceleration processes with softer intrinsic spectra would instead likely yield an asymmetry between the rising side of a gamma-ray flare (where the steep particle acceleration spectrum would be visible) and the decaying phase (where the shallower Klein-Nishina-decaying spectrum would probably dominate).

In the face of such degeneracy, observations of temporally resolved spectra provide precious information. Already in the present study, such spectra provide crucial and obvious distinctions between Thomson and Klein-Nishina reconnection -- two regimes whose time-averaged spectra have practically identical photon indices -- and, in the case of FSRQs (section~\ref{sec:discussion}), supply compelling evidence for the operation of the latter over the former. In the future, time-resolved spectra may be necessary to distinguish among different possible Klein-Nishina-coupled particle acceleration mechanisms. Hence, the present work strongly motivates collection of time-resolved (versus simply time-averaged) spectra whenever possible, as well as new instruments for which such observations are more frequently feasible.

This study also lays the groundwork for a broad range of further theoretical exploration. As detailed in section~\ref{sec:discussion}, future efforts can target the~M87$^*$ magnetosphere by explicitly treating synchrotron radiative cooling; firmer connections to gamma-ray binaries can be made by considering global properties of irradiated striped pulsar winds. Additional mode\ling[]may also verify the potential existence, outlined here (section~\ref{sec:pairyield}), of a Klein-Nishina reconnection regime featuring copious electron-positron pair production, which would have important implications for systems with a high fiducial cold magnetization,~$\sigc \gg \gamppmax / 4$. However, even with a more modest order-unity pair yield (r\ealiz[ed,]in our simulations, when~$\gamppmax / 4 \sigc \lesssim 0.25$), Klein-Nishina reconnection could still function as an important source of \textit{in situ} antimatter, taking, for example, an initially electron-proton plasma and injecting a prominent positron count. Such potential modifications to the composition of the reconnection plasma environment can be rigorously probed by applying the Klein-Nishina radiative physics studied here to the case of an initial electron-ion plasma.

More broadly, the theoretical framework established by this work (and the preceding article, \citealt{mwu21}), especially the systematic recasting of Klein-Nishina physics as dimensionless energy scales, provides a paradigm for coupling the same physics to other particle acceleration processes (e.g.\ turbulence and shocks). Furthermore, the developed numerical technology provides an infrastructure within which numerical experiments concerning such processes can be carried out. Thus, this study serves as one of a growing number of theoretical and technological stepping stones \citep[e.g.][]{hps19, hrp23, ccp20, ccd21, ccd22, sgu19, sgu23, cgc21, cud23, ghb23} toward a richer understanding of QED-coupled plasma physics in high-energy astrophysical environments.

\section*{Acknowledgements}
The authors gratefully acknowledge Guillaume Dubus, Gilles Henri, and Hayk Hakobyan for stimulating discussions.
This project has received funding from the European Research Council (ERC) under the European Union’s Horizon 2020 research and innovation programme (grant agreement No 863412).
This work was also supported by NASA and the NSF, grant numbers NASA ATP NNX17AK57G, NASA ATP 80NSSC20K0545, NASA ATP 80NSSC22K0828, NSF AST-1806084, and NSF AST-1903335.

\section*{Data Availability}
The simulation data underlying this article were generated at the XSEDE/TACC Stampede2 supercomputer and are archived at the TACC/Ranch storage facility. As long as the data remain in the archive, they will be shared upon reasonable request to the corresponding author.

\bibliographystyle{mnras}
\bibliography{ref}

\appendix
\section{Likelihood of limit cycles}
\label{sec:pairreg}
Here we expand on section~\ref{sec:origprodcompare}, using measurements from our IC(KN)+PP run (section~\ref{sec:setup}) to fill in the main quantitative uncertainties from the work of \citet{mwu21} on the pair-loading efficiency -- the parameter~$\xi$ in what follows -- including its implications for the pair-loaded magnetization,~$\sighgen^*$, and on the possibility of~$\sighgen$-mediated limit cycles. We report first the expressions for the produced-particle energy density,~$U_{\rm prod.}$, and the pair-loaded hot magnetization,~$\sighgen$, flowing into the reconnection layer from the upstream region (thus, in the context of Fig.~\ref{fig:spatialpairdom}, both quantities should be evaluated along, or perhaps just upstream of, the reconnection separatrix). \citet{mwu21} found that~$U_{\rm prod.}$ evaluated at this location can be written as
\begin{align}
    U_{\rm prod.} = \xi \mathcal{F} \frac{B_0^2}{8 \pi} \, .
    \label{eq:uprodloaded}
\end{align}
Here,~$\mathcal{F}$ is the fraction of the IC power, emitted by particles in the reconnection layer, that is radiated above pair-production threshold, and~$\xi$ is the pair-loading efficiency, which we refer to from here onward according to a more precise name signifying its role in equation~(\ref{eq:uprodloaded}), the \textit{energy recapture efficiency}: how much of the above-threshold radiation is recaptured by the reconnection layer from the inflow region as hot pairs. Accounting for the newborn pair energy density~$U_{\rm prod.}$, the hot magnetization,~$\sighgen$, can be written in terms of~$\mathcal{F}$ and~$\xi$ as
\begin{align}
    \sighgen = \frac{\sigh}{1 + 2 \mathcal{F} \xi \sigh / 3} \simeq \frac{3}{2 \mathcal{F} \xi} \, ,
    \label{eq:sighloaded}
\end{align}
where the approximation holds when~$\mathcal{F} \xi \sigh \gg 1$. Provided that the reconnection layer responds only to the present -- as opposed to the past -- magnetization of the plasma feeding it, equation~(\ref{eq:sighloaded}) encodes a universal~($\sigh$-independent) \textit{pair-loaded magnetization},
\begin{align}
    \sighgen^* \equiv \frac{3}{2 \mathcal{F} \xi} \, ,
    \label{eq:sighuniv}
\end{align}
which we first introduced qualitatively in section~\ref{sec:origprodcompare}. The pair-loaded magnetization~$\sighgen^*$ is determined self-consistently by the (potentially~$\sighgen^*$-dependent) values of~$\mathcal{F}$ and~$\xi$. As discussed in section~\ref{sec:origprodcompare}, any Klein-Nishina reconnection layer with initial magnetization~$\sigh > \sighgen^*$ will try to self-regulate to~$\sighgen^*$, but, if the self regulation is too efficient, the system will enter a limit cycle between a high and low magnetization on either side of~$\sighgen^*$.

One major uncertainty in the model developed by \citet{mwu21} was the energy recapture efficiency,~$\xi$. This parameter is particularly important for two reasons. First, it is one of the key quantities deciding the pair-loaded magnetization,~$\sighgen^*$. Second, it determines how strongly the reconnection layer is coupled, via the upstream plasma, to its own NTPA, and, in particular, whether this coupling is strong enough to drive the system into a limit cycle. For a wide range of assumptions,~\citet{mwu21} found that limit cycles require~$\xi$ of order unity -- at least~$0.3$ or so in the most lenient case, but often even higher. In order to make contact with these two issues -- the value of~$\sighgen^*$ and the existence (or not) of limit cycles -- we now proceed to measure~$\xi$ and~$\sighgen^*$ from our simulation.

We estimate the above-threshold power fraction,~$\mathcal{F}$, as the part of the average IC emission spectrum in Fig.~\ref{fig:icemitdists} radiated above pair-production threshold, measuring~$\mathcal{F} \simeq 0.5$. This agrees with the prediction one arrives at by taking, from Fig.~\ref{fig:ptcldists}, a particle distribution power-law,~$\dif N / \dif \gamma \propto \gamma^{-2}$, with a \cutoff[]at~$200 \gkn$ and consulting the corresponding~$\mathcal{F}$-value from fig.~$10$ of \citet{mwu21}. Armed with~$\mathcal{F}$, we can estimate~$\xi$ by measuring~$U_{\rm prod.}$ along the reconnection separatrix in Fig.~\ref{fig:spatialpairdom} and plugging the result into equation~(\ref{eq:uprodloaded}). From the figure, we see that, once the pair coat is built up around the reconnection layer, it presents an inflowing produced-particle energy density of about~$0.01 B_0^2 / 8 \pi$ at the separatrix crossing. This implies that~$\xi \mathcal{F} \sim 0.01$ and, hence,~$\xi \sim 0.02$. Finally, inserting~$\xi \mathcal{F} \sim 0.01$ into equation~(\ref{eq:sighuniv}) yields~$\sighgen^* \sim 150$. This is about a factor of~$10$ smaller than~$\sigh = 1250$ in the IC(KN)+PP run, and, hence, the contour~$\sighgen=0.1\sigh$ in Fig.~\ref{fig:spatialpairdom} nearly overlaps the~$U_{\rm prod.} = 0.01 B_0^2/8\pi$ contour.\footnote{In fact, the~$\sighgen=0.1\sigh$ contour lies slightly farther from the main reconnection X-point than the one for~$U_{\rm prod.} = 0.01 B_0^2/8\pi$. The small discrepancy originates from the in-plane magnetic field slightly weakening near reconnection X-points, an effect visible in the cold magnetization maps of Fig.~\ref{fig:spatialpairdom}. This effect owes to the inflowing magnetic field lines draping themselves across plasmoids, hanging from them like the cables of a suspension bridge, and, hence, thinning out as they sink toward the X-points in between.}

According to \citet{mwu21}, our measured value of~$\xi$ is much too low (by about an order of magnitude) for limit cycles to occur: the coupling between the inflow region and the layer is too mild. This is consistent with our simulations, from which we identify no evidence of cyclic behavi\spellor[,]neither in the reconnection-powered NTPA, nor in the pair loading and resulting magnetization,~$\sighgen$, presented to the reconnection layer.

Let us now examine how our measured value of~$\xi$, including its implications on the existence of limit cycles, may g\eneraliz[e]under changes of the reconnection parameters (such as to those of real astrophysical systems). To inform this discussion, we s\ummariz[e]here the basic physics that determines~$\xi$. \citet{mwu21} explain that~$\xi$ can be written as the product,~$\xi = f_{\rm inj} f_{\rm nocool} f_{\rm noesc}$, where the three factors on the right-hand-side correspond to the three main loss mechanisms that inhibit above-threshold radiation from being recaptured by the reconnection layer from the upstream region. First,~$f_{\rm inj}$ is the fraction of the layer-produced above-threshold radiation that successfully traverses the reconnection separatrix into the upstream region. The factor~$f_{\rm inj}$ falls below unity when some photons -- for example, in the case of extreme beaming of reconnection-accelerated particles along the reconnection layer \citep[e.g.][]{cub12, cwu12, cwu13, cwu14b, cwu14a, mwu20} -- produce pairs in the downstream region.\footnote{\citet{mwu21} did not include~$f_{\rm inj}$ explicitly in their model, but discuss its effect in their appendix~C.} Second,~$f_{\rm nocool}$ is that part of the energy deposited into the upstream region as newborn pairs that is not radiated away while those pairs are readvected toward the layer. Finally,~$f_{\rm noesc}$ is the fraction of newborn upstream particles that do not escape the system (e.g.\ by traveling along a field line in the~$\pm x$-directions) before being swept (in the~$\pm y$-directions) into the layer.

To determine how~$\xi$ may change with the reconnection parameters, we sketch the dependence of the three governing factors~$f_{\rm inj}$,~$f_{\rm nocool}$, and~$f_{\rm noesc}$, on these parameters as revealed both by analytic theory \citep{mwu21} and by our simulations. Given the periodic boundaries of our setup, particle escape is impossible. Hence,~$f_{\rm noesc}$, perhaps artificially, equals~$1$ in the simulations, which are then restricted to probing~$f_{\rm nocool}$ and~$f_{\rm inj}$. Of these two, \citet{mwu21} provide analytic estimates of~$f_{\rm nocool}$, showing that, similarly to~$\mathcal{F}$, it depends only on the shape of the reconnection-energized particle distribution -- for a power-law,~$\dif N / \dif \gamma \propto \gamma^{-p}$, on the index,~$p$, and on the high-energy \cutoff[,~$\gamc$,]n\ormaliz[ed]by~$\gkn$. Those authors found that, for~$p > 2$,~$f_{\rm nocool}$ is contro\led[]by the low-energy particles, and its value,~$f_{\rm nocool} \sim 0.1$, is therefore independent of~$\gamc$. For~$p<2$,~$f_{\rm nocool}$ acquires a weak dependence on~$\gamc$ -- signa\ling[]the enhanced importance of particles in the high-energy tail -- but even then does not reach order unity unless~$\gamc \gg 10^3\gkn$ and~$p\simeq1$. In the context of our IC(KN)+PP run, the \citet{mwu21} estimates (e.g.\ their fig.~B2) suggest that~$f_{\rm nocool} \sim 0.1$ for~$p\simeq 2$ and~$\gamc \sim 200\gkn$ (as in Fig.~\ref{fig:ptcldists}). If we adopt this value, we can derive the empirical measurement,~$f_{\rm inj} \sim \xi / f_{\rm nocool} \sim 0.2$.

Now, while~$f_{\rm nocool}$ depends on the details of reconnection-powered NTPA,~$f_{\rm noesc}$ and~$f_{\rm inj}$ are dictated instead mostly by the large-scale geometry and kinematics of reconnection. For example,~$f_{\rm noesc}$ depends on the free-streaming time,~$\leq L/c$, for a newborn upstream particle to vacate the system by following an unreconnected magnetic field line (a more complete discussion of the factors influencing~$f_{\rm noesc}$ can be found in the appendix~C of \citealt{mwu21}). As another example, the factor,~$f_{\rm inj}$, depends on the effective width of the reconnection layer and on beaming, both that associated with bulk (e.g.\ plasmoid-chain) motion and that stemming from kinetic effects near reconnection X-points. These remarks suggest that~$f_{\rm noesc}$ and~$f_{\rm inj}$ are constants (at least for~$\sigh \gg 1$), because the processes deciding them are either generic byproducts of relativistic reconnection \citep[in the case of kinetic beaming;][]{cwu12, mwu20} or tied to the large-scale evolution (e.g.\ of the plasmoid chain), which seems rather insensitive to Klein-Nishina radiative physics, as discussed in sections~\ref{sec:globaldynamics} and~\ref{sec:recrate}.

This, then, makes clear the utility of introducing the individual factors~$f_{\rm inj}$,~$f_{\rm nocool}$, and~$f_{\rm noesc}$: two of them,~$f_{\rm noesc}$ and~$f_{\rm inj}$, are expected to be roughly constant regardless of the exact reconnection parameter values (as long as we are in the Klein-Nishina reconnection regime), and the remaining factor,~$f_{\rm nocool}$, is one that we can estimate based on a substantially reduced set of parameters -- the NTPA power-law index,~$p$, and \cutoff[]energy,~$\gamc$. This enables us, as we seek to extrapolate our simulation results to reason about~$\xi$ and~$\sighgen^*$ in astrophysical Klein-Nishina reconnection, to skip an exhaustive exploration of the high-dimensional radiative reconnection parameter space (e.g.\ all orderings of the energy scales,~$\sigcgen$,~$\gmax$,~$\gradt$,~$\gcool$,~$\gkn$, etc.). Instead, we can focus on just~$p$ and~$\gamc$. While it is true that these two are still determined self-consistently by the much larger underlying parameter space, we leave a quantitative c\haracteriz[ation]of this dependence to future work, having already noted several general trends in section~\ref{sec:ptcldists}.

The established logical framework in terms of~$f_{\rm inj}$,~$f_{\rm nocool}$, and~$f_{\rm noesc}$ -- with just the two independent variables~$p$ and~$\gamc$ -- equips us to estimate a global upper bound on~$\xi$. In the above-mentioned case of extreme NTPA wherein~$p\simeq1$,~$\gamc\gg10^3 \gkn$ and, thus,~$f_{\rm nocool} \sim 1$, the combined factor,~$\xi = f_{\rm nocool} f_{\rm inj} f_{\rm noesc}$, reduces to~$\xi \sim f_{\rm inj} f_{\rm noesc} \leq f_{\rm inj} \sim 0.2$. Thus, even in the most favorable circumstances,~$\xi$ is still smaller than the most lenient minimum required value,~$0.3$, found by \citet{mwu21} to produce limit cycles. We e\mphasiz[e]the critical role played by simulations in reaching this result, for they provide the necessary bound on the factor,~$f_{\rm inj}$, that caps~$\xi$ to beneath the limit cycle value range. In view of this combined input from analytic theory and simulations, we speculate that~\textit{$\sighgen$-mediated limit cycles are unlikely, even in astrophysical instances of Klein-Nishina reconnection.}

Finally, let us comment on how varying system parameters may impact the pair-loaded magnetization,~$\sighgen^* = 3 / 2 \xi \mathcal{F}$. The discussion thus far can be translated into an expected lower bound on~$\sighgen^*$ as follows. We have already seen that extreme NTPA~($p\simeq1$ and~$\gamc \geq 10^3 \gkn$) yields~$f_{\rm nocool} \sim 1$ and, if one also favorably posits~$f_{\rm noesc} = 1$, a global maximum,~$\xi \sim 0.1$. In the same NTPA regime, the above-threshold power fraction,~$\mathcal{F}$, attains order unity (fig.~10 of \citealt{mwu21}). Thus, with~$\xi \mathcal{F}$ as large as possible given numerical measurements and analytic expectations, we arrive at the \textit{minimum possible} pair-loaded magnetization,~$\sighgen^* \simeq 3 / (2 \times 0.1 \times 1) \sim 15$.

If we now relax these extreme assumptions and allow for the (likely) possibility that NTPA is not quite as efficient, we may estimate a more typical value of~$\sighgen^*$. As explained previously, once the power-law index of the particle distribution softens and/or the high-energy \cutoff[,~$\gamc$,]falls below~$\sim 10^3 \gkn$,~$f_{\rm nocool}$ loses dependence on these two parameters and becomes closer to~$0.1$. In the same regime (again referring to fig.~10 of \citealt{mwu21}),~$\mathcal{F}$ becomes more strongly dependent on NTPA, but still attains~$\sim 0.5$ when~$p\leq2$ for a broad range of~$\gamc$. This gives a more typical pair-loaded magnetization (r\ealiz[ed]in our simulation) of~$\sighgen^* \sim 150 (0.5 / \mathcal{F})(1/f_{\rm noesc})$.

To s\ummariz[e,]our IC(KN)+PP run allows us to measure the all-important energy recapture efficiency,~$\xi$, revealing a value that is too low, in the context of the work by \citet{mwu21}, to expect violent limit cycles mediated by pair regulation of~$\sighgen$. Such limit cycles are consequently unlikely in real astrophysical systems. The expected evolutionary pathway taken by astrophysical Klein-Nishina reconnection (as long as the newborn pairs remain few in number) is then the same as that of our main IC(KN)+PP run: provided~$\sigh > \sighgen^*$, the system will self-regulate directly to the~$\sigh$-independent value,~$\sighgen^*$, without overshooting. However, if~$\sigh < \sighgen^*$, the system remains at its initial magnetization.

\bsp	
\label{lastpage}
\end{document}